\documentclass{article}

\usepackage{arxiv}

\usepackage[utf8]{inputenc} 
\usepackage[T1]{fontenc}    
\usepackage{hyperref}       
\usepackage{url}            
\usepackage{booktabs}       
\usepackage{amsfonts}       
\usepackage{amssymb}
\usepackage{amsthm}
\usepackage{amsmath}
\usepackage{newtxtext,newtxmath}
\usepackage{textcomp}
\usepackage{mathcomp}

\usepackage{nicefrac}       
\usepackage{microtype}      
\usepackage{cleveref}       
\usepackage{lipsum}         
\usepackage{graphicx}
\usepackage{natbib}
\usepackage{doi}
\usepackage{subfigure}
\usepackage{tabularx}
\usepackage{rotating}
\usepackage{color}
\usepackage{soul}

\title{Aerodynamic loads on groups of offshore wind turbine towers stored on quaysides during the pre-assembly phase}

\author{\href{https://orcid.org/0000-0002-3866-879X}{\includegraphics[scale=0.06]{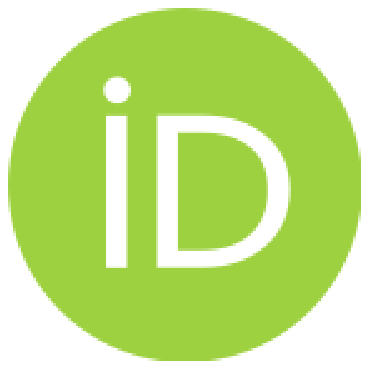}\hspace{1mm}Claudio Mannini}\\
	CRIACIV/Department of Civil and Environmental Engineering\\
	University of Florence\\
	50139 Florence, Italy \\
	\texttt{claudio.mannini@unifi.it} \\
	\And
	\href{https://orcid.org/0000-0002-6817-5910}{\includegraphics[scale=0.06]{orcid.eps}\hspace{1mm}Tommaso Massai} \\
	CRIACIV/Department of Civil and Environmental Engineering\\
	University of Florence\\
	50139 Florence, Italy \\
	\texttt{tommaso.massai@unifi.it} \\
	\And
	 \href{https://orcid.org/0000-0001-5984-7124}{\includegraphics[scale=0.06]{orcid.eps}\hspace{1mm}Andrea Giachetti} \\
	 CRIACIV/Department of Civil and Environmental Engineering\\
	 University of Florence\\
	 50139 Florence, Italy \\
	 \texttt{andrea.giachetti@unifi.it} \\
	 \And
	 \href{https://orcid.org/0000-0002-6148-1704}{\includegraphics[scale=0.06]{orcid.eps}\hspace{1mm}Alessandro Giusti} \\
	 Siemens Gamesa Renewable Energy A/S\\
	 Brande, Denmark \\
	 \texttt{alessandro.giusti@siemensgamesa.com } \\
}

\renewcommand{\shorttitle}{Aerodynamic loads on groups of offshore wind turbine towers stored on quaysides}

\hypersetup{
pdftitle={Aerodynamic loads on groups of offshore wind turbine towers stored on quaysides during the pre-assembly phase},
pdfsubject={physics.flu-dyn},
pdfauthor={Claudio Mannini, Tommaso Massai, Andrea Giachetti, Alessandro Giusti},
pdfkeywords={Wind turbine towers, tower groups, high Reynolds number, circular cylinder, finite-length cylinders, wind tunnel tests},
}

\begin{document}
\maketitle

\begin{abstract}
Offshore wind turbine towers are pre-assembled and temporarily held in close proximity to each other in group arrangements on port quaysides during which time they are highly sensitive to wind action. Accurate estimates of the aerodynamic loads on the individual towers and on the overall group are therefore essential for the safe and economic design of the quayside’s supporting structures and foundations. Given the many possible group configurations, the key role played by wind direction and the limited literature on this topic, wind tunnel tests represent the main way to address this issue. The problem is that such tests tend to lead to overconservative designs due to inevitable mismatches in the Reynolds number, which is typically subcritical in experiments and transcritical at full scale.
This crucial issue is dealt with here using an original engineering solution based on concentrated but discontinuous surface roughness, which allows, for the first time in the case of finite-height towers arranged in groups and subjected to an atmospheric boundary layer flow, the successful simulation of the target high Reynolds number regime.
This case study assumes slender wind turbine towers with a height of 115~m, and for the sake of generality, the investigations focus principally on a cylindrical shape rather than the more complex real-world geometry. The constant diameter of the towers is determined based on the theoretical equivalence of the mean overturning moment. The rationality of this procedure is verified a posteriori using a set of measurements on the real-shape towers.
The experiments show a regular behavior of the maximum mean base shear force and moment for towers arranged in double-row groups, while the results are more complicated for the heavily loaded single-row groups; indeed, despite the simulated transcritical regime and the turbulent wind profile, biased flow sometimes occurs in symmetric or nearly-symmetric configurations.
Dynamic loads are also inspected, and gust factors in good agreement with Eurocode~1's prescriptions are found.
Several parametric studies are carried out, the most extensive of which is devoted to assessing the role of tower height. A complicated non-monotonic pattern of the load coefficients with the tower height is encountered; therefore, the use of simple correction coefficients for practical design purposes must be handled with care. Moreover, the analysis reveals end-effect factors non-negligibly higher than those suggested by Eurocode~1 and ESDU's recommendations.
\end{abstract}


\section{Introduction}
\label{Intro}

Offshore wind turbine towers are usually temporarily placed on freestanding foundations on port quaysides for the pre-assembly operations during which the components (tower sections, nacelles, blades) are prepared for their final installation in offshore wind farms using special installation vessels. Tower sections are typically assembled at the quayside in groups in a square grid arrangement at small center-to-center distance to facilitate crane operations, especially during loading onto the vessels. Typical examples of arrangements are groups of $2 \times 2$, $2 \times 3$, $2 \times 4$ towers, or a line of a number of towers. Pre-assembly activities usually last 6 to 12 months. Towers can be over 100~m tall, weigh more than 600~tonnes, and have base diameters in excess of 7~m.

Although towers are generally designed for the much more severe conditions, found in offshore wind farms, and a working lifetime of at least 20 years, it goes without saying that aerodynamic interaction between towers in groups is highly relevant to the design of the foundation on which each tower is erected on the quayside. This is typically a steel gravity-based foundation that guarantees its reusability for other projects and avoids permanent changes to the quay structures (ports are usually rented and must be left in the condition in which they were rented out). The resultant static and dynamic wind actions at the base of the towers are therefore of great engineering interest.

Besides the complexity of wind-structure interaction of groups of finite-length cylinders associated with the properties of the wind flow, e.g., mean wind profile and turbulence intensity/spectrum, and their variability from site to site, the problem is even more complex due to the variability of the geometry. Indeed, wind turbine towers are usually tapered structures (with relatively small taper angle, often in addition to some cylindrical parts) with reference diameter and height that change from project to project and from generation to generation. The center-to-center spacing is another parameter to be considered. Moreover, as the towers are assembled section by section, additional temporary configurations with incomplete towers and incomplete groups increase the number of possibilities.

In terms of Reynolds number, typical effective values are close to or greater than $10^7$ (based on a characteristic tower diameter), which makes wind tunnel campaigns very challenging because of the practical impossibility of reproducing the correct Reynolds regime without the use of special stratagems, such as the use of technical surface roughness, to anticipate the transition from critical to transcritical Reynolds regimes. As a valid completion, full-scale measurements could be used for the validation and calibration of wind tunnel tests on a limited number of measured cases, whereas the sole use of full-scale measurements seems unrealistic because of the uncontrolled boundary conditions and the consequent practical impossibility to test all the cases.

The literature does not adequately cover this important engineering problem, as most of the available results refer to groups of infinite circular cylinders in smooth flow (e.g., \cite{Zdravkovich1983, Price84, Sayers88, Sumner2010, Schewe2019, Schewe2021, Andrianne2022}) rather than finite towers in turbulent shear flow.
Studies on isolated finite-length circular cylinders have been performed in various flow conditions, often changing the slenderness ratio (e.g., \cite{Farivar81,Sarode81,Ayoub82,Kawamura84,Baban91}).
\cite{Sarode81} considered two finite-length cylinders in tandem subjected to a boundary layer flow. They found that the drag on the second cylinder can be negative for a very small center-to-center distance, and that the influence of the windward cylinder is still visible when it is eight diameters upstream. 
Experiments on finite-length tandem cylinders in uniform flow and subcritical Reynolds number regime were also performed in \cite{Luo1996}.
\cite{Kareem98} extensively studied in the wind tunnel the interference effects in groups of cylindrical towers at subcritical Reynolds number exposed to an open country terrain wind profile. Pressures were measured on models with a slenderness ratio (i.e., height-to-diameter ratio) equal to 10. Single rows of either two or three cylinders were considered, along with a group of three towers in a staggered arrangement. A variable center-to-center spacing between two and seven diameters was tested. For the line arrangements, the maximum mean drag was found for a four-diameter spacing. Large lift and drag fluctuations were also observed when a cylinder is buffeted by the wake of the upstream cylinder, especially for azimuthal inclination of the flow of about $20$~deg with respect to the tower line and spacings between three and four diameters.
\cite{Mitra2006} considered scale models of three cylindrical stacks in an along-wind linear, a side-by-side
and a triangular arrangement. The stacks had a slenderness ratio of 11, a center-to-center spacing of 1.14 times the diameter, and were exposed to a sub-urban terrain wind profile in a subcritical Reynolds number regime. The distribution of the loads between the cylinders was studied in the various configurations.
\cite{Sun2020} investigated the interference effects between two tall chimneys with circular cross section, a slenderness ratio of 17, and a subcritical Reynolds number. A low-turbulence wind profile was reproduced in the wind tunnel, and pressures were measured for different wind directions, varying the distance between the chimneys from two to six diameters. The increase of the across-wind loads compared to the isolated chimney was found to be higher than that of the along-wind loads.

In general, it is clear that the number of studies on groups of finite-length cylinders in boundary-layer flows are very limited and, to the authors' best knowledge, the experiments are all performed in the subcritical Reynolds number regime. 
The present work reports for the first time the results of a broad experimental wind tunnel campaign carried out on a large number of configurations of towers arranged in groups, successfully simulating the target high Reynolds number regime through an original technical surface roughness solution. To increase the generality of the investigation, equivalent cylindrical towers are mainly addressed, although some tests were also devoted to real-shape towers. The wind loads are characterized in terms of resultant shear-force and overturning-moment coefficients at the base of each tower in a group and of the overall group.
An extensive parametric study is also discussed, focusing not only on the shape of the tower but also on the impact of a slight change in the oncoming turbulent wind profile, of a variation in tower height, and of the presence of a certain number of partially assembled towers.
In contrast, though very relevant for this engineering problem, the possible narrow-band dynamic excitation phenomena and aeroelastic instabilities are outside the scope of the current paper.

\section{Case study}

\subsection{Reference tower and tower groups}

The problem was addressed based on a specific case study, considering as a reference the geometry of a modern offshore wind turbine tower. This presents a height of 115~m and a diameter varying from 7~m at the base to slightly less than 5~m at the top, alternating cylindrical segments to linearly tapered ones. The center-to-center distance $d$ between the towers is the same for all the group arrangements considered and equal to 10~m.

\begin{figure}
	\centering
	\includegraphics[angle=0, width=0.9\textwidth]{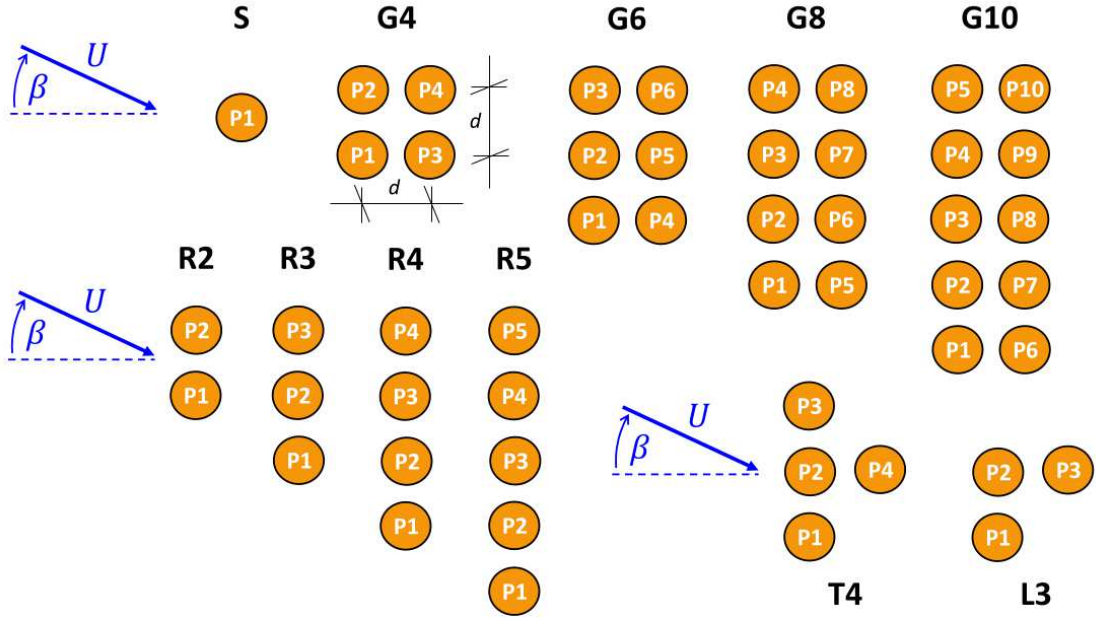}\vspace{0.5cm}%
	\caption{Tower group configurations tested in the framework of the present work and adopted nomenclature.\\ $U$ indicates the mean wind velocity and $\beta$ its direction.}
	\label{fig:configurations}
\end{figure}

As previously mentioned, many tower configurations are possible during the pre-assembly phase on the quayside; those shown in Fig.~\ref{fig:configurations} were considered in this study.
Double-row groups with up to ten towers are common configurations (G4, G6, G8 and G10) because they guarantee the optimization of crane operations, stability and ground pressures. Nevertheless, for very large wind turbines, cranes with a remarkable lifting capacity may be necessary for the second-row towers. Single-row group arrangements (R2, R3, R4 and R5) can then become convenient, although larger wind loads are expected in the most loaded tower of the group and foundations are less efficient against stability and ground pressures (keeping fixed the center-to-center spacing and the foundation area per tower position). Finally, more complicated L- or T-shape configurations can be envisaged in some cases (either as transitory configurations or reflecting the layout of the grillage of the installation vessel). For the sake of brevity, only a selection of results are reported in Sections~\ref{baseline_results}-\ref{parametric_results}, focusing on the more common and representative tower arrangements.

\subsection{Equivalent cylindrical tower}
\label{Equivalent_diameter}

For the sake of generality, the vast majority of tests were carried out on cylindrical towers with an \textquotedblleft equivalent\textquotedblright \,diameter, although some results for the real tower shape will also be presented in Section~\ref{tower_shape}.
The determination of the equivalent diameter $D_{eq}$ is based on two simplifying hypotheses, namely the strip assumption and the assumption that the mean drag force per unit length along the tower is proportional to the square of the mean wind speed at the considered height. Moreover, since the resultant moment at the base of the tower is the quantity of major interest for design purposes, the following expression is obtained for $D_{eq}$:
\begin{equation}
 D_{eq} = \frac{\int_0^H U^2(z) D(z) z dz}{\int_0^H U^2(z) z dz}
\label{eq:equivalent_diameter}
\end{equation}
where $H$ is the tower height, $D$ is the tower diameter variable along the $z$-axis, and $U$ is the mean wind velocity. Moreover, it is assumed that the base of the tower is at the ground level (the foundation is not modeled in the wind tunnel tests).
Assuming Eurocode~1's mean wind profile for a terrain category I \citep{Eurocodice1}, one obtains $D_{eq} = 6.55$~m. Obviously, a slightly larger equivalent diameter (6.70~m) would have been obtained if one focused on the resultant shear force at the base of the tower. The validity of the procedure followed here will be verified a posteriori based on the experimental results for the equivalent cylindrical and real-shape towers.

Based on the equivalent diameter of the tower, the aspect ratio of the tower $H/D_{eq}$ is 17.56, while the nondimensional center-to-center distance between the towers is $d/D_{eq}$ = 1.527.

\section{Wind tunnel experiments}

\subsection{Facility}

The experimental campaign was carried out in the CRIACIV (Inter-University Research Center on Building Aerodynamics and Wind Engineering) boundary layer wind tunnel in Prato, Italy.
The overall length of the open-circuit facility is about 22~m, while 11~m is the  fetch available to develop the boundary layer flow. The closed test section is 2.4~m wide and 1.6~m high.
Air is drawn by a motor with a nominal power of 156~kW, and the flow speed can be varied continuously up to about 30~m/s.
In the absence of turbulence generating devices, the residual turbulence is slightly lower than 1\%.

\subsection{Physical models}
\label{models}

Based on the blockage ratio associated with the largest tower group configurations, the atmospheric boundary layer profiles that can be reproduced in the CRIACIV wind tunnel, and the need to work with the largest possible models for manufacturing reasons and above all not to reduce too much the nominal Reynolds number, a geometric scale of 1:187 was chosen for the tests. This means that the models of the towers had a height of 615~mm, and the scaled equivalent diameter was 35~mm.
The corresponding maximum blockage ratio, calculated considering the gross projected area of the group (voids in between towers are considered solid) for the worst configuration, is 4.1\%.

\begin{figure}
	\centering
	\subfigure[\label{fig:carbon-fibre model}]
  {\includegraphics[angle=0, height=5cm]{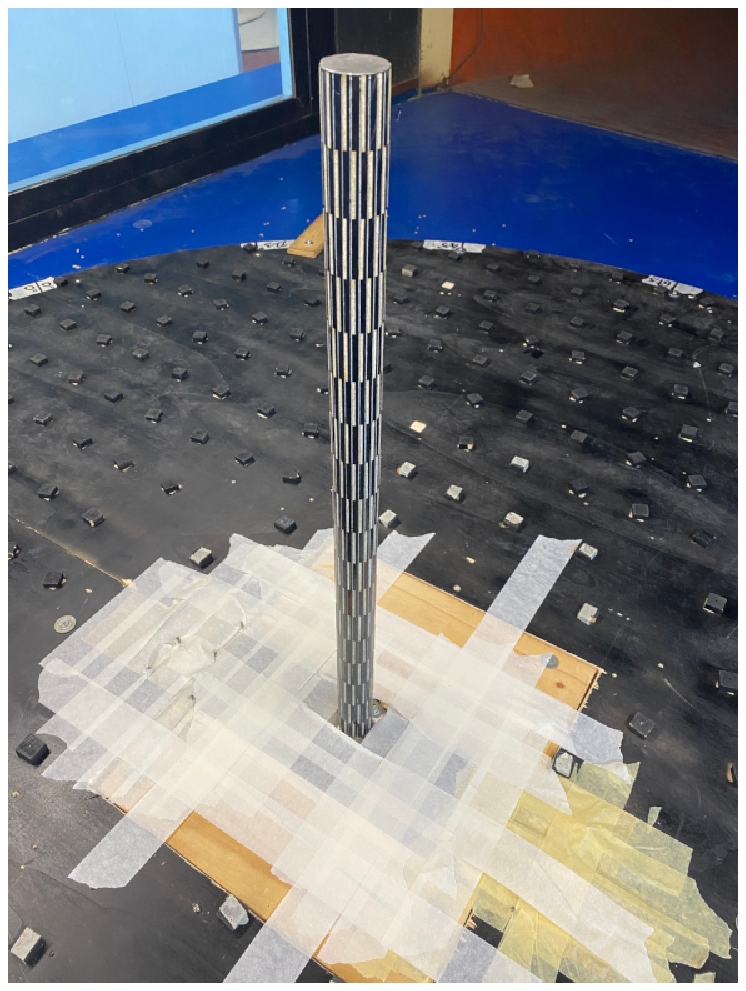}}\hspace{0.5cm}%
	\subfigure[\label{fig:ABS model}]
  {\includegraphics[angle=0, height=5cm]{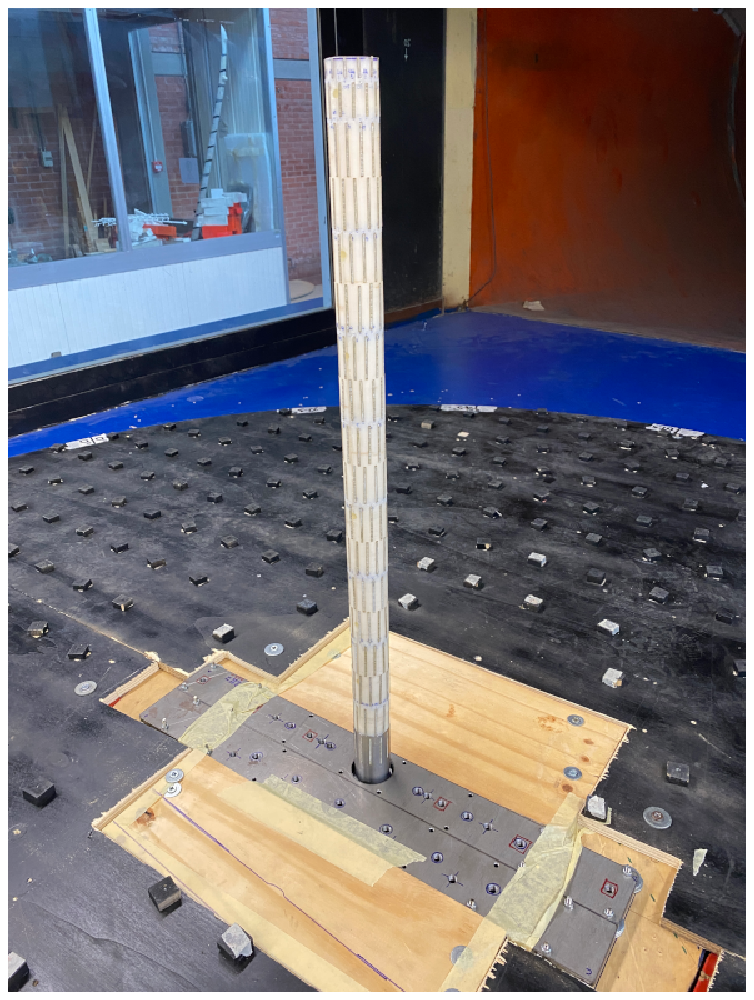}}\hspace{0.5cm}%
	\subfigure[\label{fig:steel model}]
  {\includegraphics[angle=0, height=5cm]{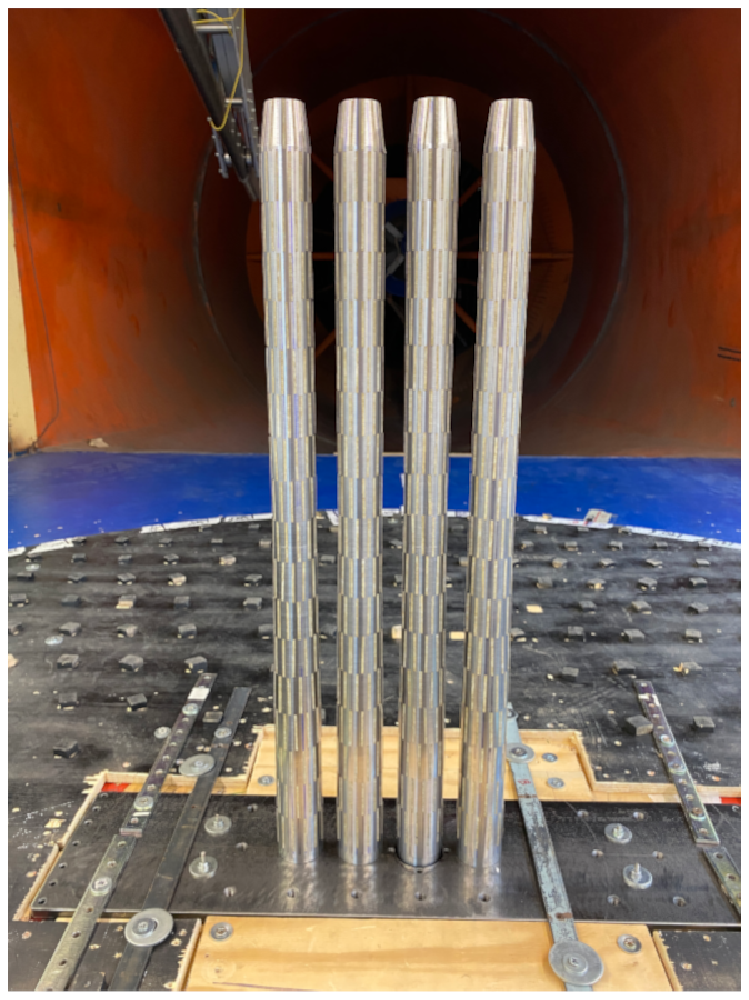}}%
	\caption{(a) carbon-fibre model for force measurements; (b) 3D-printed ABS model for pressure measurements; (c) steel models of real-shape towers.}
	\label{fig:model_pictures}
\end{figure}

Given that not only force measurements at the base of the towers are required in the present study but in some cases also pressure measurements along their height (see Section~\ref{exp_validation}), different types of models were made. Indeed, monolithic models of the towers were constructed in carbon fibre (Fig.~\ref{fig:carbon-fibre model}) to obtain a high first natural vibration frequency, thus allowing for dynamic force measurements. Moreover, a two-part model made of acrylonitrile butadiene styrene (ABS) was 3D-printed and equipped with pressure taps and Teflon tubes (Fig.~\ref{fig:ABS model}). In this case, the model natural frequency was significantly lower, thus preventing accurate dynamic force measurements with a force balance.
All of the dummy models (non-instrumented towers) in the group configurations were always made of carbon fibre.

Finally, very stiff lathed steel models (Fig.~\ref{fig:steel model}) were made for the tests on real-shape towers (see Section~\ref{tower_shape}).


\subsection{Set-up for single-tower measurements}
\label{single_tower_setup}

\begin{figure}
	\centering
	\subfigure[\label{fig:single-tower set-up}]
  {\includegraphics[angle=0, height=6cm]{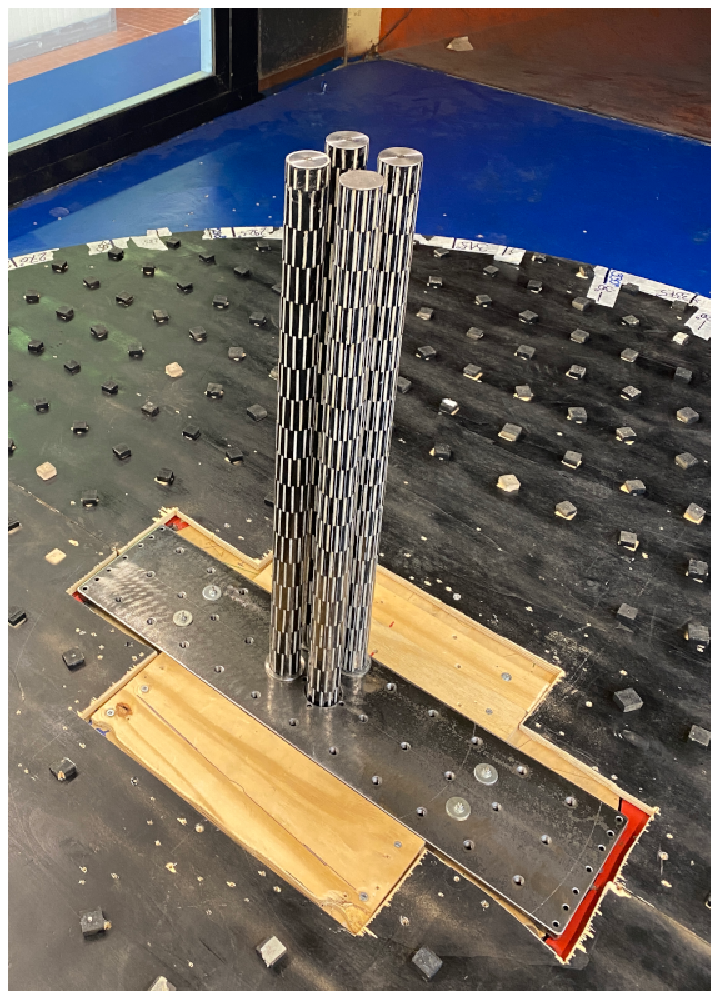}}\hspace{1.0cm}%
	\subfigure[\label{fig:tower-group set-up}]
  {\includegraphics[angle=0, height=6cm]{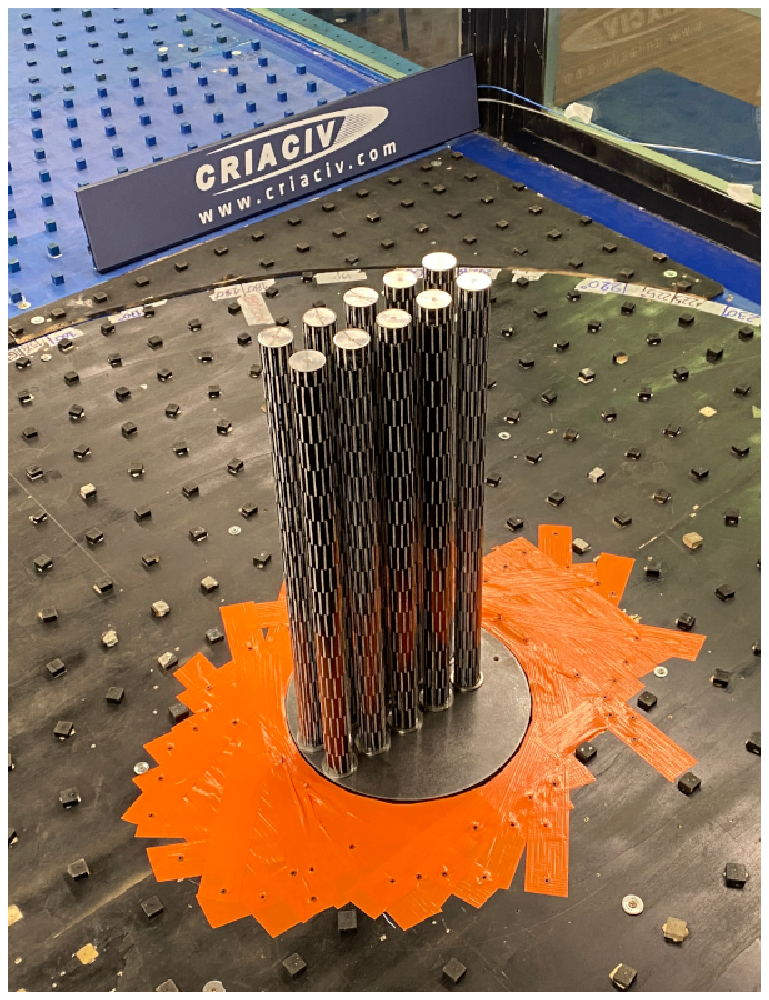}}\\%
	\caption{Set-ups for measurements of the loads either on each tower in a group (a) or on the overall tower group (b).}
	\label{fig:set-ups}
\end{figure}

In this study, the loads acting on each tower in a group need to be determined. The measuring tower was therefore mounted on a high-frequency force balance placed below the wind tunnel floor. This tower model was kept fixed during the tests, while the dummy towers were screwed onto a steel plate rigidly connected to the wind tunnel turntable. The plate presented a circular hole encompassing the measuring tower without touching it (Fig.~\ref{fig:single-tower set-up}). Exploiting the polar symmetry of the structures, the dummy towers were rotated about the measuring tower to mimic the variation of the mean wind direction. Different tower group configurations could be obtained without moving the measuring tower but by changing the arrangement of the dummy towers around it. 

The resultant forces and moments at the base of the towers were measured through a six-component strain-gauge high-frequency force balance (ATI FT-Delta SI-165-15). The raw measurements of overturning moments were corrected accounting for the small distance between the wind tunnel floor and the reference plane of the force balance. The natural frequency of the carbon-fiber models connected to the force balance was 49~Hz, thus requiring low-pass filtering of the recorded signals with a cut-off frequency of 40~Hz.

The set-up for pressure measurements was the same except for the measuring tower model (see Section~\ref{models}). The ABS model was equipped with nine arrays (or rings) of 24 pressure taps distributed over the height of the tower; however, it was only possible to connect 189 pressure taps to the pressure sensors (see Section~\ref{exp_validation}).
The pressure signals were transferred to the sensors through Teflon tubes with a nominal internal and external diameters of 0.8~mm and 1.3~mm, respectively. The length of the tubes varied between 300~mm for the lower arrays and 825~mm for the uppermost one.
The pressure data were recorded at a sampling rate of 500~Hz with seven 32-channel miniaturized piezoelectric scanners (having a full scale of 2.5~kPa) and the Pressure Systems’ Digital Temperature Compensation (PSI DTC) Initium system. The nominal accuracy of the pressure sensors is $\pm$2.45~Pa.
In this set-up, the natural frequency of the measuring tower was significantly lower (24~Hz), requiring low-pass filtering with a cut-off frequency of 17~Hz.

\subsection{Set-up for group-of-towers measurements}
\label{group_setup}

In order to determine the dynamic loads acting on the foundation slab on which the towers are placed at the quayside, a second experimental set-up was conceived, in which the overall loads acting on the groups of towers were measured.
In this case, the carbon-fibre models were screwed onto a circular steel plate with a diameter of 280~mm, flush with the wind tunnel floor. The plate was connected to the same force balance used for the single-tower measurements.
In this case, the entire rig (models, plate and force balance) was rotated to account for different wind directions.

Obviously, the lower natural frequency of the system depended on the number of towers installed on the plate, and the minimum cut-off frequency for low-pass filtering was 23~Hz.

\subsection{Wind modelling}
\label{wind_modelling}

A crucial point in such an experimental campaign on scale models was the reproduction of the turbulent wind profile. The Eurocode~1 terrain category~I profile was assumed as a target for the present study.

\begin{figure}[b!]
  \centering
	\includegraphics[angle=0, width=0.35\textwidth]{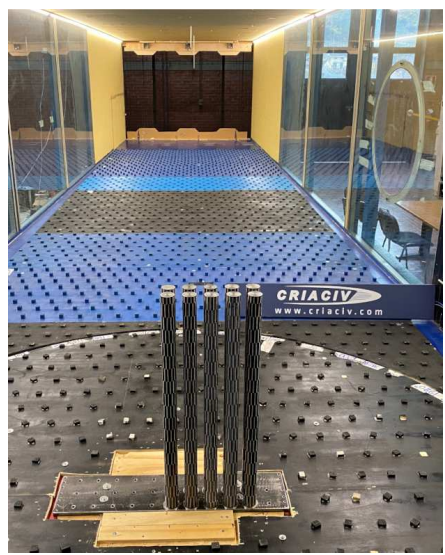}
	\caption{View of a group of ten towers placed in the wind tunnel with in the background all the devices placed upstream to simulate the turbulent boundary layer.}
	\label{fig:ABL}
\end{figure}

The oncoming turbulent wind field was simulated by placing several wooden panels on the wind tunnel floor with randomly distributed square prisms whose height reduced slightly as they approached the test chamber (Fig.~\ref{fig:ABL}). Moreover, at the inlet of the tunnel, immediately after the nozzle, two different castellated barriers were placed on the floor and on the ceiling.

\begin{figure}
	\centering
	\subfigure[\label{fig:U_UH}]
  {\includegraphics[angle=0, width=0.495\textwidth]{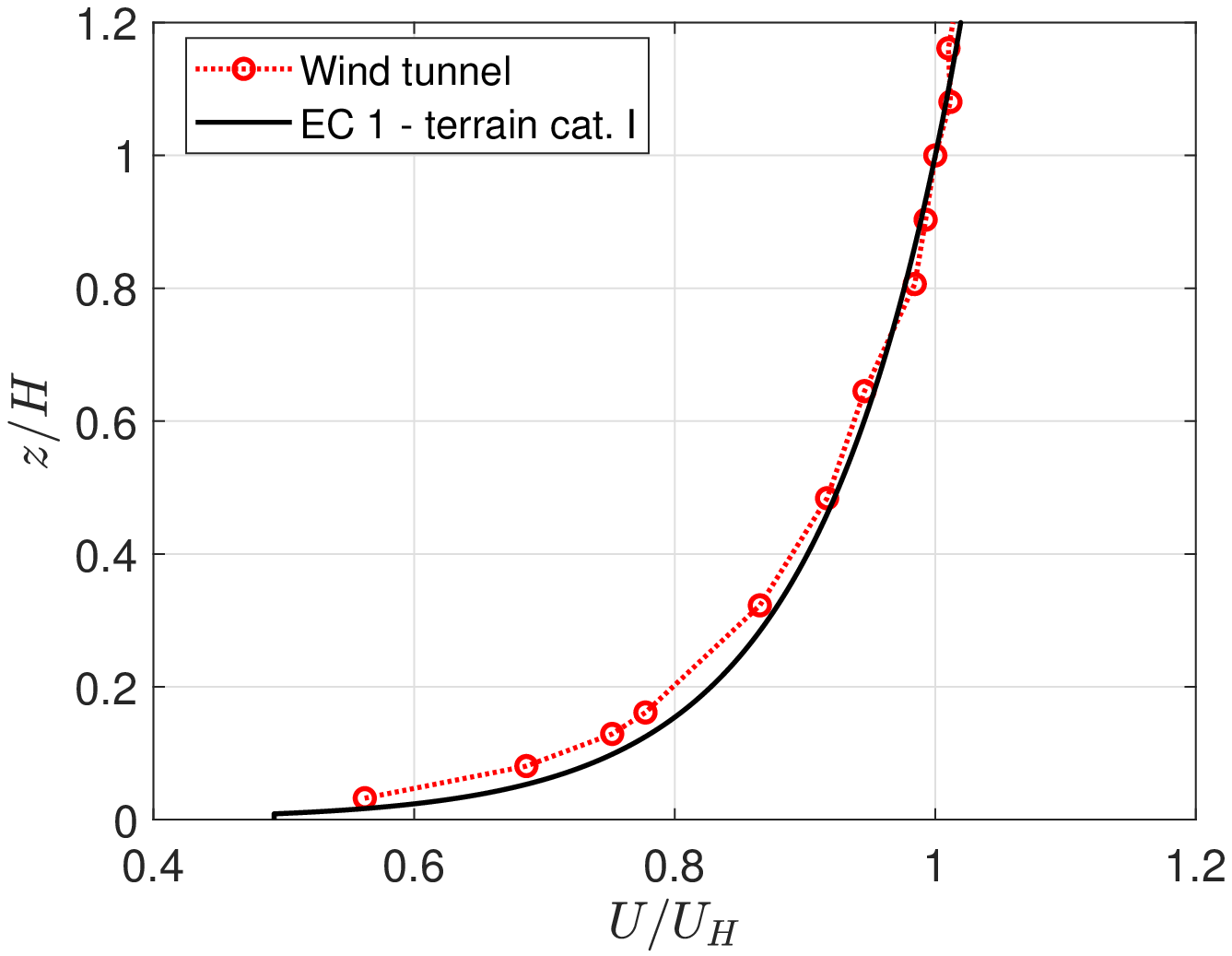}}\hspace{0.0cm}%
	\subfigure[\label{fig:Iu}]
  {\includegraphics[angle=0, width=0.495\textwidth]{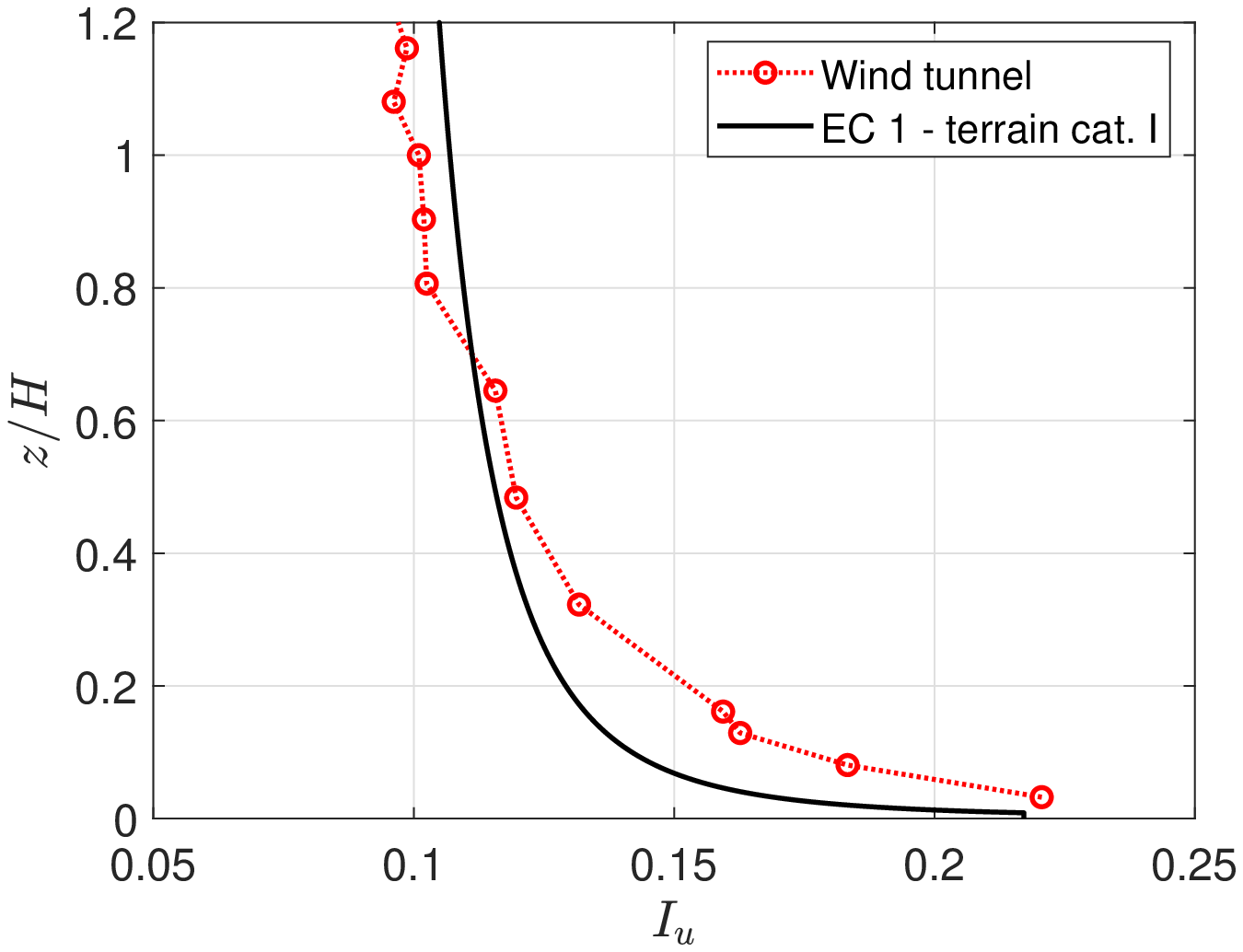}}\\%
	\caption{(a) profiles of the mean wind speed $U$ (normalized) and (b) longitudinal turbulence intensity $I_u$: comparison with the target Eurocode~1 profile (for terrain category I). $H$ and $U_H$ denote the height corresponding to the top of the tower and the mean wind speed at that height, respectively, while $z$ indicates the elevation above the ground.}
	\label{fig:wind profiles}
\end{figure}

The statistical properties of the atmospheric boundary layer were measured with both a single-component (Dantec 55P11) and a two-component (Dantec 55P61) hot-wire anemometer, connected to Dantec Mini-CTA 54T42 modules. A sampling frequency of 10~kHz was set.
The comparison between the simulated and the target wind profile is reported in Fig.~\ref{fig:wind profiles} in terms of mean wind speed and longitudinal turbulence intensity. The agreement is clearly satisfactory.

\begin{figure}
	\centering
	\subfigure
  {\includegraphics[angle=0, width=0.33\textwidth]{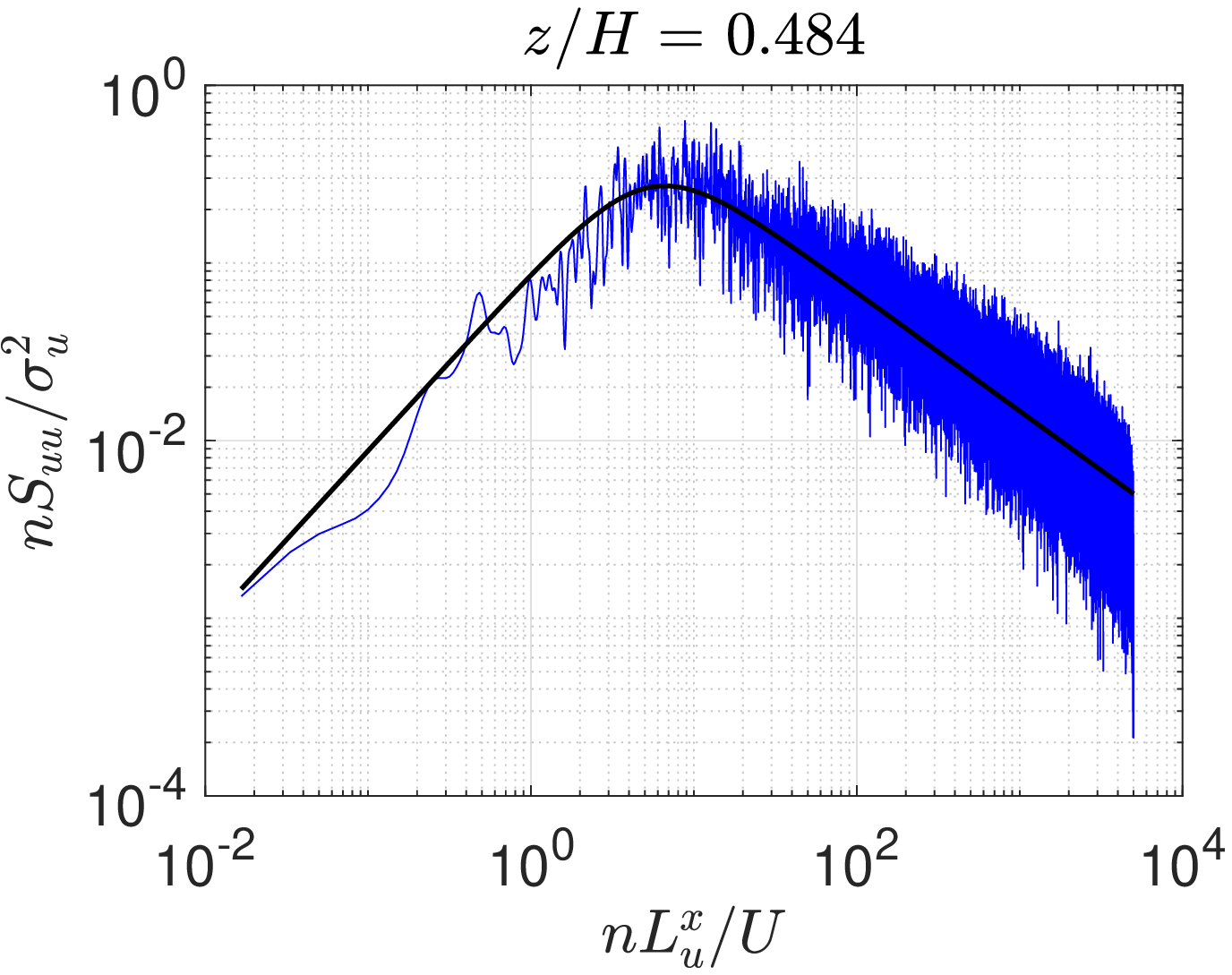}}\hspace{0.0cm}%
	\subfigure
  {\includegraphics[angle=0, width=0.33\textwidth]{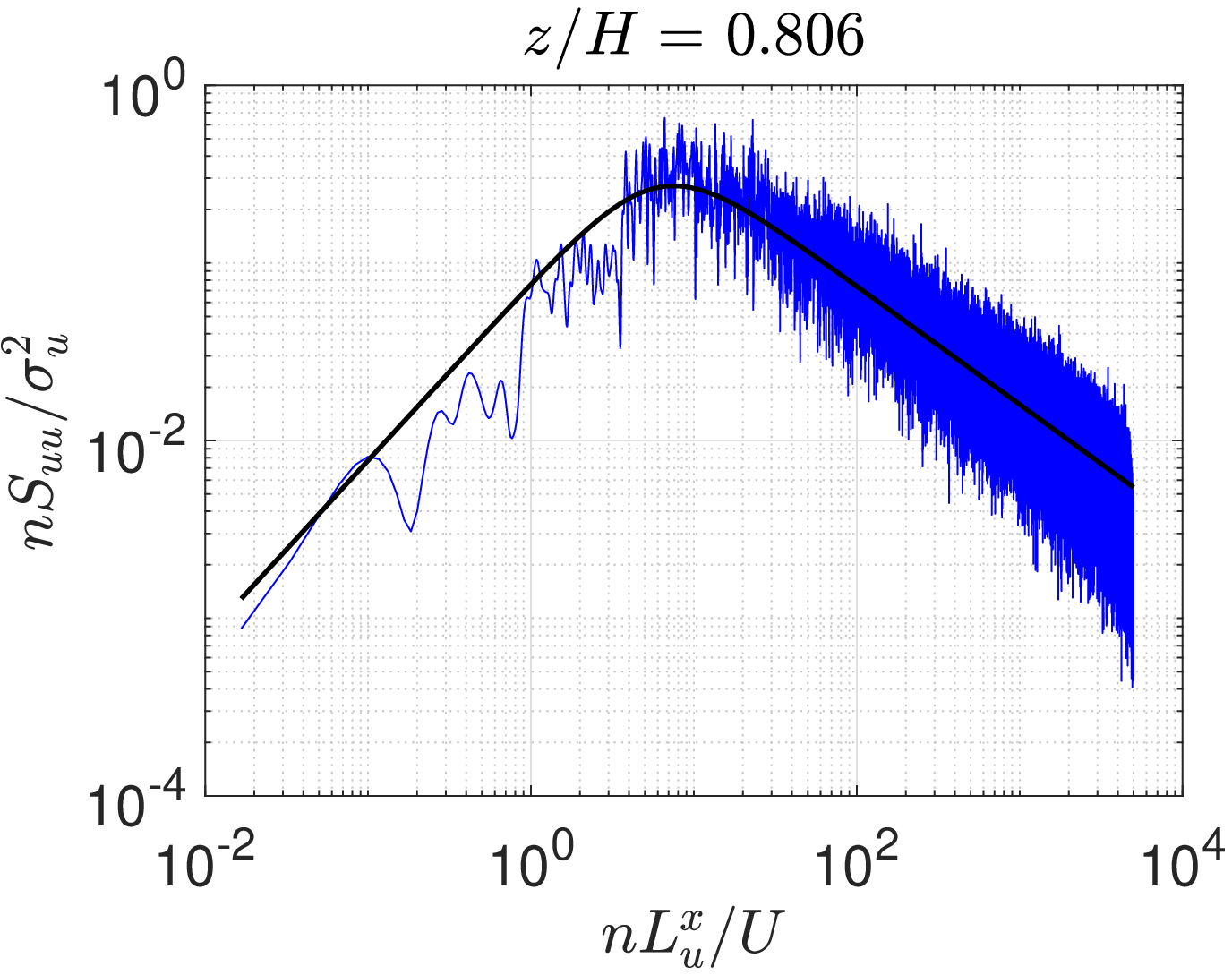}}\hspace{0.0cm}%
	\subfigure
  {\includegraphics[angle=0, width=0.33\textwidth]{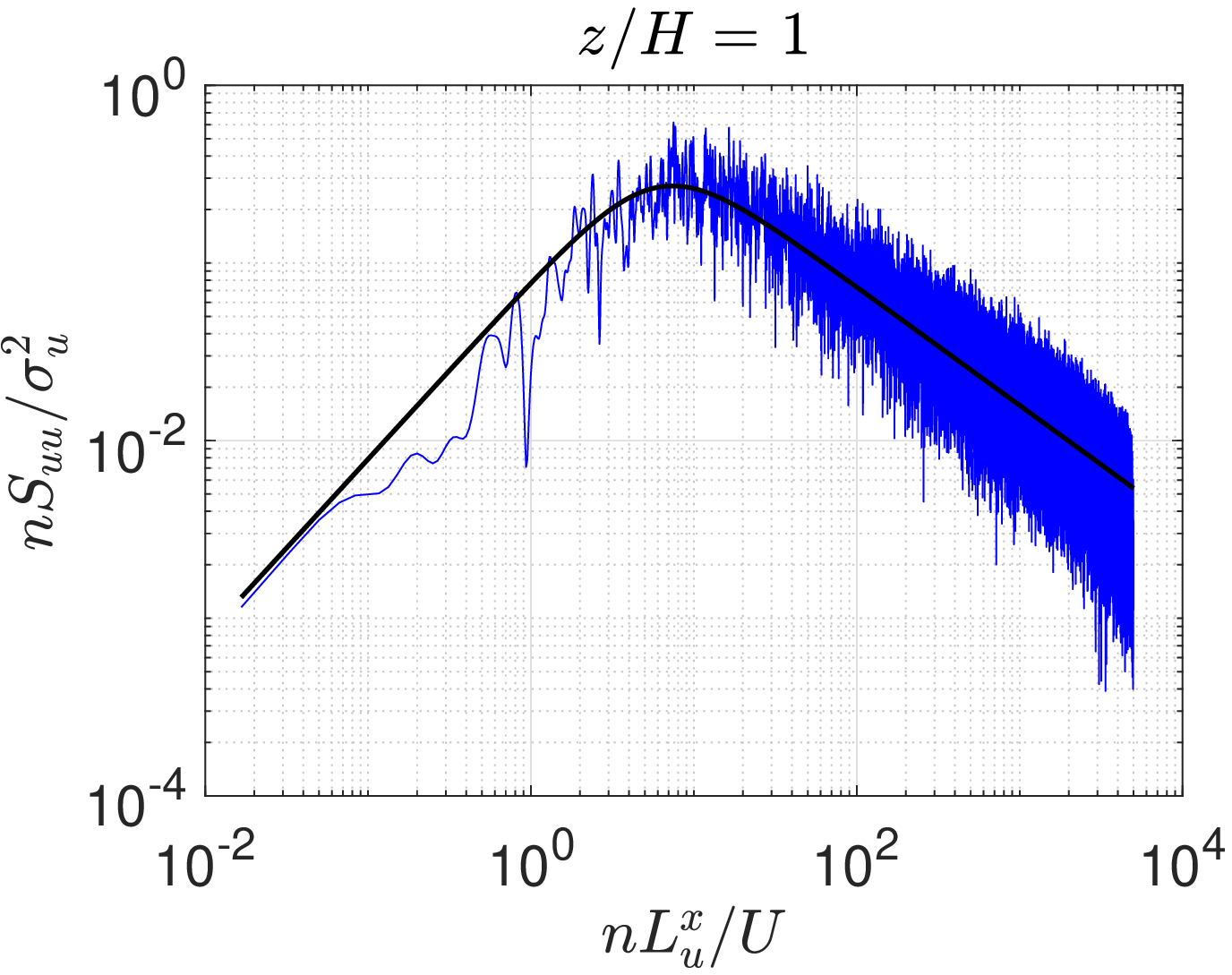}}\hspace{0.0cm}%
	\caption{Longitudinal turbulence spectra measured in the wind tunnel at three heights $z$ above the ground. Comparison with the Eurocode~1 spectrum (thick black line). $L_u^x$ denotes the longitudinal integral length scale of turbulence, $\sigma_u^2$ the variance of the longitudinal wind velocity fluctuations, $U$ the mean wind speed, $n$ the generic frequency in Hz, and $H$ the height corresponding to the top of the towers.}
	\label{fig:turbulence_spectra}
\end{figure}

\begin{figure}
  \centering
	\includegraphics[angle=0, width=0.495\textwidth]{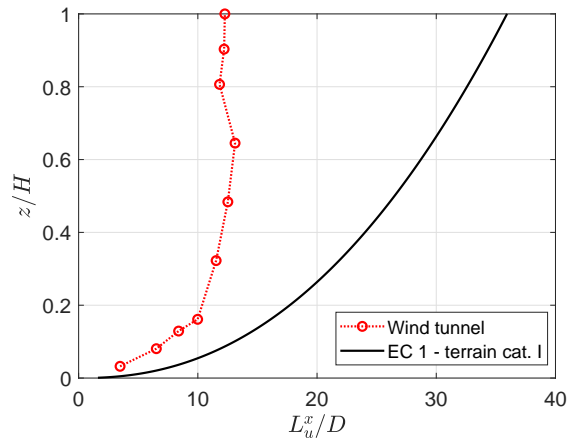}
	\caption{Profile of the longitudinal integral length scale of turbulence $L_u^x$ (normalized with the tower diameter $D$): comparison with the target Eurocode~1 profile (for terrain category I).}
	\label{fig:Lux}
\end{figure}

Fig.~\ref{fig:turbulence_spectra} reports the normalized spectra of longitudinal turbulent velocity fluctuations at three different heights. The very good fit obtained with the Eurocode~1 spectral law demonstrates the correct development of turbulence in the wind tunnel. Nevertheless, as clearly shown in Fig.~\ref{fig:Lux}, the longitudinal integral length scale of turbulence $L_u^x$ simulated in the wind tunnel is significantly lower than the target values provided by Eurocode~1. This mismatch often affects wind tunnel tests, especially when the geometric scale of the model cannot be reduced too much either for manufacturing issues or to have the highest possible Reynolds number. In the present case, this mismatch is not expected to significantly influence the mean aerodynamic coefficients of the towers, since the total energy of the turbulent fluctuations (turbulence intensity) was correctly reproduced and the order of magnitude of $L_u^x$ was matched. In contrast, a modest impact is expected on the dynamic wind loads, and this will be discussed in Sections~\ref{gust_factors} and \ref{overall_group_loads}.

During force and pressure measurements, the mean wind speed was monitored through a Pitot tube connected to a differential membrane pressure transducer of Setra Systems (AccuSense model ASL with a full scale of 0.62~kPa), placed in a low turbulence region of the test section. Prior to placing the tower models in the wind tunnel, the conversion coefficient between the mean wind speed registered by the reference Pitot tube and that measured at the reference height ($U_H$, corresponding to the top of the towers) was determined.

In order to maximize the nominal Reynolds number in the experiments and the accuracy of force and pressure measurements, the tests were carried out with a reference wind speed $U_H$ slightly lower than 31~m/s. Assuming a design full-scale wind speed at the top of the towers of 47.6~m/s, the velocity scale was set to 1:1.55, and the resultant time scale for the experiments was 1:120.7.

\section{Simulation of the high Reynolds number regime}

The key role played by the Reynolds number in this engineering problem has already been emphasized in Section~\ref{Intro}. In the present work, a long preliminary study was carried out to simulate in the best possible way the so-called ‘‘transcritical’’ regime ($\rm{Re} \gtrsim 3.5 \cdot 10^6$, where the drag coefficient reaches a plateau, according to \cite{Roshko1961}; also known as ‘‘postcritical’’ regime)  expected for the full-scale towers. Indeed, based on the equivalent diameter and the abovementioned design wind speed at the top of the tower, the Reynolds number is estimated to be about $2 \cdot 10^7$, however a subcritical value of only about $7 \cdot 10^4$ can be reached in the wind tunnel at the considered geometric scale.
Since detailed high-Reynolds number data for finite-length cylinders in shear flows are very rare in the literature, this preliminary investigation was addressed to the infinite two-dimensional cylinder, for which a certain number of data can be collected.

\subsection{Literature data}
\label{High_Re_data}

An important point in the current work was the choice of a reasonable target for the transcritical Reynolds number regime for the circular cylinder. Since the experimental campaign had to comply with the practical design of the supporting structures for the towers at the quayside, the pressure distribution and the drag coefficient provided by Eurocode~1 for very high Reynolds numbers were eventually assumed as a reference. Prior to this, however, an extensive literature review was conducted to put these data into context in the state-of-the-art knowledge. Unfortunately, just a limited number of data sets are suitable for our purposes, as the majority of results are below or up to the so-called ‘‘supercritical’’ regime, often below $\rm{Re} = 2 \cdot 10^6$.

In addition to the Reynolds number, the key role played by surface roughness in governing the aerodynamics of the circular cylinder was already highlighted in the pioneering works by \cite{Fage_Warsap29}, carried out in a military laboratory, and by \cite{DrydenHill30}, who measured the wind pressures on a full-scale tall power plant chimney. Later, the impact of this parameter was extensively investigated by many researchers \citep{Achenbach71, Szechenyi75, Guven80, Buresti81}, and it is now clear that one must always account for it while comparing the experimental results available in the literature.
As for the oncoming flow turbulence, at first this seemed to play a role only in the Reynolds number range $10^5$ to $5 \cdot 10^5$ \citep{DrydenHill30}, whereas its importance has been better investigated some decades later \citep{Batham73, Bruun&Davies75}. The turbulence effects were also reviewed by \cite{Bell79, Bell83}, revealing how the results in the literature are complicated and controversial.
Both turbulence and surface roughness contribute to anticipate the critical Reynolds number range, although the influence on the overall aerodynamic behavior of the circular cylinder is different in the two cases (e.g., \cite{Buresti2012}).

The high Reynolds number data available in the literature are collected in Table~\ref{tab:CD_CP_table}, highlighting the most influential experimental parameters. The target here is the behavior of an infinite circular cylinder with a smooth surface, experimental studies on very rough cylinders are therefore excluded.

\begin{sidewaystable}
	\caption{Experimental data characteristics for the reference literature employed for the comparisons. $L$ denotes the spanwise length of the model, $D$ is the cylinder diameter, $\rm{Ma}$ indicates the Mach number, $k$ the surface roughness characteristic length, $I_u$ the longitudinal turbulence intensity, $C_D$ the mean drag coefficient, and $\rm{Re}$ the Reynolds number (based on $D$).}
	\label{tab:CD_CP_table}
	\centering	
	\begin{small}
		\begin{tabular}{lccccccc}
			\toprule
Authors          &$L/D$&Blockage [\%]&$\rm{Ma}$ &$k/D$&$I_{u}$ [\%]&$\rm{Re_{max}}$&$C_{D}(\rm{Re_{max}})$			\\ \hline                
			\noalign{\smallskip}\noalign{\smallskip}
\cite{DrydenHill30}    & 12.03 & in situ & $\sim 0.131$ & $8.34 \cdot 10^{-4}$ &  $-$      &$1.70 \cdot 10^{7}$&       0.67   \\
                                                 &  4.76 & 10.5 & $\leq 0.087$ & $-$                  &  $-$      &$1.82 \cdot 10^{5}$&     $-$      \\
\cite{DelanySorensen53}&  9.0 & 14.3 & $\leq 0.34$  & $-$                  &  $-$      &$2.30 \cdot 10^{6}$&       0.52   \\
\cite{Roshko1961} &  5.67 & 13.6 & $\leq 0.25$  & $1.11 \cdot 10^{-5}$ &  $-$      &$8.60 \cdot 10^{6}$&       0.74   \\          
\cite{Achenbach68} &  3.33 & 16.7 & $\leq 0.1$   & $1.33 \cdot 10^{-5}$ &  $-$      &$4.70 \cdot 10^{6}$&       0.73   \\                
\cite{Jones69} & 5.34 & 18.7 & $\leq 0.6$   & $10^{-5}$ &  $0.17$ &$1.80 \cdot 10^{7}$ &       0.54   \\
\cite{Szechenyi75}&  4.38 & 22.6 & $\leq 0.29$  &$10^{-4}$& $-$      &$6.50 \cdot 10^{6}$&  $-$         \\                       
                                   &  4.0 & 17.9 & $\leq 0.29$  &$2.86\cdot10^{-4}$&$-$ &$4.20 \cdot 10^{6}$&$-$    \\                     
\cite{Christensen78}& 32.5 & in situ& $\sim 0.062$&$2.5 \cdot 10^{-3}$ &  $-$ &$1.08 \cdot 10^{7}$& 0.64   \\               
\cite{James80} &  11.56 &  8.7  & $< 0.3$  & $3 \cdot 10^{-6}$                &  $-$       &$1.09 \cdot 10^{7}$&       0.45    \\
\cite{AchenbachHeinecke81} &  6.75  &  16.7 & $< 0.3$  &$< 10^{-5}$          &$0.45$ &$4.20 \cdot 10^{6}$ &0.69\\  
\cite{Schewe1983}& 10.0  & 10.0  & $0.112$  & $\sim 10^{-5}$           &$0.15\div0.4$&$7.10 \cdot 10^{6}$&  0.52      \\            
\cite{Shih_Roshko92}& 10.76  &  9.0  & $< 0.3$  &$10^{-5}$             &0.49&$8.00 \cdot 10^{6}$&    $-$    \\
\cite{Adachi97} &  $-$   &  $-$     & $< 0.3$  &$4.5\cdot10^{-6}$& $-$    &$\sim 10^{7}$&0.66\\
\cite{vanHinsberg2015}&  10.0 & 10.0  & 0.071    & $1.2 \cdot 10^{-3}$              &$0.15\div0.4$ &$1.20 \cdot 10^{7}$&0.84         \\  
\bottomrule  
   		\end{tabular}
	\end{small}
\end{sidewaystable}

\begin{figure}[t]
	\centering
	\includegraphics[angle=0, width=0.7\textwidth]{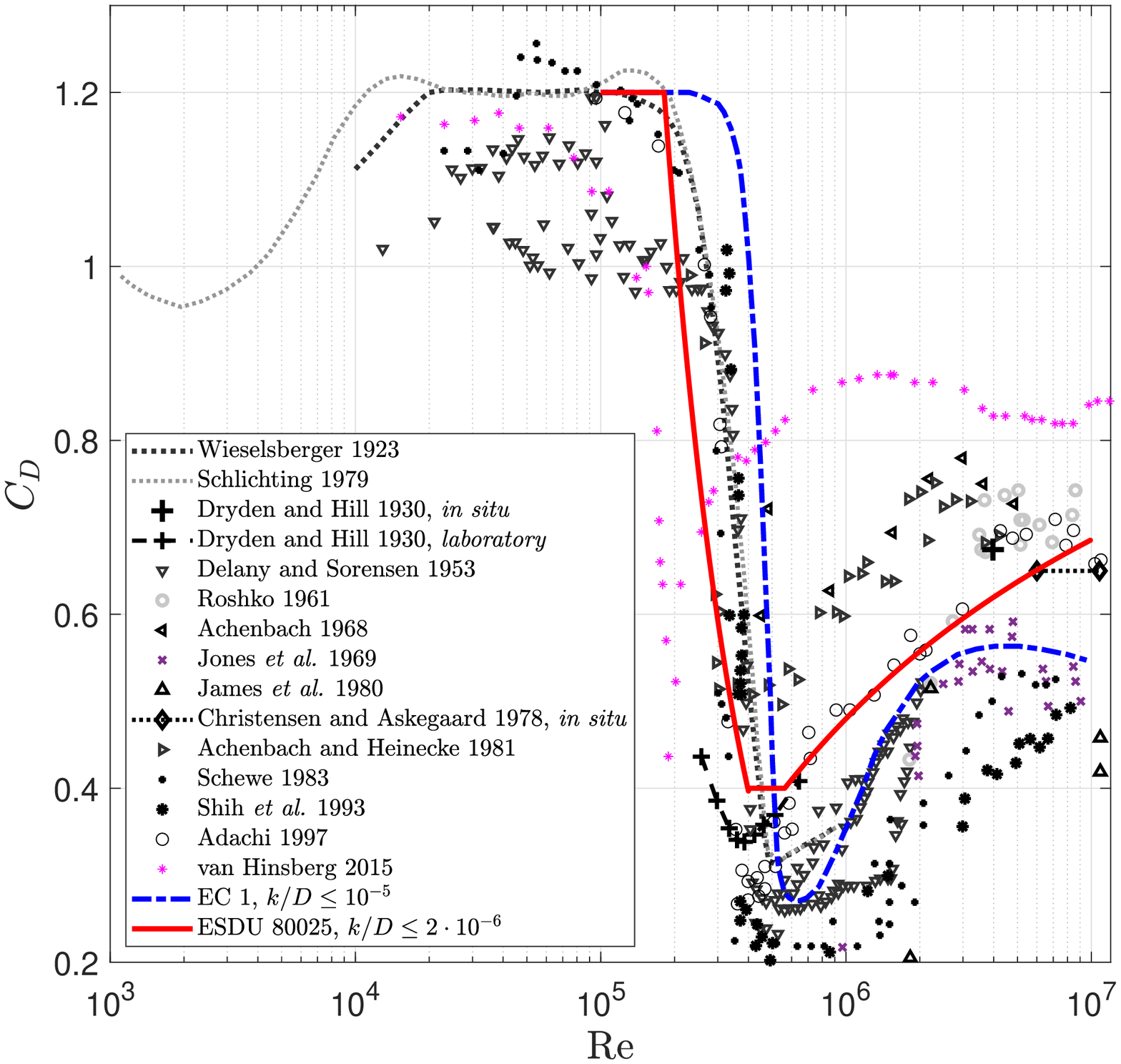}
	\caption{Collection of drag coefficients $C_D$ from the literature as a function of the Reynolds number $\rm{Re}$. EC~1 stands for Eurocode~1 \cite{Eurocodice1}. See Table~\ref{tab:CD_CP_table} for the references.}
	\label{fig:Cd_literature}
\end{figure}

Fig.~\ref{fig:Cd_literature} collects the main contributions in terms of mean drag coefficient ($C_D$) associated with high Reynolds numbers. The drag coefficient values proposed by Eurocode~1 are in line with the data set of \cite{Roshko1961}, \cite{Adachi97} and with the full-scale measurements of \cite{DrydenHill30} and \cite{Christensen78}. Larger values of $C_D$ were reported by \cite{Achenbach68} and \cite{AchenbachHeinecke81}. 
In contrast, the ESDU recommendations \citep{ESDU1986} provide lower values of the drag coefficient, which agree well with the data sets by \cite{Jones69}, \cite{Schewe1983} and \cite{Shih_Roshko92}.
It is worth noting that the drag coefficients measured by \cite{James80} are particularly low, whereas the high drag obtained by \cite{vanHinsberg2015} may be ascribed to the higher surface roughness (see Table~\ref{tab:CD_CP_table}).

\begin{figure}[t]
	\centering
	\includegraphics[angle=0, width=0.7\textwidth]{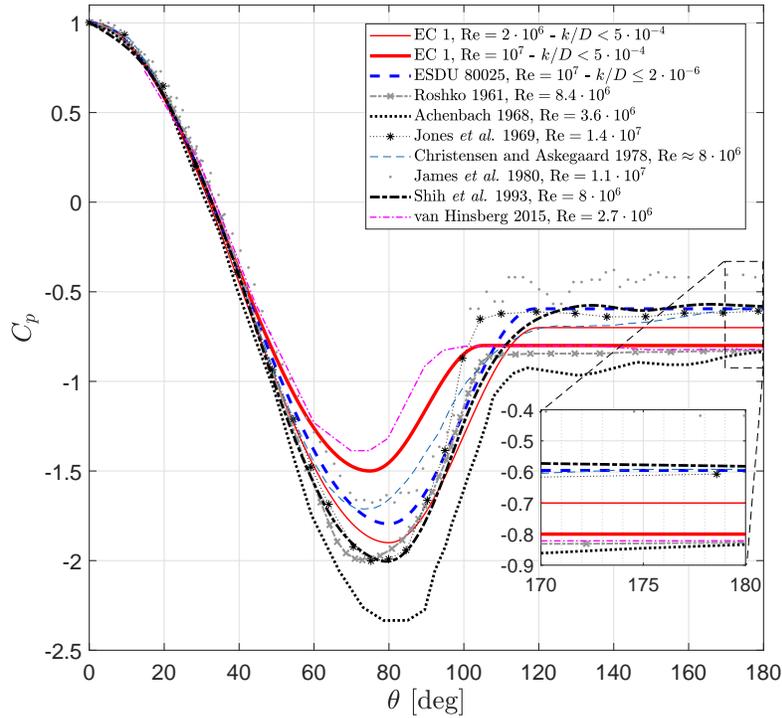}
	\caption{Pressure coefficient ($C_p$) distributions of interest for the present study. See Table~\ref{tab:CD_CP_table} for the references.}
	\label{fig:Cp_literature}
\end{figure}

Moreover, a selection of results in terms of pressure coefficients is proposed in Fig.~\ref{fig:Cp_literature}. The variability in the $C_p$-patterns complies with the previously observed spread in the drag coefficient. In particular, the figure highlights the curves reported by Eurocode~1 for the transcritical ($\rm{Re} = 10^7$) and supercritical ($\rm{Re} = 2 \cdot 10^6$) regimes. These curves are qualitatively the same as those recommended by ESDU, which are based on a semi-empirical approach based on the theoretical wake-source model of \cite{Parkinson70}. The discrepancy between the $C_p$-distributions for $\rm{Re} = 10^7$ provided by Eurocode~1 and ESDU is probably due to the different drag coefficient (see Fig.~\ref{fig:Cd_literature}) and, therefore, to the different base pressure coefficient.
Coherently with the results for the drag coefficient, the base pressure associated with Eurocode~1's transcritical curve is in very good agreement with Roshko's data \citep{Roshko1961}, although the latter present a lower pressure prior to separation.
Despite the very large drag coefficient (Fig.~\ref{fig:Cd_literature}), the outcome of the recent experimental campaign conducted by \cite{vanHinsberg2015} is also in good agreement with the Eurocode~1 transcritical curve, though separation seems to occur slightly before.
\cite{Achenbach68} obtained very low base pressure (coherently with the large drag coefficients reported in Table~\ref{tab:CD_CP_table} and Fig.~\ref{fig:Cd_literature}) and minimum pressure.
The other collected data sets are characterized by a higher base pressure (even higher than the Eurocode~1 supercritical curve, $\rm{Re} = 2 \cdot 10^6$), even though those by \cite{Jones69} and by \cite{James80} suggest the same position of the mean separation point as the Eurocode~1 transcritical curve.

Finally, one may remark that, to reach very high Reynolds numbers, the data sets considered in Table~\ref{tab:CD_CP_table} are often associated with a high blockage ratio (though \cite{Szechenyi75} used a tolerant wind tunnel and extensively investigated the effect of blockage), a Mach number close to the incompressibility limit (conventionally 0.3), and model aspect ratio values that are not so large. All these factors contribute to the variability of the experimental data in the literature and make the identification of target aerodynamic quantities for the transcritical Reynolds number regime for circular cylinders rather uncertain.

\subsection{Proposed surface roughness solution}

\begin{figure}
  \centering
	\includegraphics[angle=0, width=0.5\textwidth]{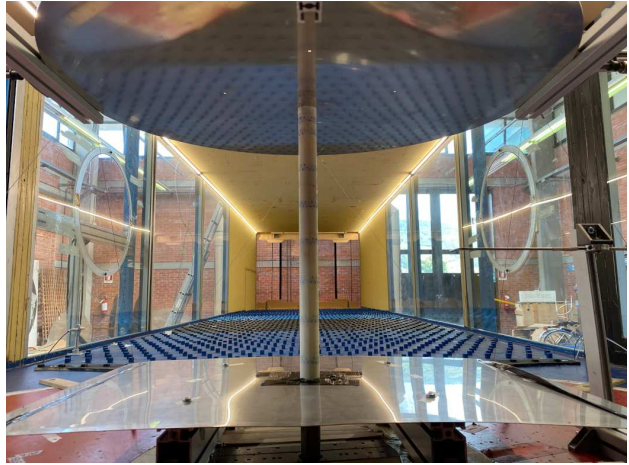}
	\caption{Set-up for the assessment of Reynolds number effects.}
	\label{fig:Setup_Reynolds}
\end{figure}

In order to use the ABS model equipped with pressure taps as an infinite circular cylinder, two steel large end-plates parallel to the wind tunnel floor were installed to confine the flow. The model support below the wind tunnel was placed as high as possible, so that the model tip resulted at 645~mm above the wind tunnel floor ($1.049 H$). The end-plates were placed at the positions $z/H = 0.245$ and $z/H = 1.049$ (reported here in normalized form for the sake of comparison with Fig.~\ref{fig:wind profiles}), as shown in Fig.~\ref{fig:Setup_Reynolds}. To minimize the thickness of the end-plates but also to avoid vibrations, especially in turbulent flow, a system of steel stays was employed. Clearly, force measurements at the base of the tower were useless with this set-up, and one could only rely on pressure measurements.

\begin{figure}
  \centering
	\includegraphics[angle=0, width=0.495\textwidth]{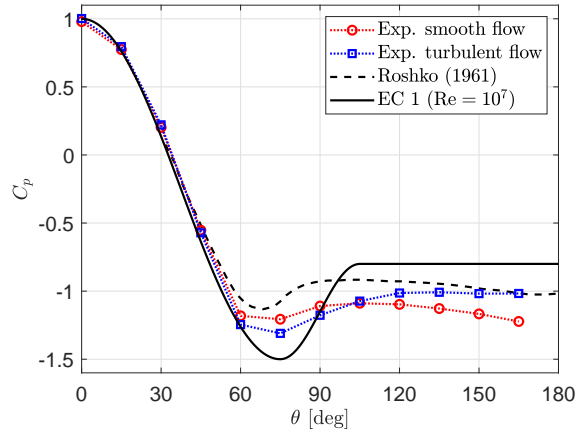}
	\caption{Pressure coefficient $C_p$ for the circular cylinder without surface roughness in smooth and turbulent flow, measured at the maximum mean wind speed ($U \cong 30$ m/s, corresponding to $\rm{Re} \cong 7 \cdot 10^4$). The results, reported in \cite{Roshko1961, Simiu2019}, and originally belonging to \cite{Fage&Falkner31}, refer to a cylinder in smooth flow for a subcritical Reynolds number ($\rm{Re} = 1.1 \cdot 10^5$).}
	\label{fig:Cp_SF_TF}
\end{figure}

Prior to testing some surface roughness solutions to simulate the high Reynolds number regime, the plain cylinder was tested in smooth and turbulent flow.
The mean wind speed gradient inside the end-plates was limited, and the turbulent flow characteristics only slightly changed (Fig.~\ref{fig:wind profiles}), and so flow conditions could be considered, at first approximation, to be two-dimensional. In addition, to exclude the effect of the end-plates, only the pressures from the fourth ($z/H = 0.537$) to the seventh array ($z/H = 0.878$) were actually used. In particular, the fifth pressure array ($z/H = 0.650$) was assumed as the reference, and the others were inspected for cross-validation.
The results are shown in Fig.~\ref{fig:Cp_SF_TF}. It is apparent that the oncoming turbulence slightly changes the pattern of the pressure coefficient on the cylinder surface; in particular, it increases the pressure in the base region, thus implying a drop in the resulting drag coefficient. Nevertheless, the pressure distribution is still a long away from the target provided by Eurocode~1 for the transcritical Reynolds number regime. Surface roughness is therefore necessary to simulate it.

A large number of surface roughness solutions were investigated, but the central intention was to avoid fully covering the tower model with sandpaper whose  drawback is a remarkable increase in drag, which would make the present tests overconservative anyway. Moreover, an excessive subtraction of momentum from the boundary layers attached to the model surface may result in an earlier separation (see e.g. \cite{Buresti80}), thus possibly leading to results that, even qualitatively, are far from the desired full-scale condition. It was therefore decided to achieve discontinuous roughness solutions using small strips of sandpaper. Moreover, another important requirement for the current investigation was that the solution should work properly for any wind direction.

47 possible solutions were tested, differing in sandpaper type and strip geometry (either straight or zig-zagged), as well as their width and spacing. The solution which was selected was that which was most promising from the theoretical point of view, employing discontinuous and alternate small strips of sandpaper P220 with a thickness of 0.25~mm (corresponding to $7.1 \cdot 10^{-3} D_{eq}$), as shown in Figs.~\ref{fig:model_pictures}-\ref{fig:set-ups}.
12 strips (33~mm long and 2~mm wide) were placed along the circumference of the model cross section (one every 30~deg).
The alternate strip arrangement is thought not to impose a preferential separation line along the cylinder.

\subsection{Experimental validation}
\label{exp_validation}

\begin{figure}
  \centering
	\subfigure[\label{fig:Cp_roughness}]
  {\includegraphics[angle=0, width=0.495\textwidth]{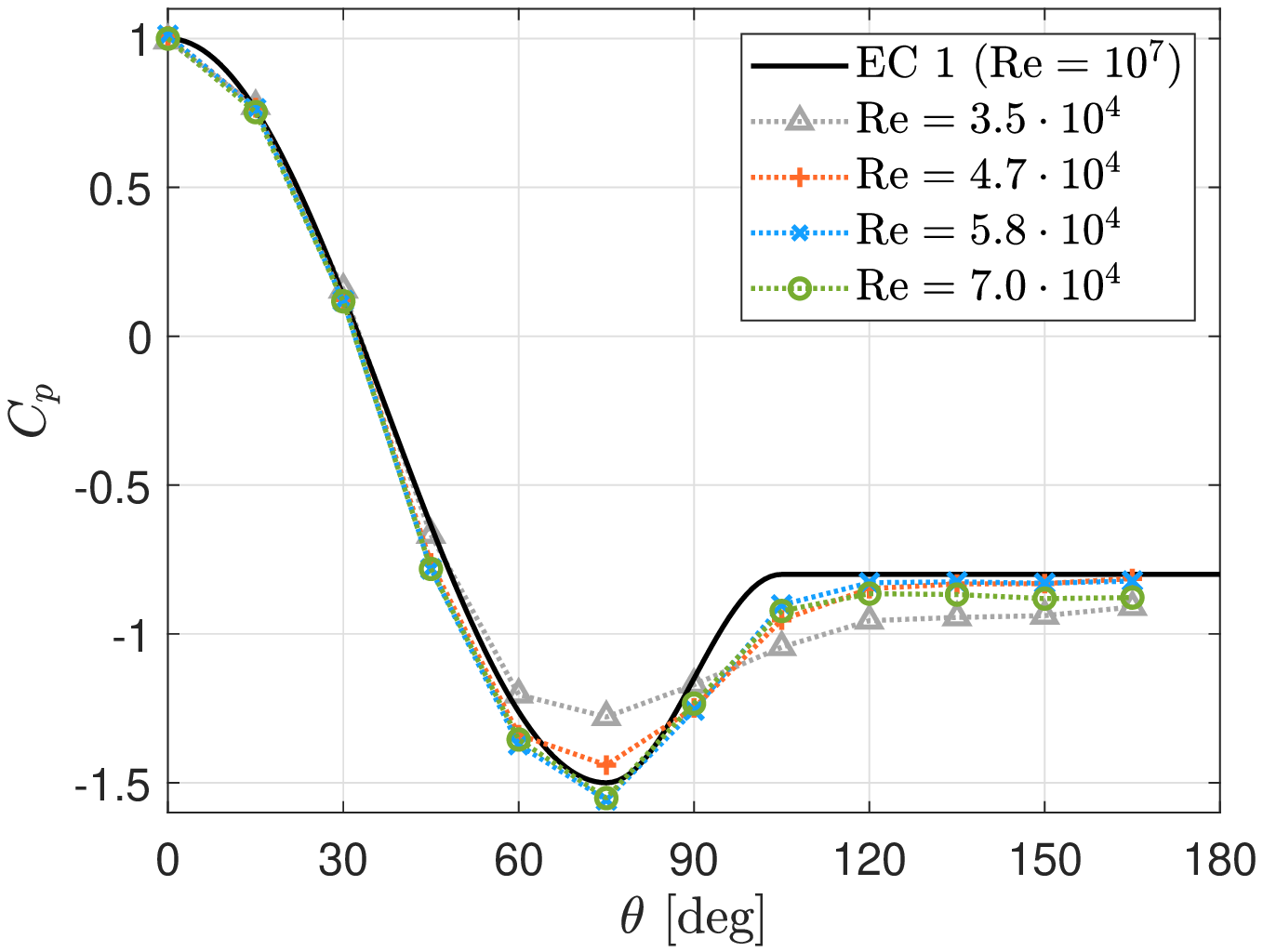}}\hspace{0.0cm}%
	\subfigure[\label{fig:CD_roughness}]
  {\includegraphics[angle=0, width=0.495\textwidth]{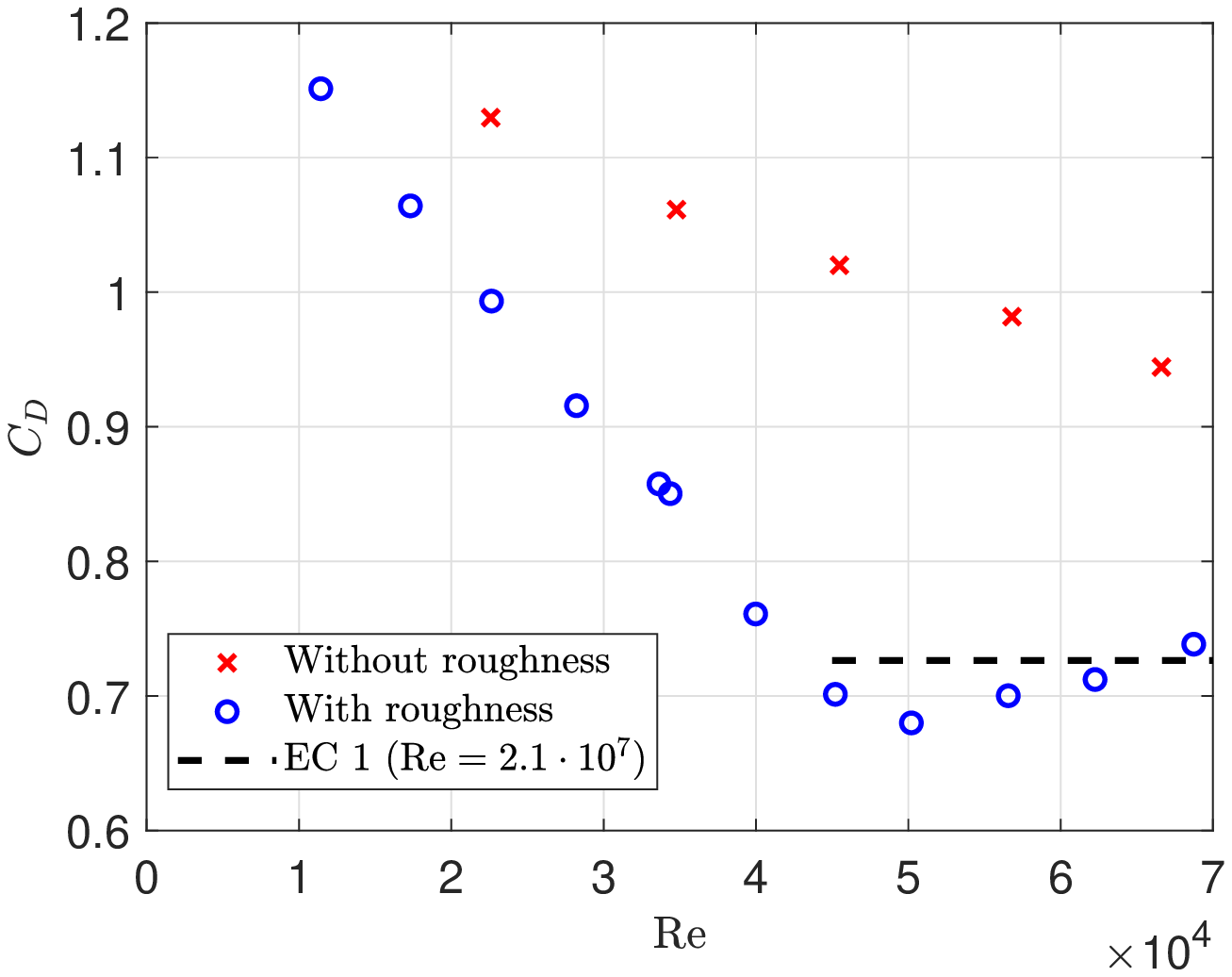}}\hspace{0.0cm}%
	\caption{Pressure coefficient distribution (a) and integrated drag coefficient (b) measured for the circular cylinder with the chosen surface roughness solution in a confined turbulent flow. Given the small variability of the mean wind speed for the considered arrays, the pressure coefficients were all normalized based on the kinetic pressure corresponding to the central array (array \#5). The Reynolds number values associated with wind tunnel tests are just nominal (simply calculated based on wind tunnel mean wind speed and cylinder diameter) and do not correspond to the simulated regime when surface roughness is employed.}
	\label{fig:Cp_CD}
\end{figure}

The effectiveness of the selected surface roughness solution was verified by measuring the pressure distribution on the circular cylinder in turbulent flow. The results are reported in Fig.~\ref{fig:Cp_roughness} along with the target pressure distribution provided by Eurocode~1. In the high wind speed range (say, beyond a wind tunnel Reynolds number of about $5 \cdot 10^4$), the $C_p$-distribution is sufficiently stable and very close to this target. This is confirmed by the drag coefficient obtained through pressure integration and reported in Fig.~\ref{fig:CD_roughness}.
Considering the turbulent flow condition, it is worth noting that without surface roughness, the drag coefficient is lower than the subcritical value of about 1.2 and tends to decrease with the Reynolds number, although it always remains significantly above the target value.
 
\begin{figure}
  \centering
	\includegraphics[angle=0, width=1\textwidth]{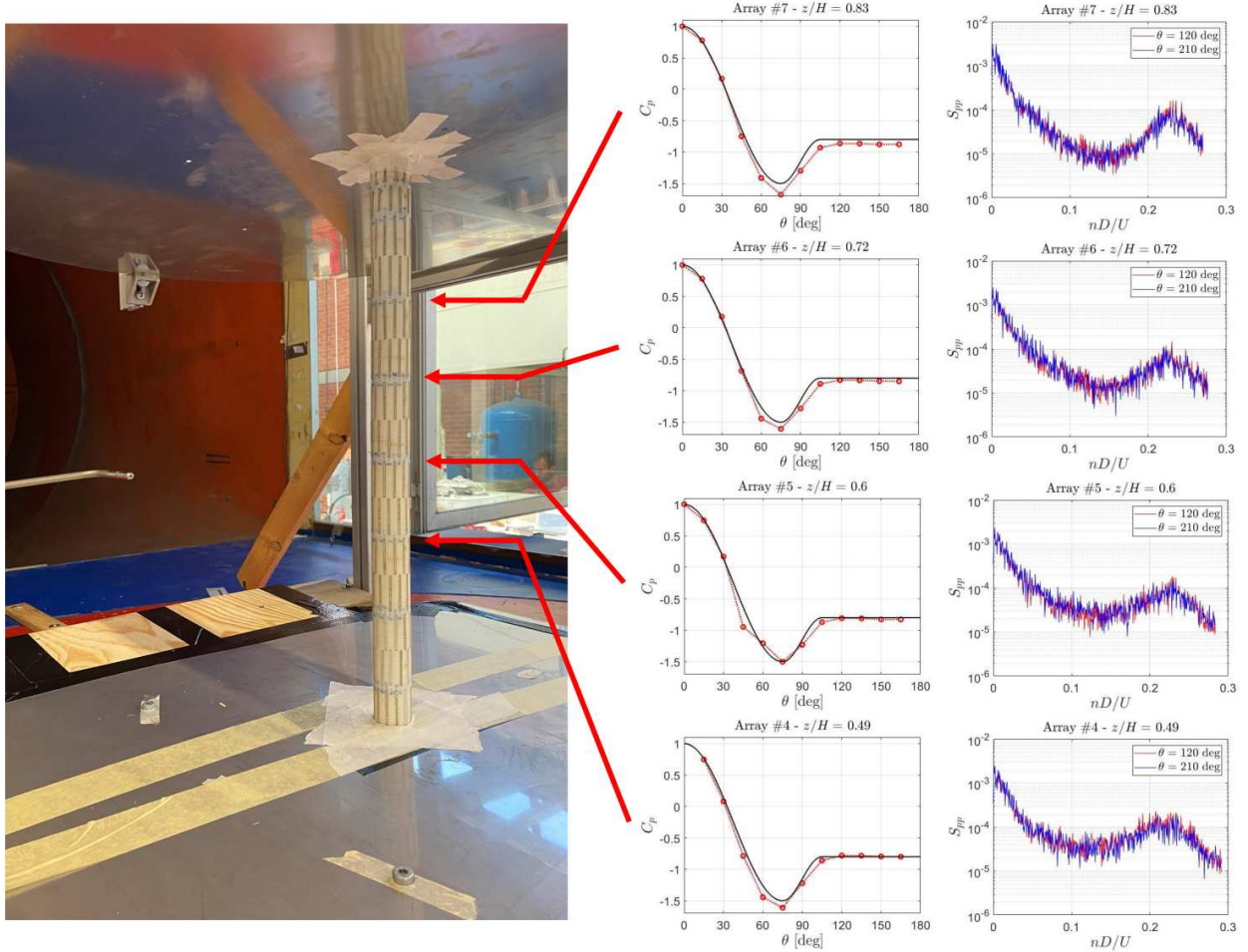}
	\caption{Mean pressure coefficient distribution and power spectral density for two representative pressure taps in correspondence of four pressure tap arrays.}
	\label{fig:Strouhal}
\end{figure}

Fig.~\ref{fig:Strouhal} shows that the previous results are confirmed by the other pressure tap arrays. Moreover, the fluctuating pressure demonstrates a clear bump in the power spectral densities corresponding to a Strouhal number of about 0.22, which is perfect agreement with the Italian Recommendations \cite{CNR-DT207-R1}. In contrast, in subcritical conditions (cylinder without surface roughness and smooth flow), the peaks in the pressure spectra were obviously sharper, and the Strouhal number was about 0.19.

Finally, it was verified that the results were independent of wind direction, thus confirming that the proposed engineering solution succeeds in simulating the transcritical Reynolds number regime for an infinitely long circular cylinder. This solution was then also assumed valid for finite-height towers and group configurations and it was therefore used for all the wind tunnel tests discussed hereinafter.

\section{Results for the baseline configurations}
\label{baseline_results}

As previously mentioned, the wind loads are described here in terms of shear-force and overturning-moment coefficients at the base of each tower and of the entire groups of towers. The results are mostly presented in terms of resultant moment coefficient, as this is the principal quantity of interest for design purposes, and because shear-force coefficient does not generally add much to the discussion. The focus is mainly on the mean coefficients, although in Section~\ref{gust_factors} gust factors are also discussed. 

Generally speaking, the base shear-force and moment coefficients are defined here, respectively, as follows:
\begin{equation}
 C_{F} = \frac{F}{\frac{1}{2} \rho U_H^2 \int_0^H D(z) dz}
\label{eq:generic_shear_coefficient}
\end{equation}
\begin{equation}
 C_{M} = \frac{M}{\frac{1}{2} \rho U_H^2 \int_0^H D(z) z dz}
\label{eq:generic_moment_coefficient}
\end{equation}
where $F$ and $M$ denote the magnitude of the vector resultants of shear force and overturning moment at the base of the towers, respectively; as previously explained, $U_H$ indicates the mean wind speed of the undisturbed flow at the top of the towers.
Hereinafter, the referred forces and moments are those measured on the carbon-fibre models with the high-frequency force balance.

Except for the results presented in Section~\ref{tower_shape}, the diameter of the towers is constant ($D(z) = D_{eq}$). The definitions of force and moment coefficients therefore simplify to:
\begin{equation}
 C_{F} = \frac{F}{\frac{1}{2} \rho U_H^2 D_{eq} H}
\label{eq:shear_coefficient}
\end{equation}
\begin{equation}
 C_{M} = \frac{2M}{\frac{1}{2} \rho U_H^2  D_{eq}^2 H}
\label{eq:moment_coefficient}
\end{equation}
From this, it is clear that $C_F = C_M$ means that the resultant of the aerodynamic force is applied at a height $z = H/2$.

\begin{figure}
  \centering
	\includegraphics[angle=0, width=0.8\textwidth]{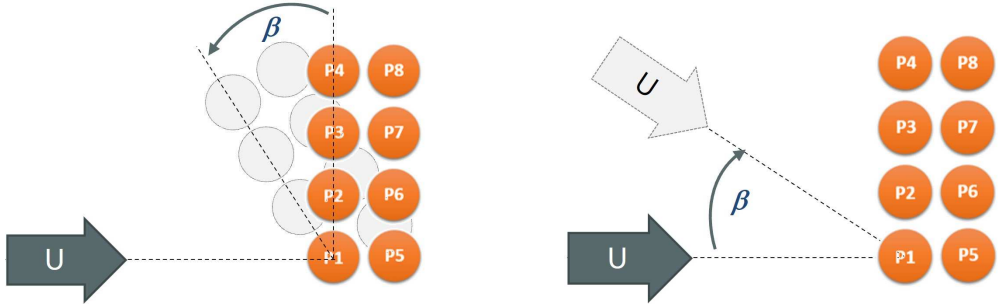}\vspace{0.5cm}%
	\caption{Definition of the wind direction $\beta$ considering either a rotation of the model in the wind tunnel or a variation of the wind direction at full scale.}
	\label{fig:wind_direction}
\end{figure}

The force and moment coefficients are reported in the following as a function of wind direction $\beta$, defined according to the schematic reported in Fig.~\ref{fig:wind_direction}.
As shown also in Fig.~\ref{fig:model_pictures}, the investigated group configurations present various symmetries, which were exploited to limit the number of towers successively connected to the force balance. For instance, for the 8-pack group (G8) depicted in Fig.~\ref{fig:wind_direction}, all possible tower positions and wind directions were accounted for by measuring the loads just on towers P1 and P2 but considering all azimuth angles in the range [0 2$\pi$).

\subsection{Isolated tower}
\label{S_tower}

\begin{figure}
  \centering
	\subfigure[\label{fig:CF_free_standing}]
  {\includegraphics[angle=0, width=0.495\textwidth]{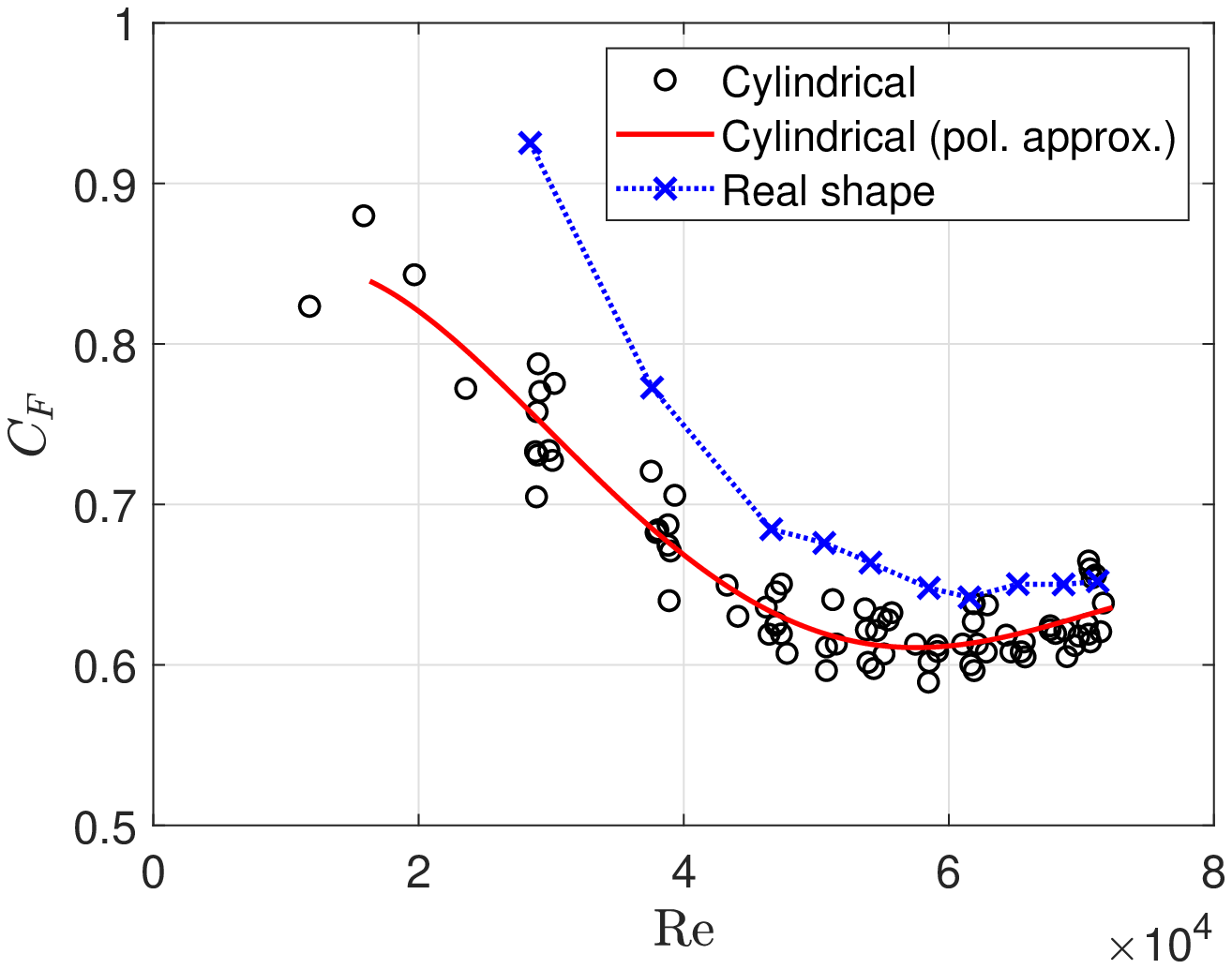}}\hspace{0.0cm}%
	\subfigure[\label{fig:CM_free_standing}]
  {\includegraphics[angle=0, width=0.495\textwidth]{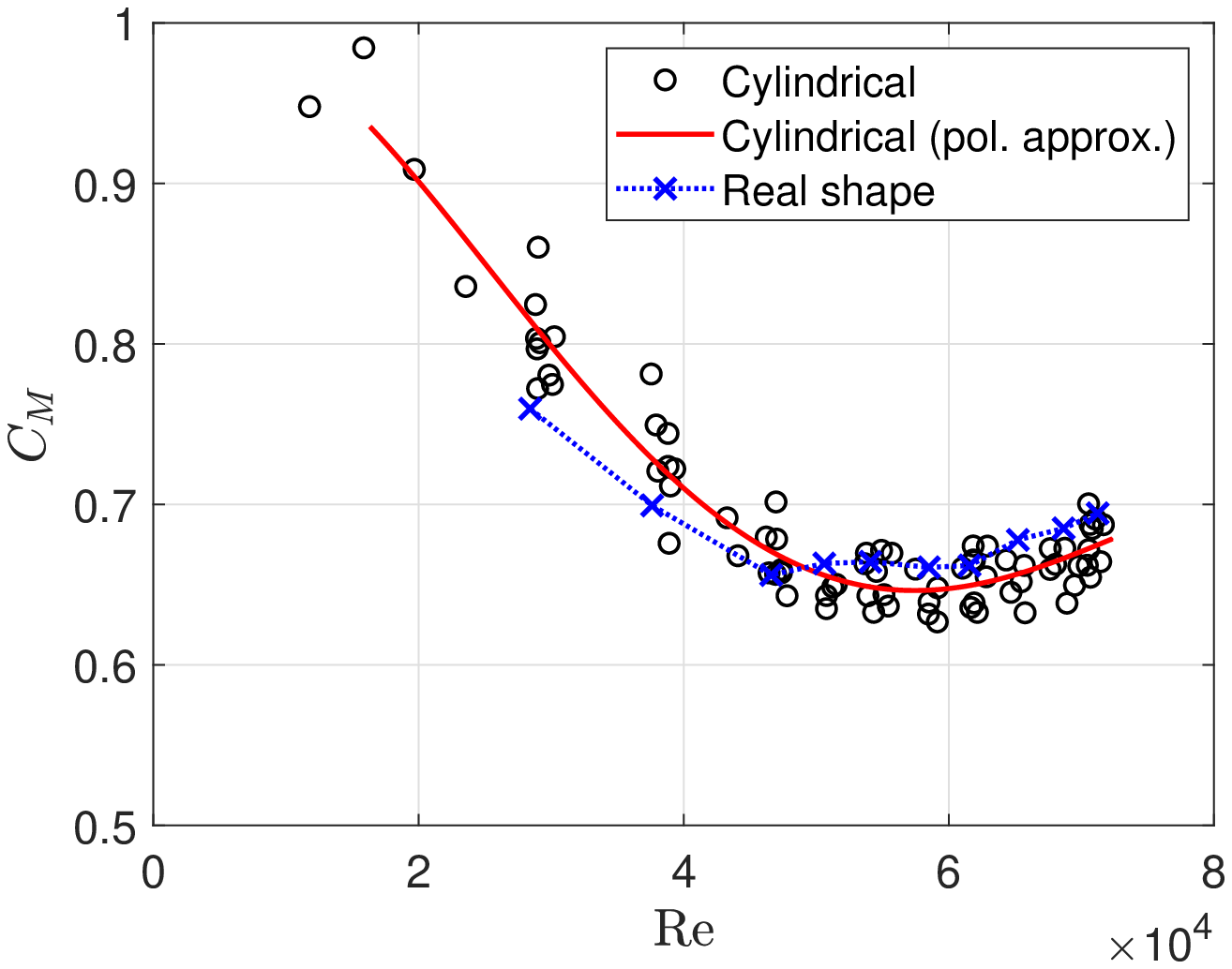}}%
	\caption{Resultant shear-force (a) and moment (b) coefficients at the base of the isolated tower. The solid red lines indicate the polynomial approximations of the data. The comparison between cylindrical and real-shape configurations is also outlined.}
	\label{fig:CF_CM_free_standing}
\end{figure}

Starting from the isolated tower, it was first verified that the results for the carbon-fibre model were reasonably the same as those for the ABS model, for which the surface roughness was calibrated. Moreover, these reference measurements were repeated several times during the experimental campaign, assembling the experimental rig in different moments, rotating the tower by various angles, and even using slightly different set-ups (e.g., both rigs described in Sections~\ref{single_tower_setup}-\ref{group_setup}). The results are collected in Fig.~\ref{fig:CF_CM_free_standing}, allowing for an estimate of the measurement uncertainty.

Mean $C_F$ and $C_M$ slightly higher than 0.6 and 0.65, respectively, were measured. It is worth noting that, for the isolated tower, the mean resultant shear-force coefficient is nearly identical to the mean drag coefficient.
This means that the resulting aerodynamic force is applied at a height slightly higher than $H/2$, in agreement with ESDU recommendations \citep{ESDU1987}.
The drag coefficient is somewhat higher than the value reported in Eurocode~1 for a tower with the same slenderness ratio (due to an overly small end-effect factor; this issue will be further discussed in Section~\ref{height_effect}), but the overturning-moment coefficient is only slightly lower than the value 0.7 provided by the \cite{CICIND2002} Model Code.

\subsection{Towers in double-row groups}

\begin{figure}
  \centering
	\subfigure[\label{fig:G-groups_corner}]
  {\includegraphics[angle=0, width=0.45\textwidth]{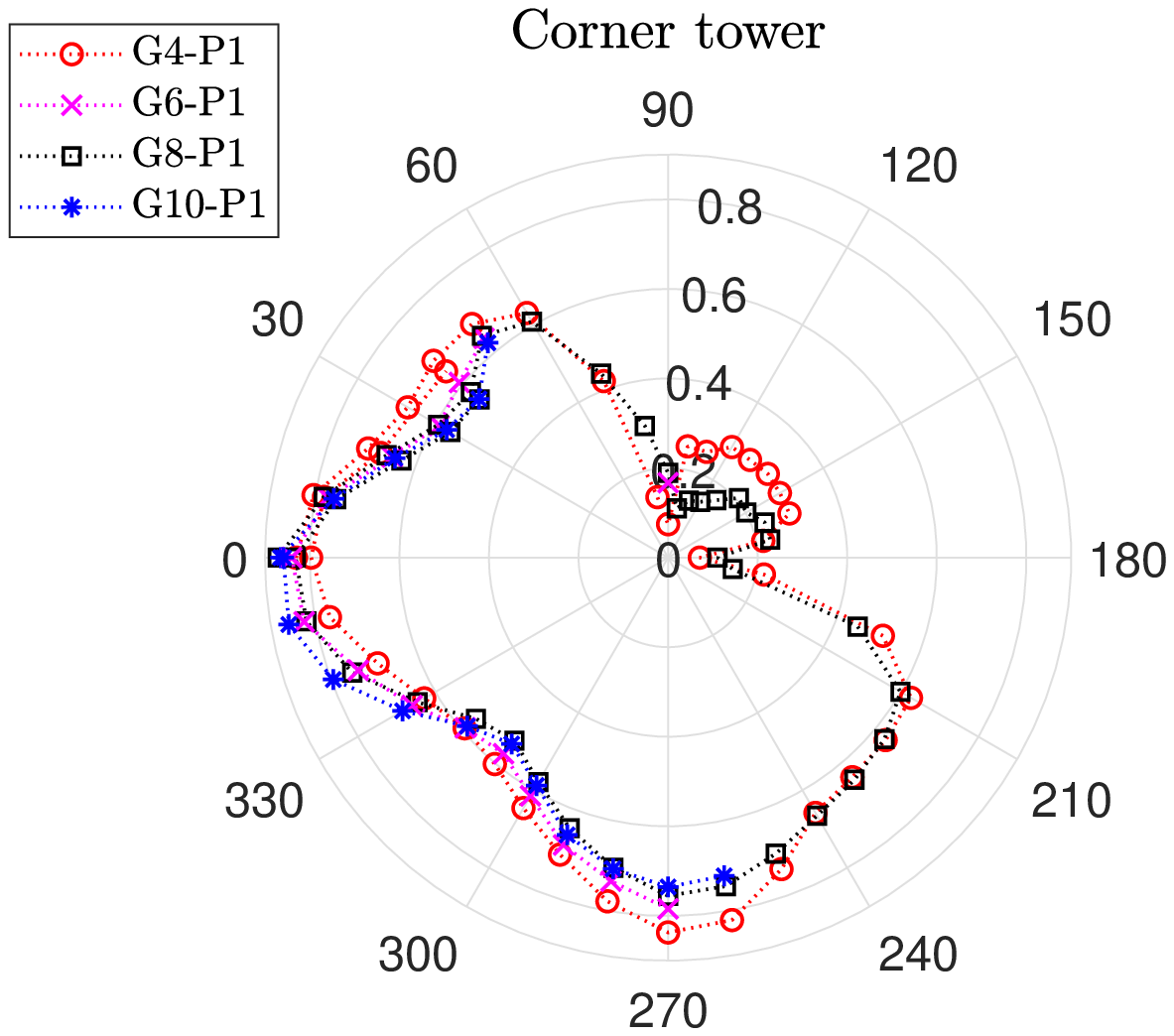}}\hspace{0.5cm}%
	\subfigure[\label{fig:G-groups_internal}]
  {\includegraphics[angle=0, width=0.45\textwidth]{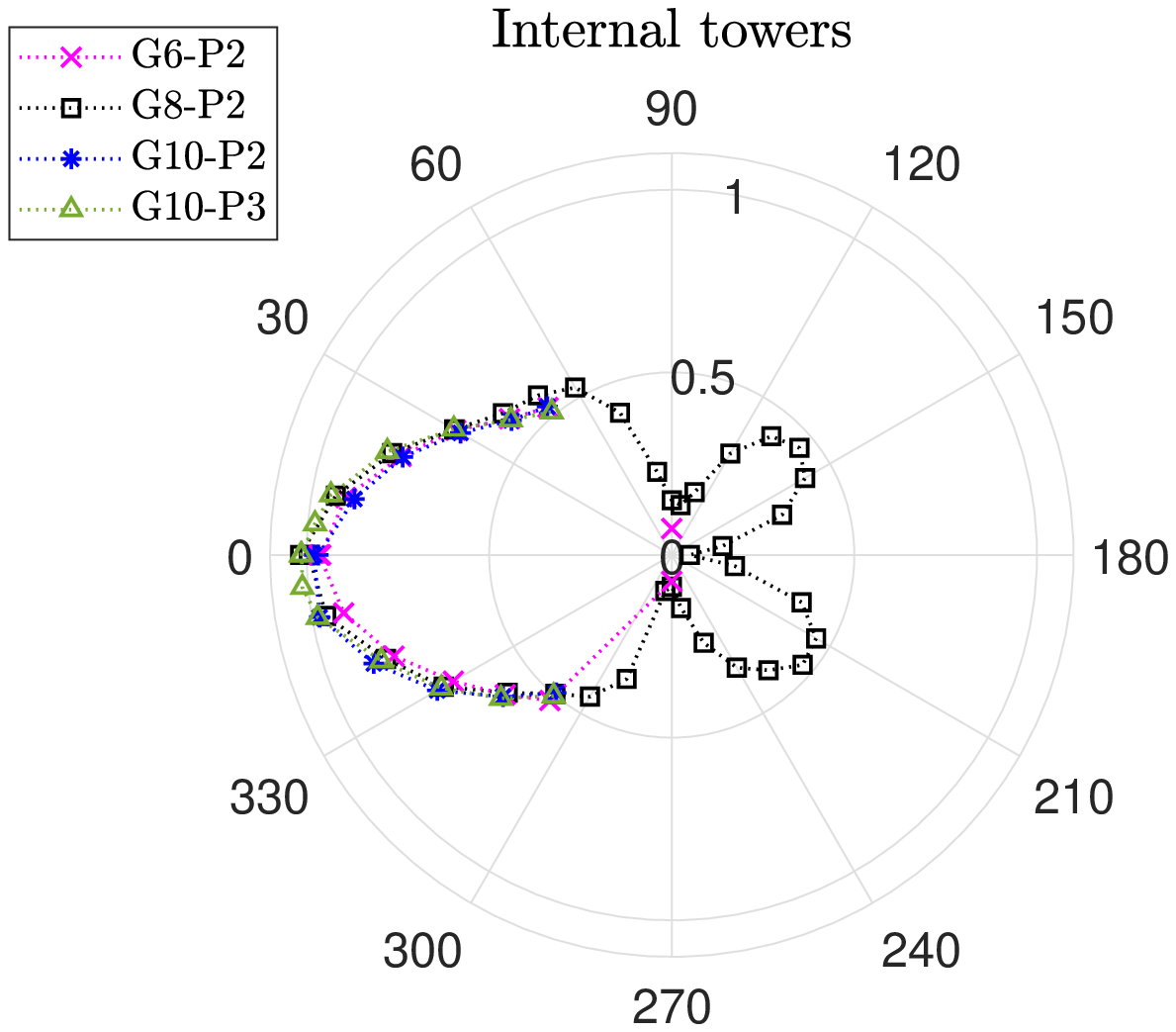}}%
	\caption{Comparison of mean resultant moment coefficients at the base of corner towers (a) and internal front towers (b) in double-row groups for different angles of attack $\beta$ (in deg). The repeatability of the results was verified in some cases by performing the same measurements more than once.}
	\label{fig:G-groups_CM}
\end{figure}

For the double-row groups of towers (G4 to G10 in Fig.~\ref{fig:configurations}), the largest loads affect the windward towers, and one can distinguish between corner and internal towers (Fig.~\ref{fig:G-groups_CM}). In general, the forces and moments acting on the latter are significantly higher. In any case, for the considered center-to-center distance between the towers, the loads are much higher than for the isolated tower (see Fig.~\ref{fig:CM_free_standing}). Noteworthy is also the good symmetry of the moment coefficient patterns, where expected.

\begin{figure}[t!]
  \centering
	\subfigure[\label{fig:G-groups_corner_theta}]
  {\includegraphics[angle=0, width=0.495\textwidth]{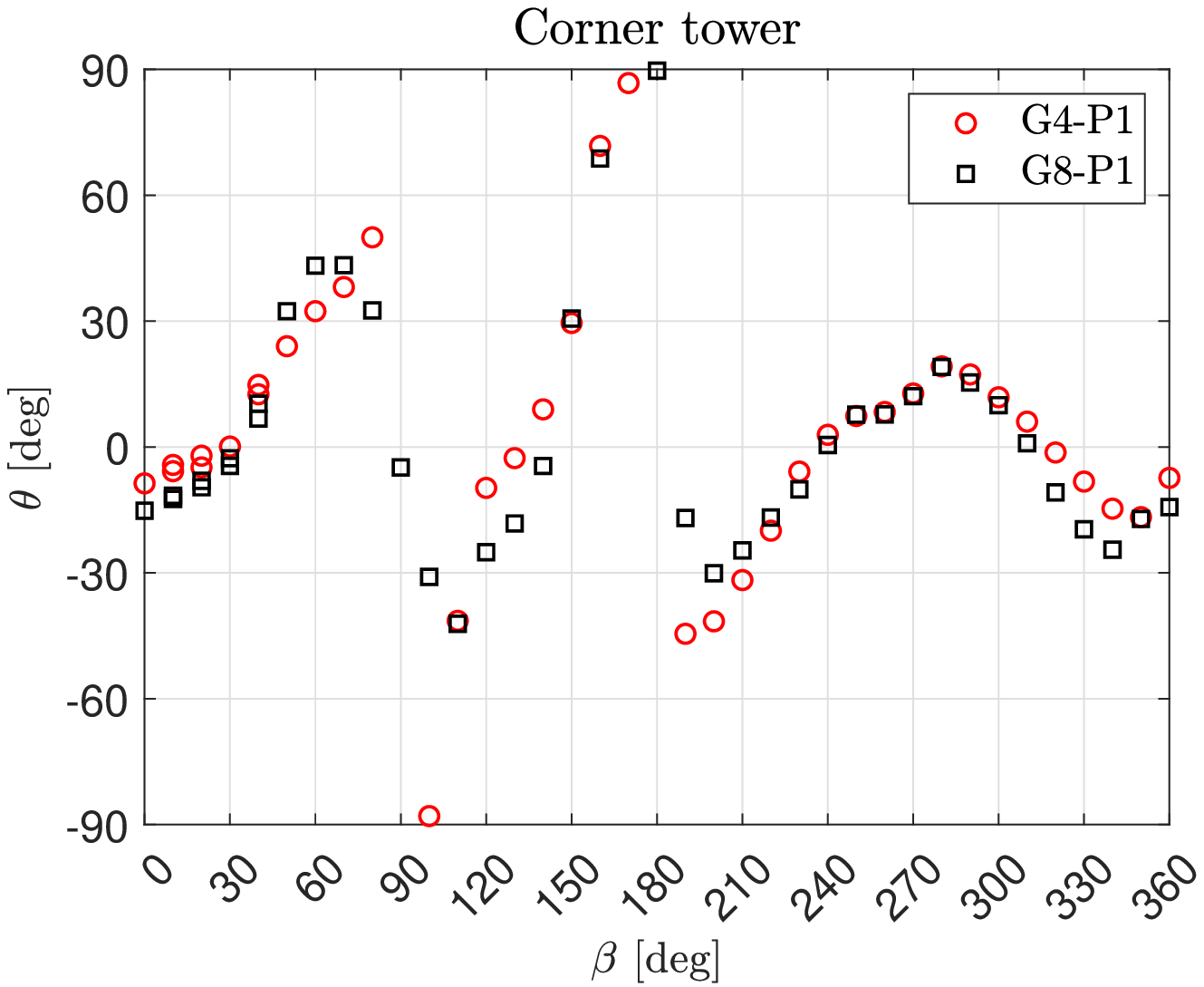}}%
	\subfigure[\label{fig:G-groups_internal_theta}]
  {\includegraphics[angle=0, width=0.495\textwidth]{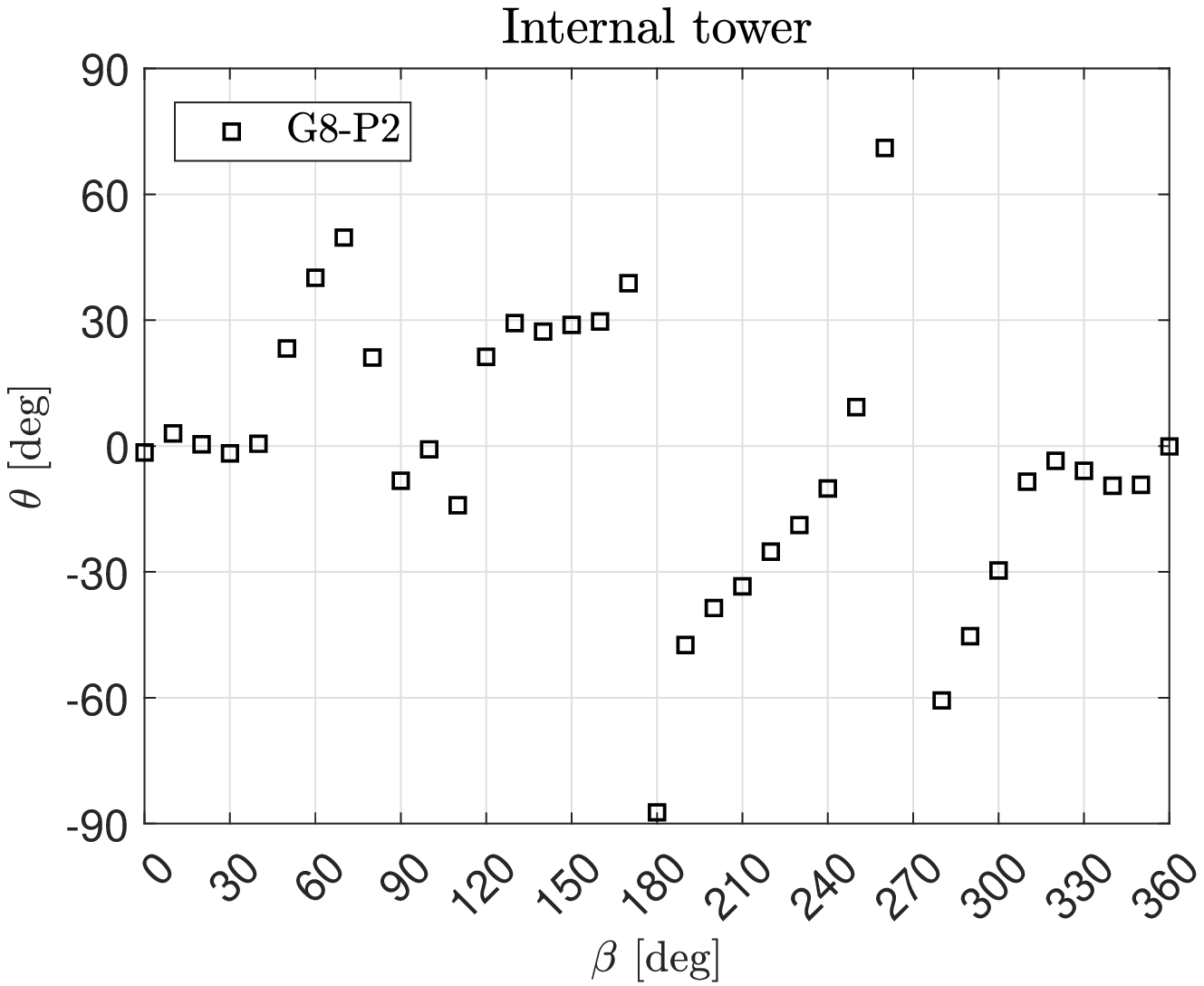}}%
	\caption{Angular deviation $\theta$ of the resultant force on the tower with respect to the wind direction $\beta$ ($\theta = 0$ means that the force is along-wind): corner towers in double-row groups G4 and G8 (a) and internal front tower for G8 (b).}
	\label{fig:G-groups_theta}
\end{figure}

Fig.~\ref{fig:G-groups_theta} shows the angular deviation $\theta$ of tower resultant force with respect to the wind direction for groups G4 and G8. In general, around the maximum mean values of base shear and moment coefficients these deviations are rather small, that is the aerodynamic force is nearly along-wind.

\begin{figure}
  \centering
	\includegraphics[angle=0, width=0.495\textwidth]{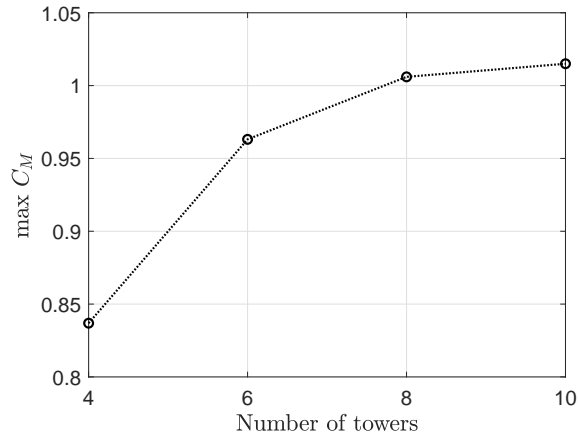}
	\caption{Evolution of the maximum mean resultant moment coefficient at the base of the most loaded tower in double-row groups.}
	\label{fig:G_groups_max_CM}
\end{figure}

Fig.~\ref{fig:G_groups_max_CM} reports the maximum mean moment coefficient measured for any tower and wind direction in the various group configurations. The load significantly increases as we pass from group G4, where there are only corner towers, to the group G6, where there are also internal towers. The maximum load increases further, by about 5\%, as the number of towers increases from six to eight. In contrast, for group G10, load increase is nearly negligible.
It is also worth noting that the maximum mean moment coefficient is always obtained for wind directions in the range $0 \leq \beta \leq -10$~deg (Fig.~\ref{fig:wind_direction}).

\subsection{Towers in single-row groups}

The loads on towers arranged in single-row groups are expected to be significantly higher than those just seen for the double-row groups. Indeed, the absence of the second row implies a lower pressure in the region behind the towers, with a consequent increase of drag and moment.

\begin{figure}
  \centering
	\subfigure[\label{fig:R4-P2}]%
  {\includegraphics[angle=0, width=0.45\textwidth]{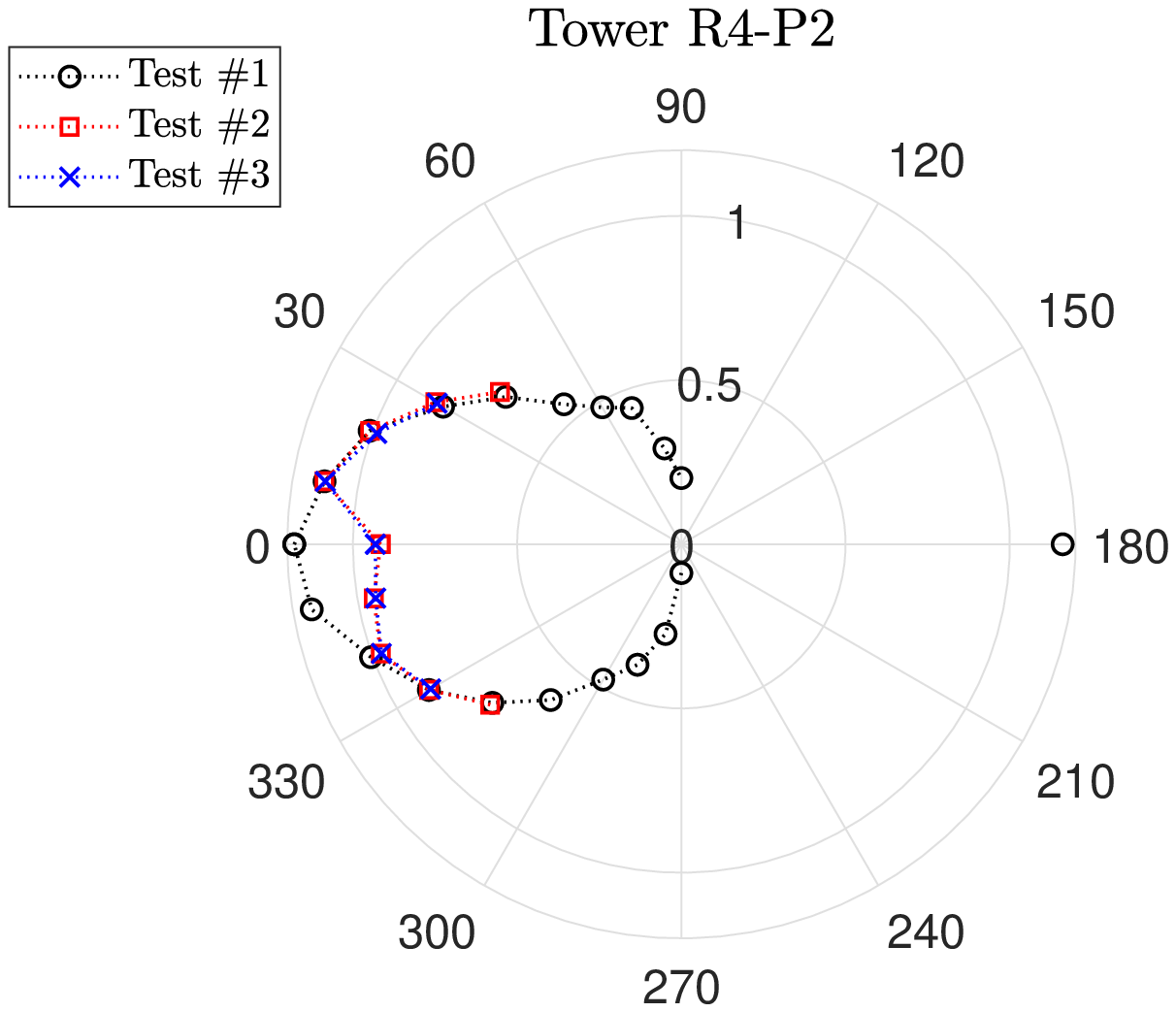}}\hspace{0.5cm}%
	\subfigure[\label{fig:R5-P3}]%
  {\includegraphics[angle=0, width=0.45\textwidth]{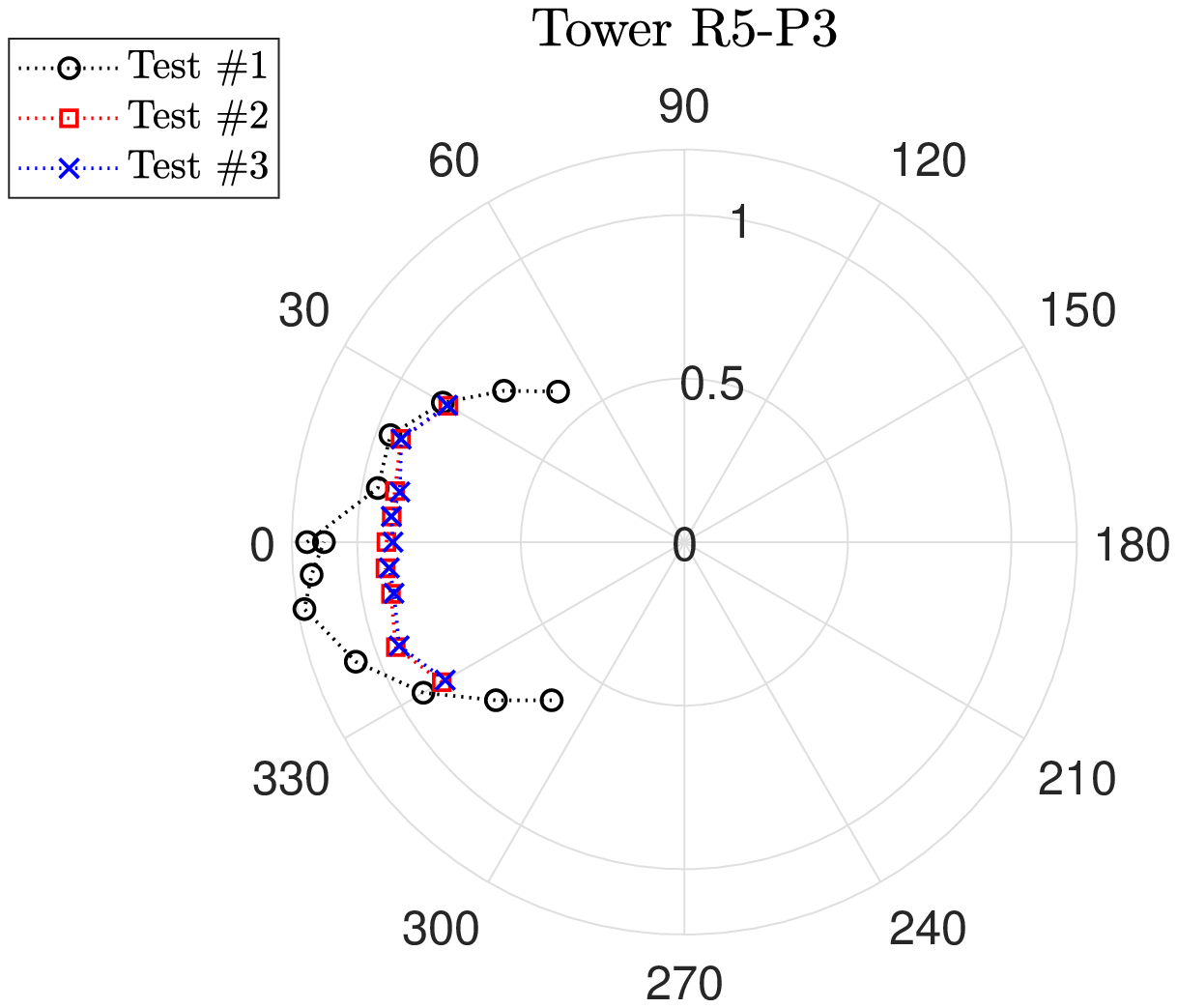}}%
	\caption{Mean resultant moment coefficients at the base of the internal tower P2 in the single-row group R4 (a) and internal tower P3 in group R5 (b) for different angles of attack $\beta$ (in deg).}
	\label{fig:R-groups_CM}
\end{figure}

Fig.~\ref{fig:R-groups_CM} depicts the resultant moment coefficient diagrams for the internal towers, which are most heavily loaded, in groups of four and five towers. In both cases, a symmetric trend of the coefficient with wind direction (or nearly symmetric for R4-P2) is obtained, as expected, in some tests. Nevertheless, imperceptible changes in tower arrangement after dismounting and reassembling the towers in the wind tunnel were sometimes sufficient to trigger different flow patterns for wind incidences around 0~deg (wind perpendicular to the tower line), resulting in strongly asymmetric trends of $C_M$. Interestingly, for the tower R4-P2 the maximum load is significantly higher in case of a nearly symmetric pattern (test~\#1 in Fig.~\ref{fig:R4-P2}), while for the tower R5-P3 the symmetrical solutions are less demanding (tests~\#2 and \#3 in Fig.~\ref{fig:R5-P3}).
The biased flow is associated with the instability of the stream between the towers, and such a behavior has already been encountered for a pair of side-by-side circular cylinders with a center-to-center distance ratio approximately in the range $1.1-1.2 < d/D_{eq} < 2-2.2$, in smooth subcritical Reynolds number flow (see e.g. \cite{Sumner1999,Sumner2010}). Interestingly, a similar result is observed here for finite-height towers in an atmospheric boundary layer flow and transcritical Reynolds number conditions.

The loads on the corner towers are significantly smaller than those on the internal towers ($C_M$ is never larger than 1.0), but the increase compared to double-row arrangements is still apparent.
In contrast to the present results (considering also the group R3, see Fig.~\ref{fig:configurations}), for three towers in a side-by-side arrangement ($\theta = $ 0~deg), \cite{Mitra2006} found that the central tower experiences higher drag but, due to the lateral force contribution, the highest loads are seen on the external towers. However, it is worth remembering that the towers studied in \cite{Mitra2006} were less slender ($H/D = 11$) and closer to each other ($d/D = 1.14$), and the Reynolds number flow regime was subcritical.

Finally, the outcome highlighted for the isolated tower, namely that the resultant aerodynamic force acts slightly above the tower mid-height, is confirmed for the group arrangements.

\begin{figure}
  \centering
	\subfigure[\label{fig:gust_factors_G4}]%
  {\includegraphics[angle=0, width=0.45\textwidth]{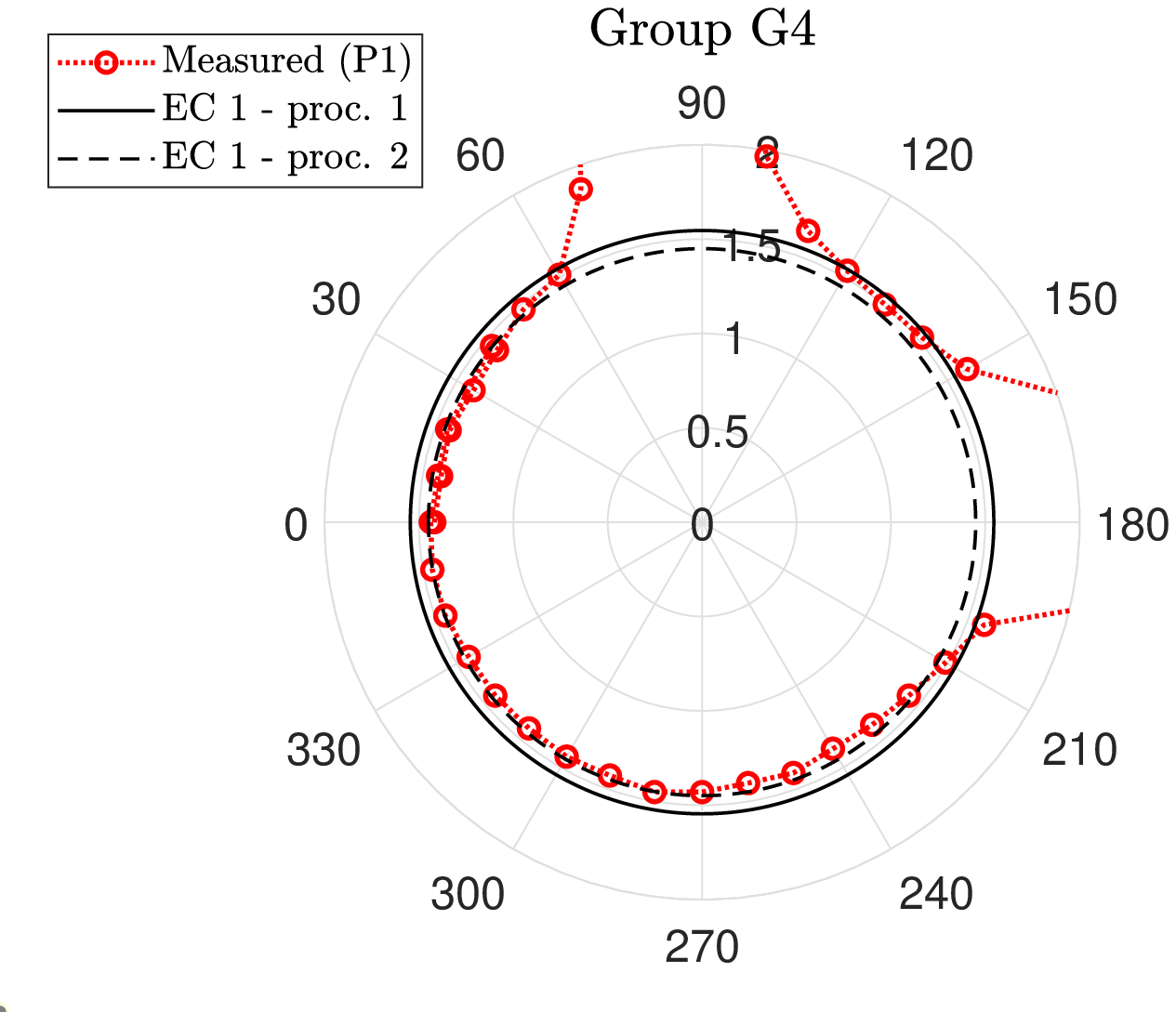}}\hspace{0.5cm}%
	\subfigure[\label{fig:gust_factors_G8}]%
  {\includegraphics[angle=0, width=0.45\textwidth]{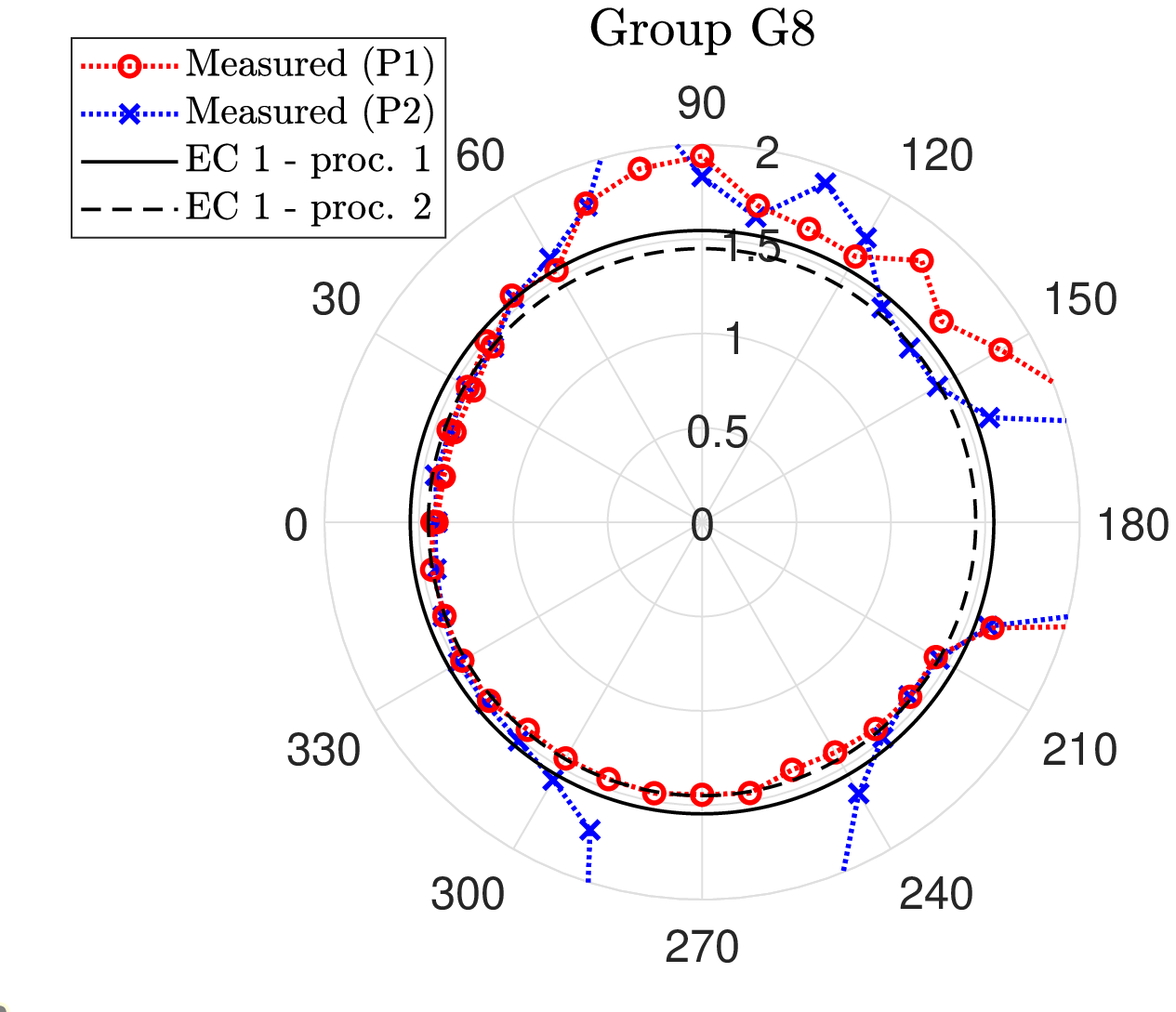}}%
	\caption{Moment gust factors measured for the towers in the double-row groups G4 (a) and G8 (b) for different angles of attack $\beta$ (in deg).}
	\label{fig:gust_factors}
\end{figure}

\subsection{Gust factors}
\label{gust_factors}

To design the supporting structures of the towers pre-assembled at the quayside, wind load peak values over time must be calculated. This can be done using the methods proposed by codes and standards (e.g. \cite{Eurocodice1}) based on the mean force and moment coefficients discussed so far. However, limiting to the quasi-static contribution, the gust load factors can also be determined directly from measurements using the high-frequency force balance. Without invoking the assumption of Gaussian processes, the peak resultant shear-force and moment coefficients ($\hat{C}_F$ and $\hat{C}_M$) are calculated as the mean of the absolute maxima obtained over time windows corresponding to 600~s at full scale \citep{Davenport1961} (based on the recording time of 120~s and the time scale of 1:120.7 mentioned in Section~\ref{wind_modelling}, 24 windows are available for each measurement). Then, the gust factors related to the base shear and moment are calculated respectively as:
\begin{equation}
 G_{F} = \frac{\hat{C}_F}{C_F}
\label{eq:gust_factor_shear}
\end{equation}
\begin{equation}
 G_{M} = \frac{\hat{C}_M}{C_M}
\label{eq:gust_factor_moment}
\end{equation}

Fig.~\ref{fig:gust_factors} shows examples of gust factors calculated for the towers in groups G4 and G8 based on the experimental measurements.
In all cases, for both base shear force and moment, except when the loads on the towers are very low (i.e., when the considered tower is sheltered by another tower), the gust factor is nearly constant and close to 1.5.

It is interesting to compare the current results with those obtained through Eurocode~1, considering the size factor but not the dynamic factor, as the towers are modeled here as rigid bodies. By applying either Procedure 1 or Procedure 2 \citep{Eurocodice1} as if the considered tower was isolated, and assuming the turbulence properties measured in the wind tunnel, one can see that the agreement is very good. In particular, Procedure 2 seems to provide the most accurate predictions. However, it is worth considering that the force and moment time histories were low-pass filtered with a cut-off frequency of 40~Hz (see Section~\ref{single_tower_setup}), and this is expected to entail a reduction of the experimental gust factor of about 2-3\% \citep{Mannini2023}. Consequently, one can probably say that the experimental data fall in between the predictions of Eurocode~1 based on the two available procedures.
Data filtering also excludes the contribution of possible narrow-band excitations, which are outside the scope of the current investigation.

At this stage, it is also possible to discuss the mismatch of the longitudinal integral length scale of turbulence, mentioned in Section~\ref{wind_modelling}. This shortcoming of the experimental set-up is expected to lead to an underestimation of the gust factors. However, given the accuracy of Eurocode~1 in estimating $G_F$ and $G_M$, one can use Procedure 1 and Procedure 2 to estimate the error associated with the mismatch of $L_u^x$. Underestimation of the gust factors of around 5\% is estimated according to Procedure 1, and between 8 and 10\% according to Procedure 2.


\begin{figure}
  \centering
	\subfigure[\label{fig:G4_overall_CM}]%
	{\includegraphics[angle=0, width=0.45\textwidth]{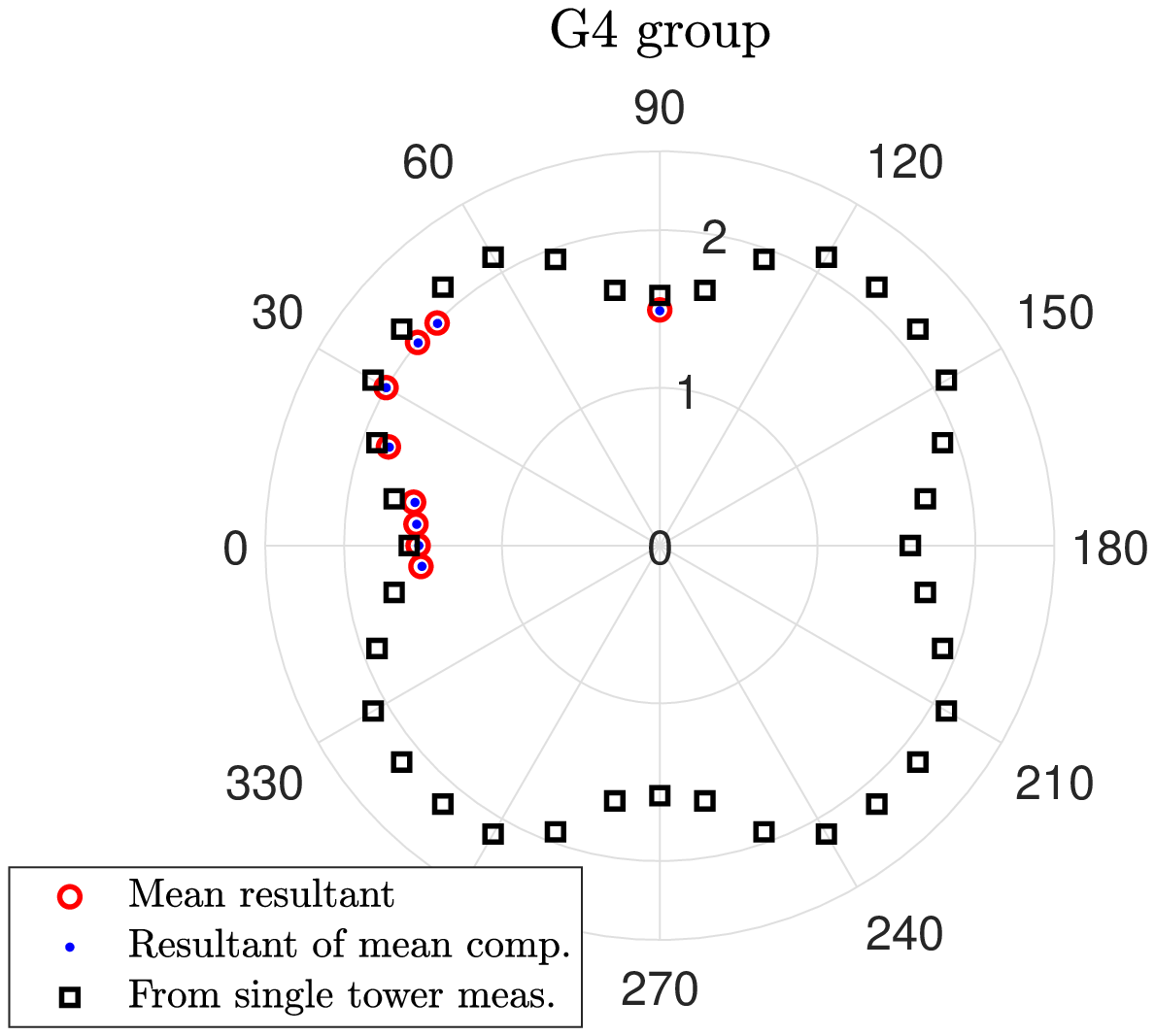}}\hspace{0.5cm}%
	\subfigure[\label{fig:G8_overall_CM}]%
	{\includegraphics[angle=0, width=0.45\textwidth]{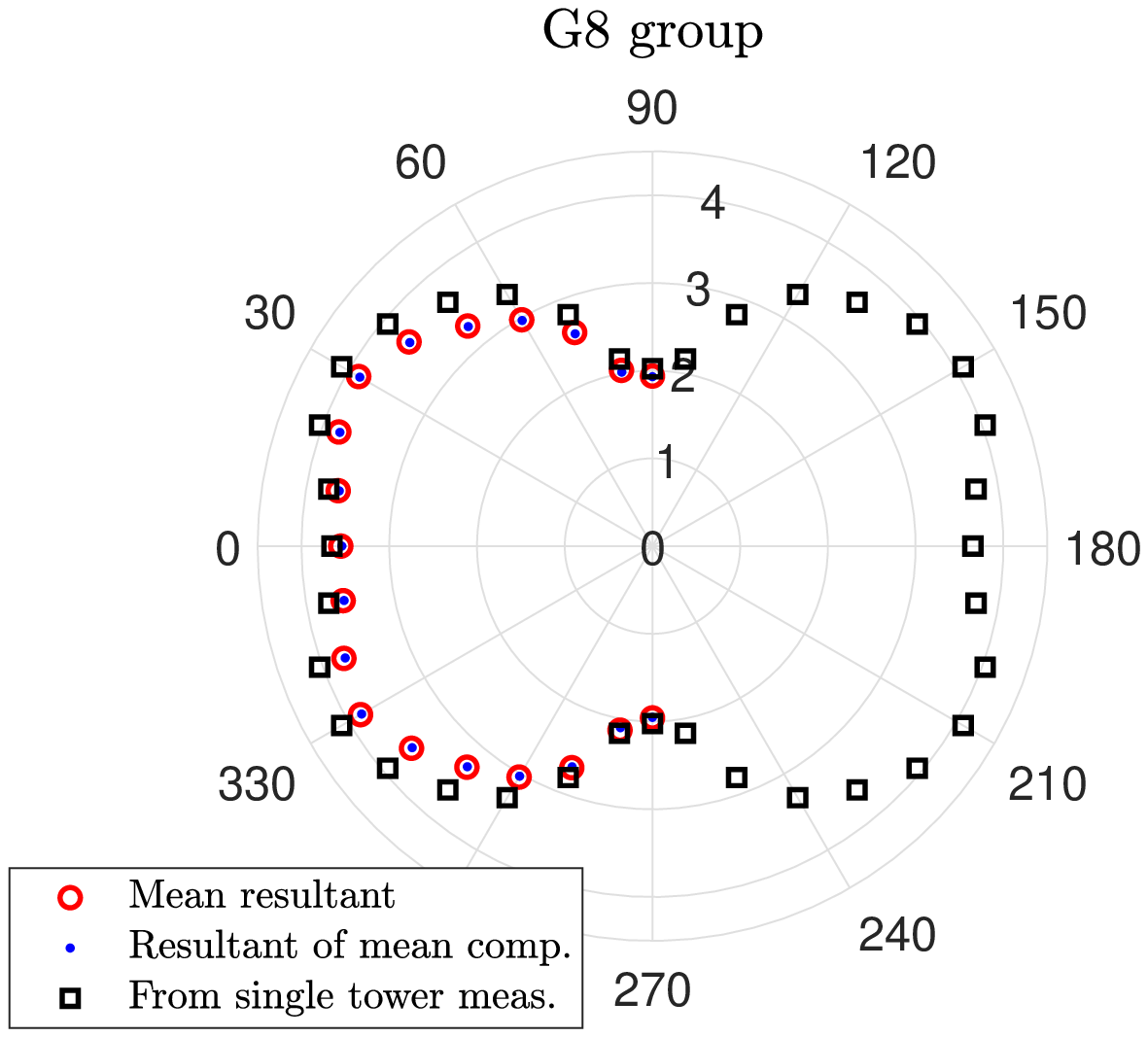}}%
	\caption{Moment coefficients associated with the mean overall resultants at the base of the groups of towers G4 (a) and G8 (b) for different angles of attack $\beta$ (in deg). A comparison is proposed with the vector sum of the mean Cartesian components of the measured moments and with the loads obtained from single-tower measurements. It is to note that also the overall load coefficients are defined according to Eqs.~(\ref{eq:shear_coefficient})-(\ref{eq:moment_coefficient}), and this explains their large values.}
	\label{fig:Overall_CM}
\end{figure}

\begin{figure}
  \centering
	\includegraphics[angle=0, width=0.45\textwidth]{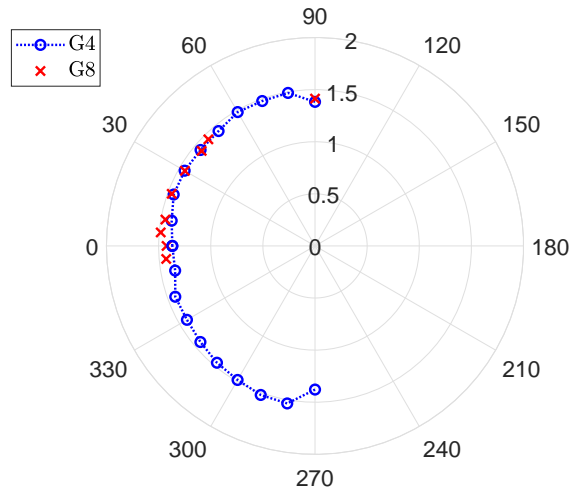}\vspace{0.5cm}%
	\caption{Gust factors for the overall moment coefficient at the base of the double-row groups G4 and G8 for different angles of attack $\beta$ (in deg).}
	\label{fig:Gust_factor_G4_G8}
\end{figure}

\subsection{Overall loads for groups of towers}
\label{overall_group_loads}

In addition to the loads at the base of each tower, the resultant shear force and overturning moment were also measured for the whole tower groups. Here too, the symmetries of the various arrangements were exploited to minimize the number of tests necessary to investigate the effect of wind direction. As an example, the mean resultant moment coefficient is reported in Fig.~\ref{fig:Overall_CM} for groups G4 and G8. It attains a maximum value of about 2 around $\beta =$ 45~deg (or 135, 225, 315~deg) for group G4 and 4 for a wind direction of about 30~deg (or 150, 210, 330~deg) for group G8.

As for single-tower measurements, theoretically the mean magnitude of the resultant vector of the base actions is not equal to the vector sum of the mean Cartesian components of forces and moments. Nevertheless, for the level of fluctuation due to turbulence, Fig.~\ref{fig:Overall_CM} clearly shows that the two results are practically the same.

This means that a comparison can be made between the overall mean loads directly measured on the groups of towers and those reconstructed from single-tower measurements exploiting the group symmetries. Obviously, such a reconstruction emphasizes measurement errors and small imperfections in the two set-ups.
The agreement between the two sets of results is satisfactory for group G4 (Fig.~\ref{fig:G4_overall_CM}), for which just the loads on tower P1 are required to calculate the total loads acting on the overall group. The same can be said for group G8 (Fig.~\ref{fig:G8_overall_CM}), for which the results for both towers P1 and P2 must be used, although larger discrepancies are found for wind directions in the ranges $20 \leq \beta \leq 70$~deg and $300 \leq \beta \leq 340$~deg.

Although the overall mean loads acting on the groups of towers can reasonably be estimated from single-tower measurements, the peak loads obviously cannot. Hereunto, the moment gust factor measured for groups G4 and G8 is shown for various wind directions in Fig.~\ref{fig:Gust_factor_G4_G8}. Interestingly, in both cases, it always assumes again values very close to 1.5.

\section{Parametric studies}
\label{parametric_results}

\subsection{Effect of the turbulent wind profile}

\begin{figure}
	\centering
	\subfigure[\label{fig:U_UH_partE}]
  {\includegraphics[angle=0, width=0.495\textwidth]{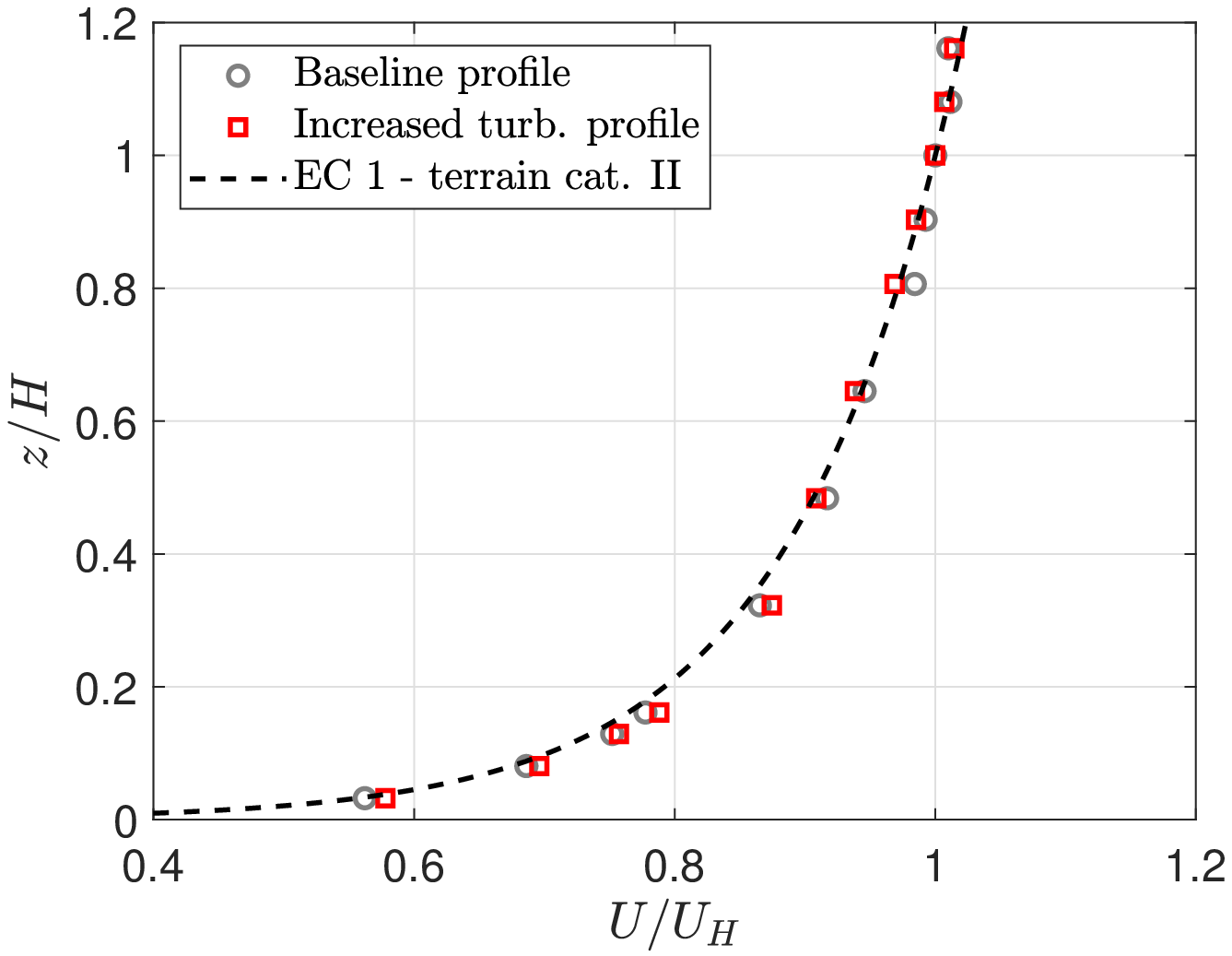}}\hspace{0.0cm}%
	\subfigure[\label{fig:Iu_partE}]
  {\includegraphics[angle=0, width=0.495\textwidth]{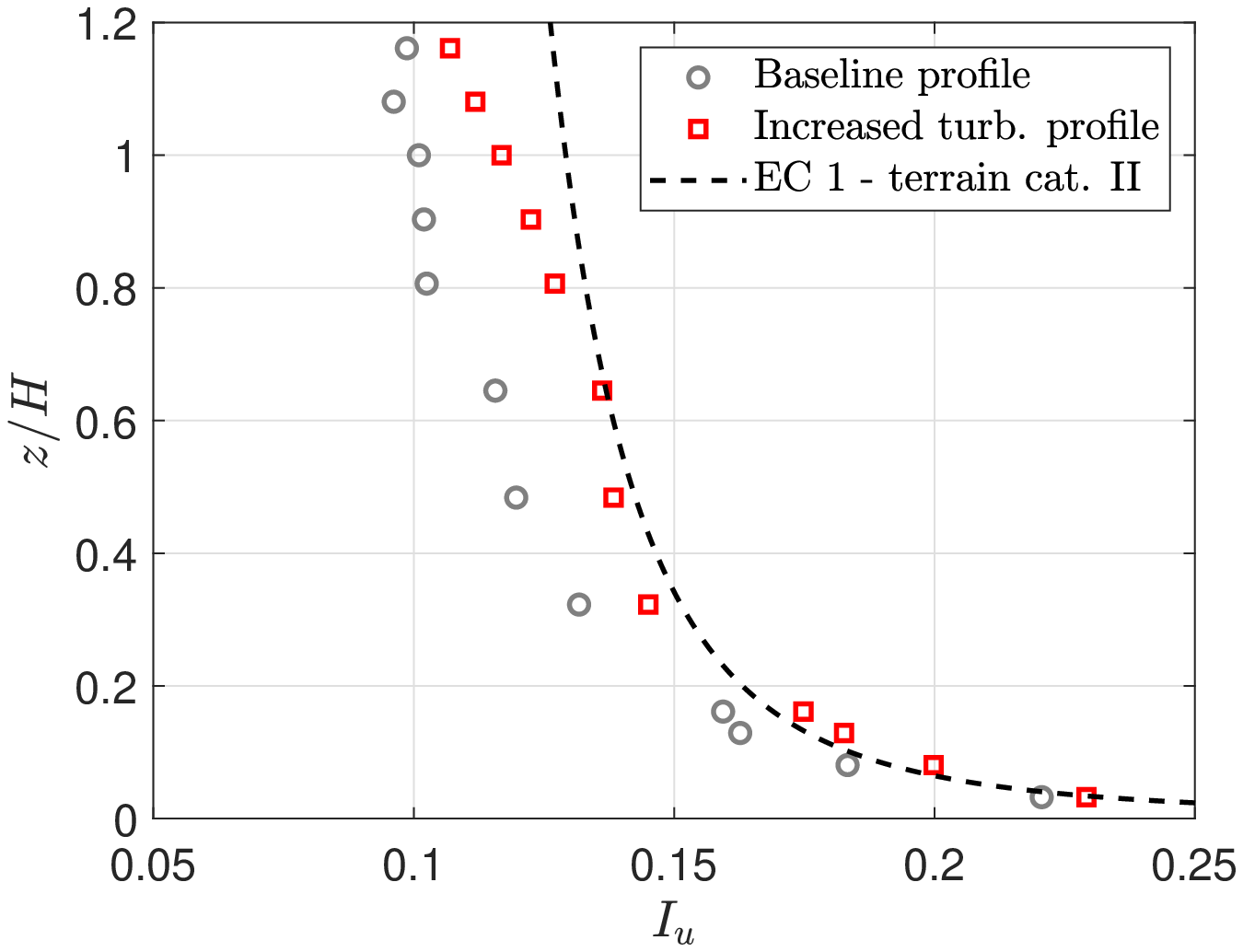}}\\%
	\caption{(a) mean wind speed (normalized) and (b) longitudinal turbulence intensity for the parametric study on the effect of the turbulent wind profile: comparison with Eurocode~1's profiles for a terrain category II and with the baseline profiles (see Fig.~\ref{fig:wind profiles}).}
	\label{fig:wind profiles_partE}
\end{figure}

One of the questions that arises after discussion of the many results of the baseline investigation concerns the sensitivity of the mean load coefficients to turbulent wind characteristics. To answer it, slight changes were made to the system of castellated barriers and roughness elements placed upstream of the models. Then, the mean wind speed and turbulence intensity profiles approached those prescribed by Eurocode~1 for a terrain category II (rather than I, used for the baseline tests; see Fig.~\ref{fig:wind profiles_partE}).

The mean base shear-force and moment coefficients measured for the isolated tower under this new wind condition fall in the uncertainty band for the baseline tests, close to the approximating polynomial curves reported in Fig.~\ref{fig:CF_CM_free_standing}.

The results for the more turbulent wind profile are also reported here for group G8. For both towers P1 and P2, we observed a small reduction in the mean moment coefficient compared to the baseline configuration for the wind directions for which the loads are maximum. However, the impact of slightly changing the turbulent wind profile is modest here, thanks too to the small difference in mean wind speed (Fig.~\ref{fig:wind profiles_partE}).

\begin{figure}
  \centering
	\subfigure[\label{fig:G8_P1_turb}]%
  {\includegraphics[angle=0, width=0.45\textwidth]{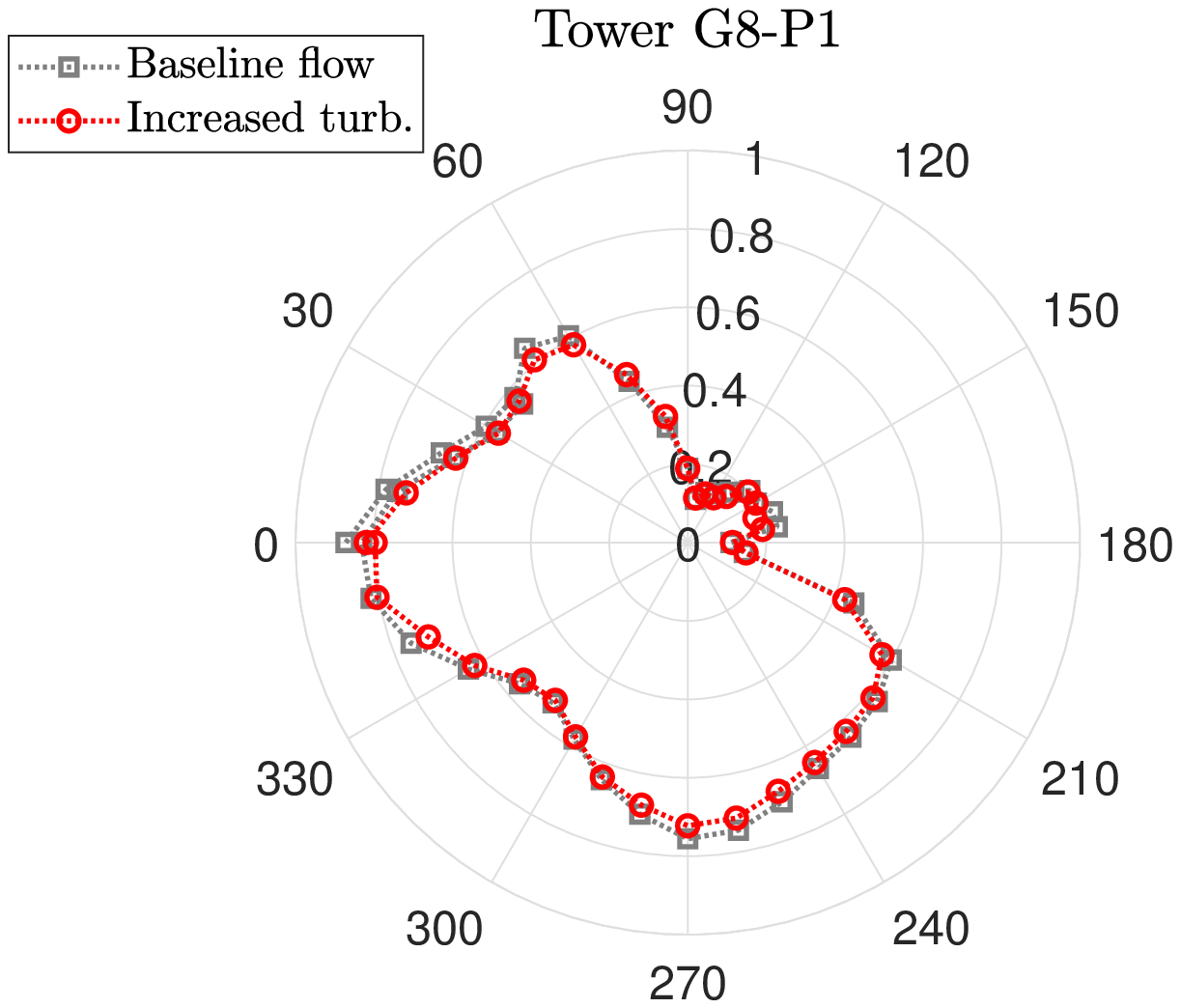}}\hspace{0.5cm}%
	\subfigure[\label{fig:G8_P2_turb}]%
  {\includegraphics[angle=0, width=0.45\textwidth]{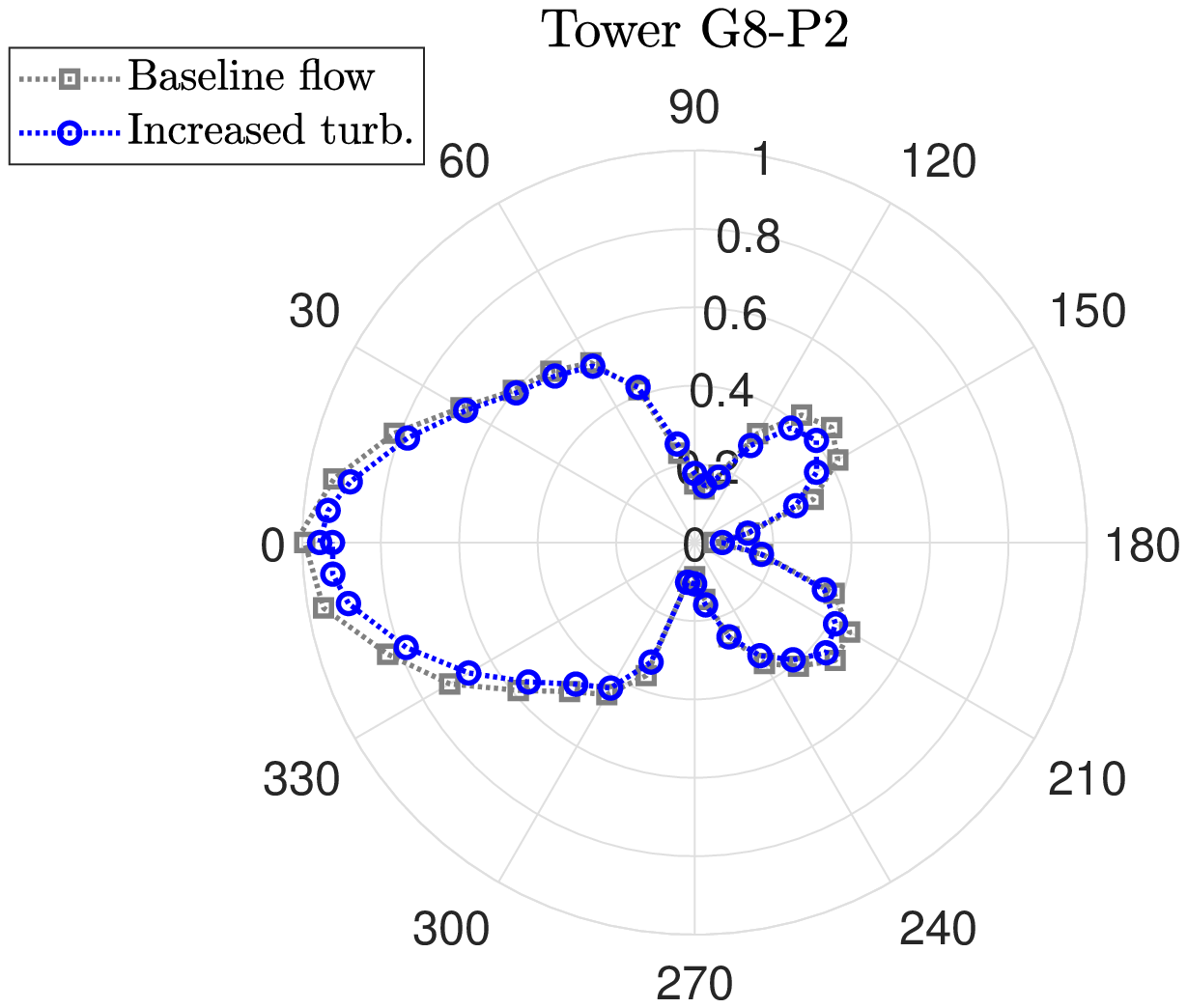}}%
	\caption{Mean resultant moment coefficient at the base of towers P1 (a) and P2 (b) in the double-row group G8 for different angles of attack $\beta$ (in deg): comparison of results for baseline and increased-turbulence profiles.}
	\label{fig:G8_turb}
\end{figure}


\subsection{Effect of tower shape}
\label{tower_shape}

\begin{figure}
  \centering
	\subfigure[\label{fig:G4_shape}]%
  {\includegraphics[angle=0, width=0.45\textwidth]{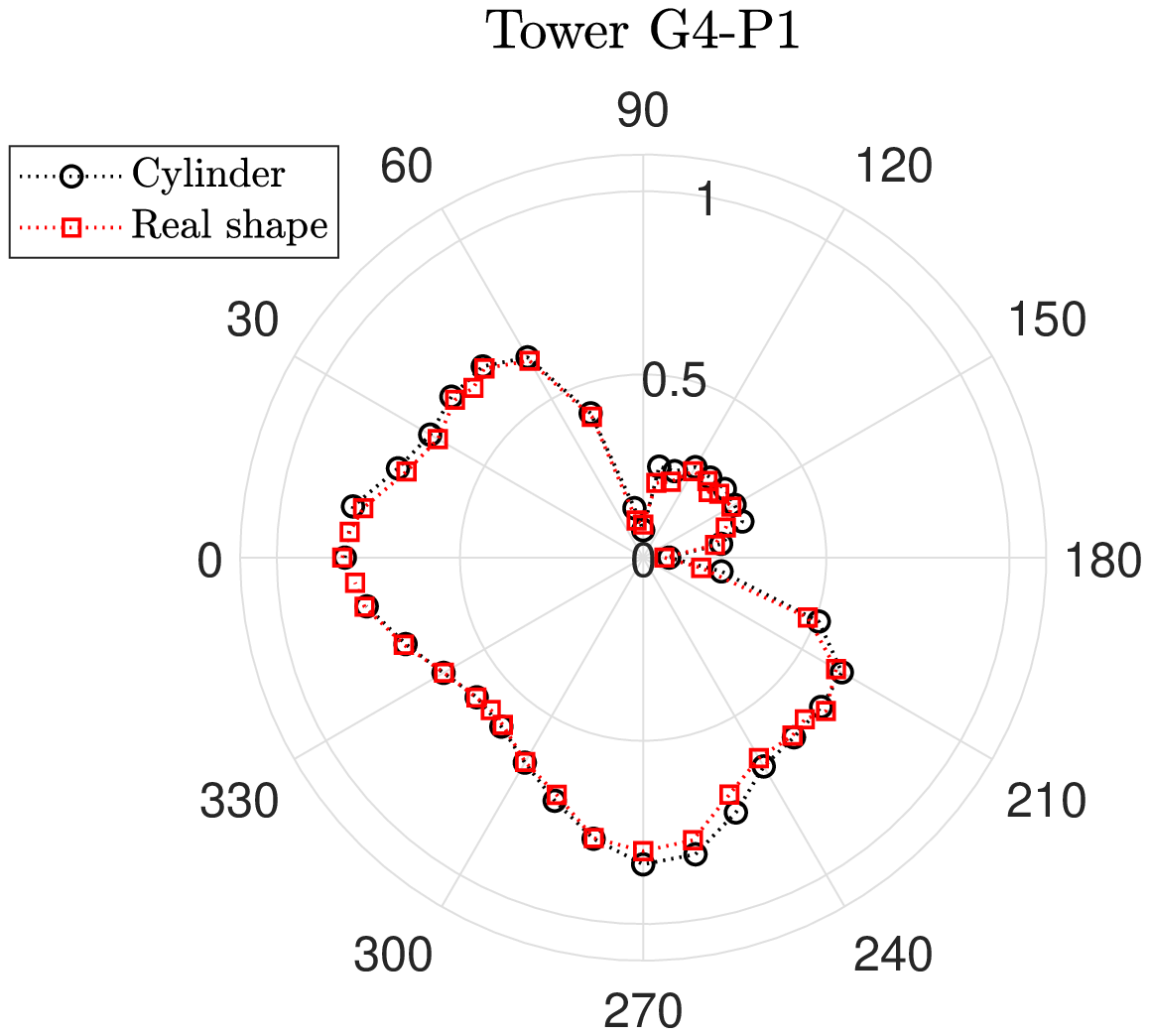}}\hspace{0.5cm}%
	\subfigure[\label{fig:R4_shape}]%
  {\includegraphics[angle=0, width=0.45\textwidth]{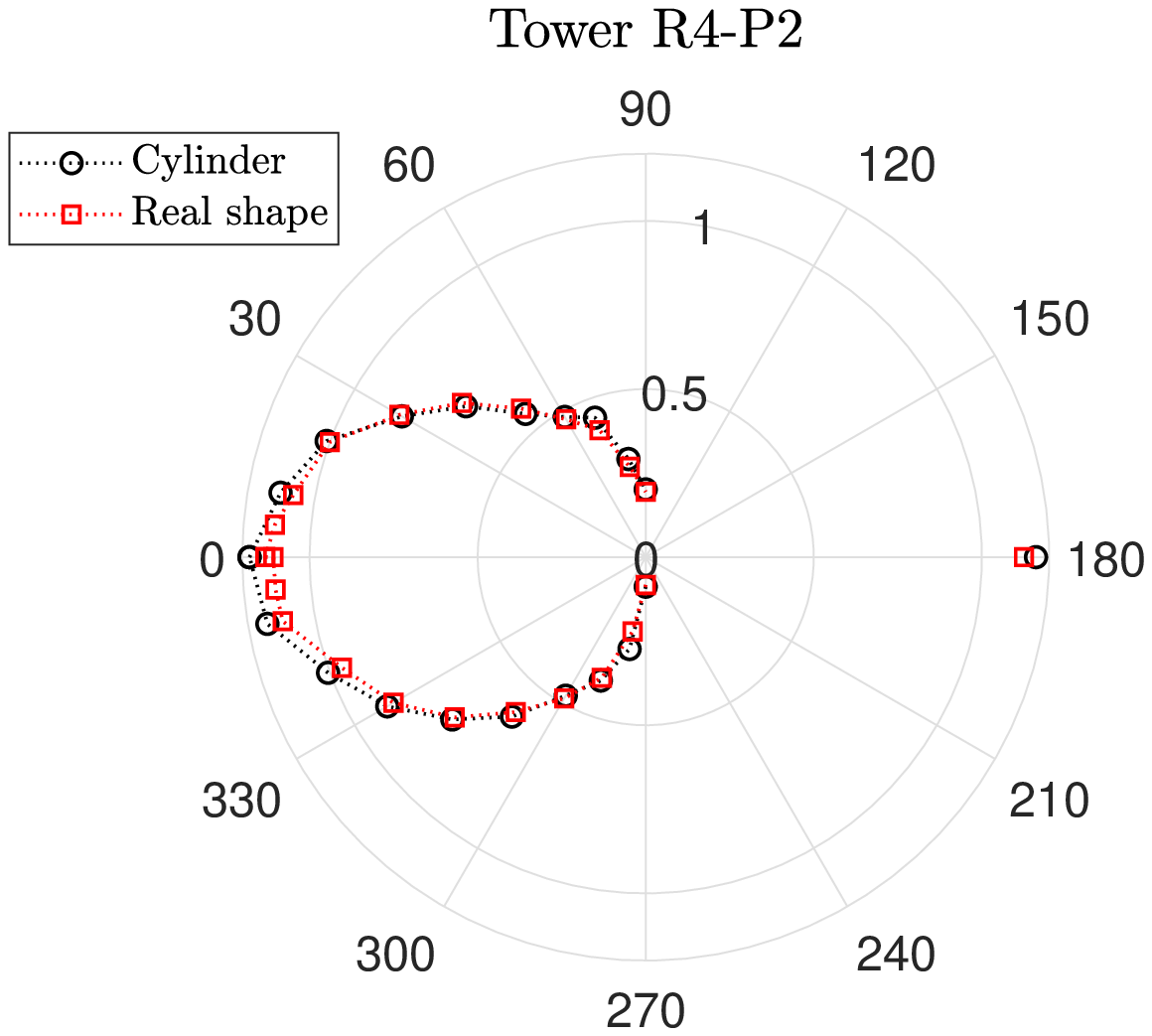}}%
	\caption{Mean resultant moment coefficient at the base of tower P1 in the double-row group G4 (a) and tower P2 in the single-row group R4 (b) for different angles of attack $\beta$ (in deg): comparison of results for equivalent cylindrical and real-shape towers.}
	\label{fig:G4_R4_shape}
\end{figure}

The effect of tower shape was investigated by comparing the results for simplified cylindrical structures with those with real geometry, including all tapered portions. The results revealed that the difference between the two cases is small both for the isolated tower (Fig.~\ref{fig:CF_CM_free_standing}) and for groups G4 and R4 (Fig.~\ref{fig:G4_R4_shape}). For the isolated configuration, this is particularly true in terms of mean base moment coefficient, thus corroborating the rationale for choosing an equivalent diameter, which was based on the moment coefficient (see Section~\ref{Equivalent_diameter}).

However, it is worth noting that in this case biased flow configurations were not observed for group R4 with real-shape towers. Therefore, the comparison outlined in Fig.~\ref{fig:R4_shape} for the tower R4-P2 refers to the results of test \#1 in Fig.~\ref{fig:R4-P2}.

\subsection{Effect of tower height}
\label{height_effect}

Another extensive study was carried out on the effect of tower height. From a practical engineering point of view, tests conducted on isolated towers with heights varying from 100~m to 135~m aimed at defining a sort of correction factor that might reasonably be used for the group configurations too (at least for the directions where the loads are higher). In addition, tests on groups G4 and R4 with towers of 105~m and 125~m were performed to validate this correction factor.

\begin{figure}[t]
   \centering
	\subfigure[\label{fig:S_height_CF}]%
  {\includegraphics[angle=0, width=0.495\textwidth]{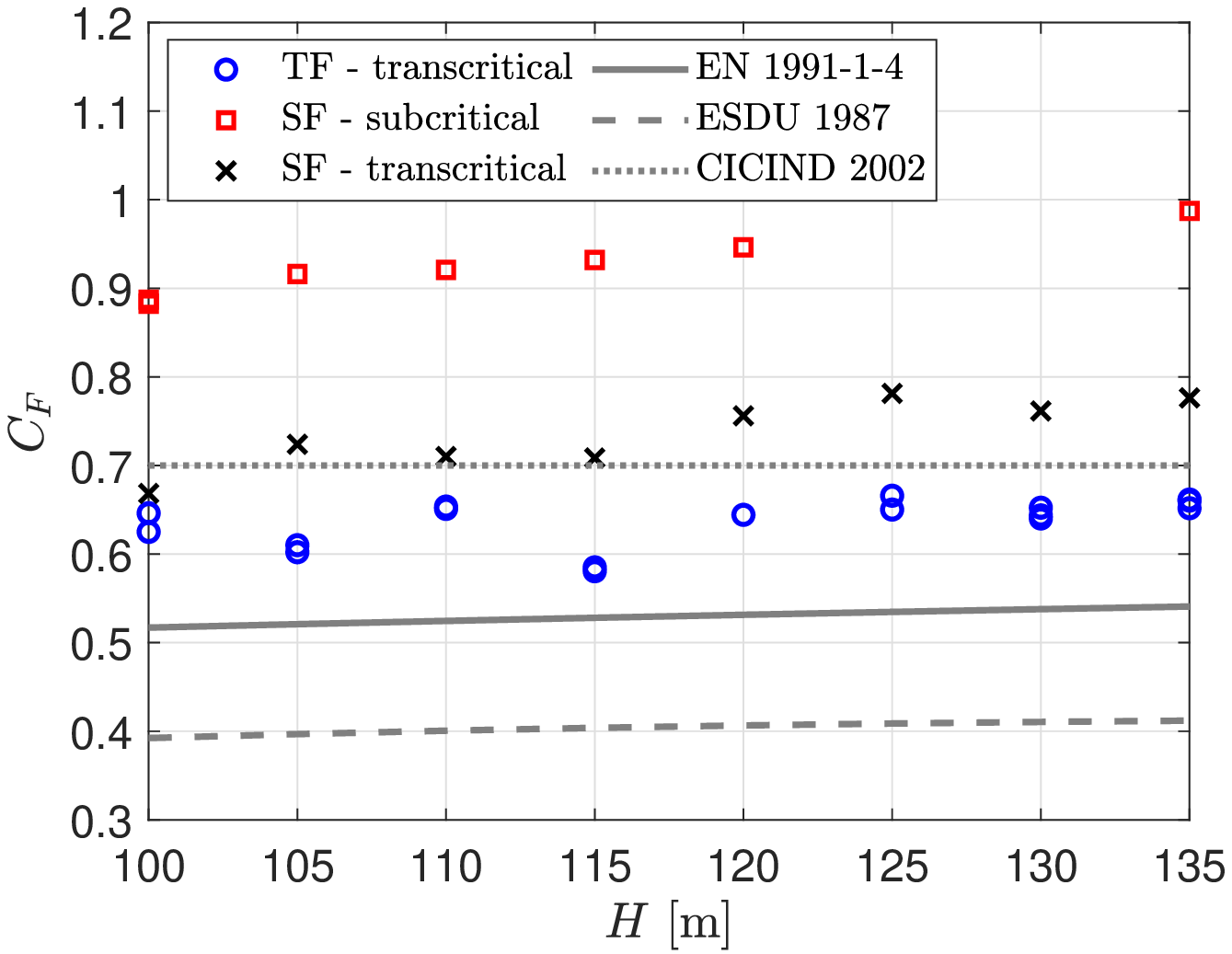}}\hspace{0.0cm}%
	\subfigure[\label{fig:S_height_CM}]%
  {\includegraphics[angle=0, width=0.495\textwidth]{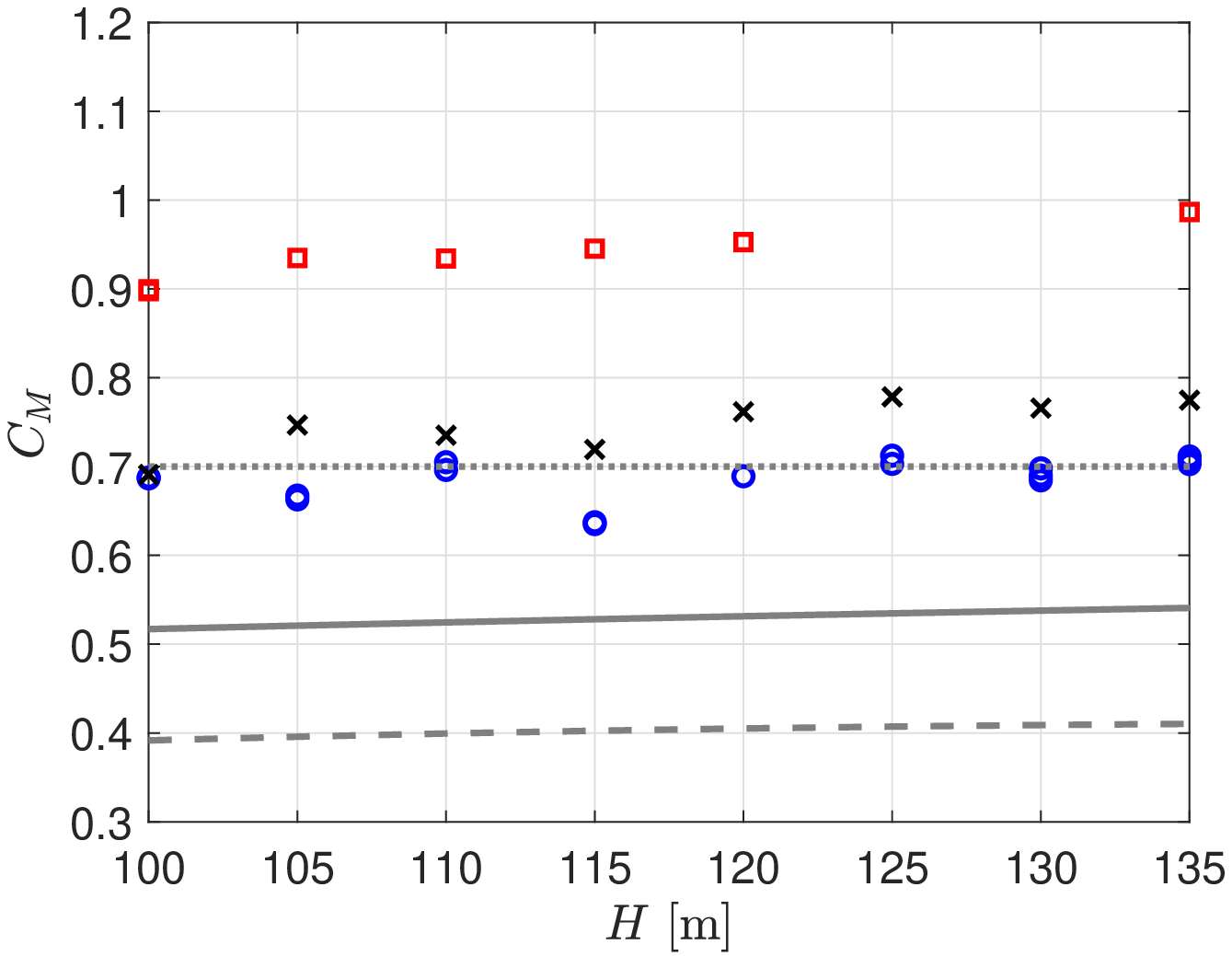}}\hspace{0.0cm}%
	\caption{Mean base force and moment coefficients for the isolated tower of different heights with and without sandpaper strips in both smooth and scale atmospheric boundary layer flow. \textquotedblleft SF\textquotedblright \,and \textquotedblleft TF\textquotedblright \,stand for \textquotedblleft smooth flow\textquotedblright \,and \textquotedblleft turbulent flow\textquotedblright, respectively. The values reported in Eurocode~1 \citep{Eurocodice1}, ESDU \citep{ESDU1986, ESDU1987} and \cite{CICIND2002} recommendations are shown for the sake of comparison. The moment coefficient associated with Eurocode~1 is calculated assuming a center of pressure located at the tower mid-height. The ESDU coefficients refer to the undisturbed wind speed at the top of the tower (like for all the other data sets).}
	\label{fig:S_height}
\end{figure}

\begin{figure}
  \centering
	\subfigure[\label{fig:CF_height_scatter}]
  {\includegraphics[angle=0, width=0.495\textwidth]{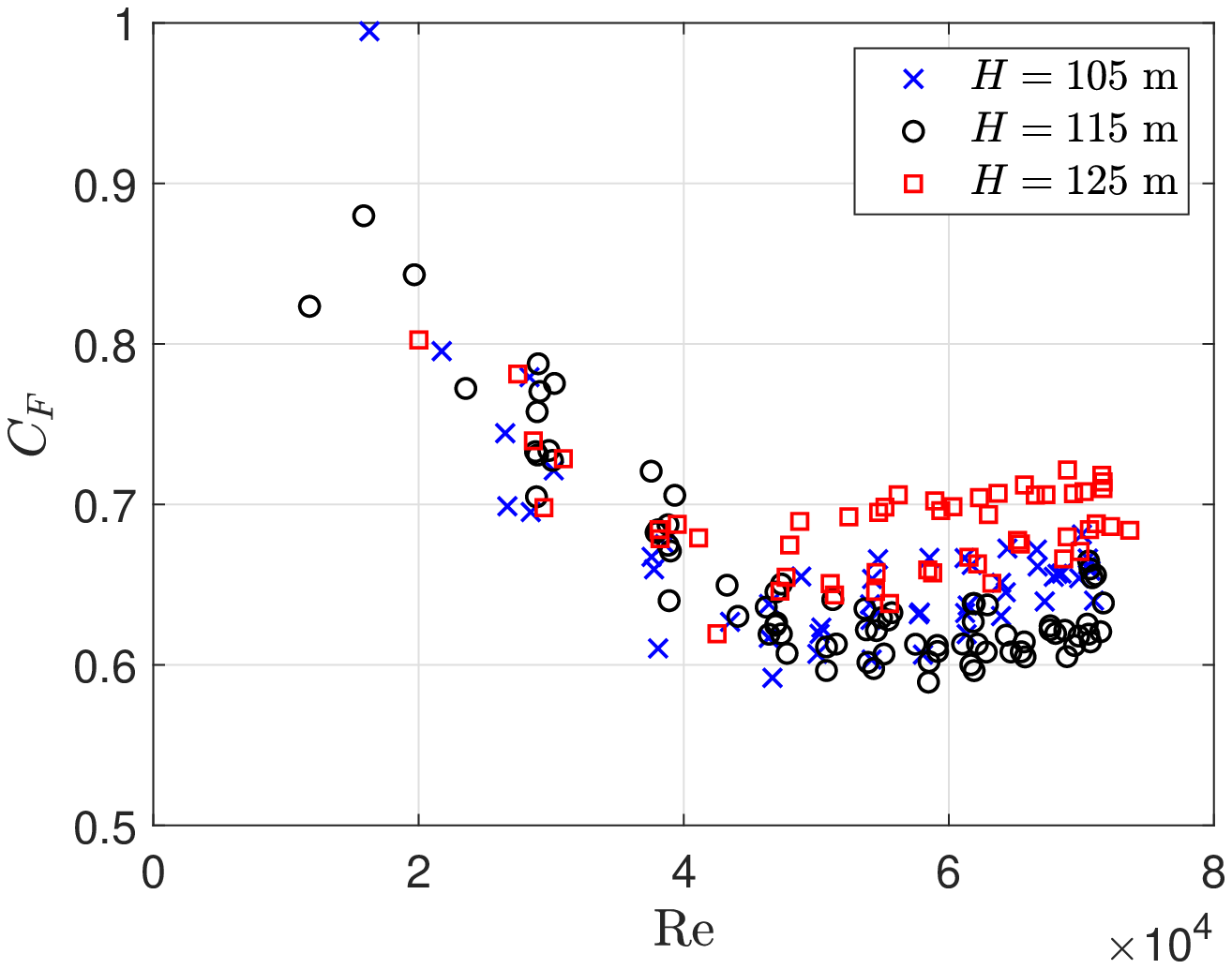}}\hspace{0.0cm}%
	\subfigure[\label{fig:CM_height_scatter}]
  {\includegraphics[angle=0, width=0.495\textwidth]{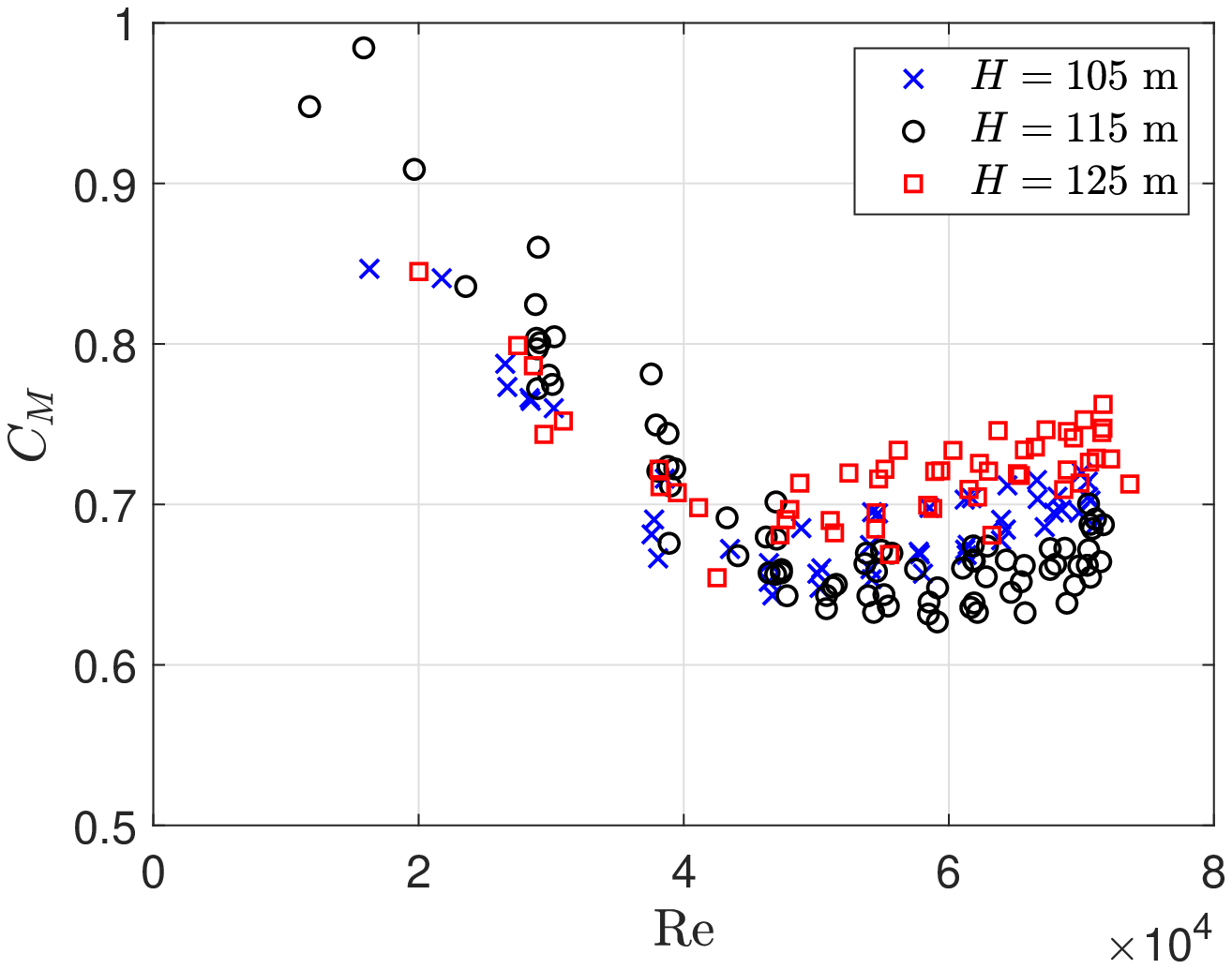}}\hspace{0.0cm}%
	\caption{Mean resultant shear-force (a) and moment (b) coefficients at the base of the isolated tower for three different heights ($H$ = 105~m, $H/D_{eq}$ = 16.0; $H$ = 115~m, $H/D_{eq}$ = 17.6; and $H$ = 125~m, $H/D_{eq}$ = 19.1).}
	\label{fig:CF_CM_height_scatter}
\end{figure}

The force-measurement results shown in Fig.~\ref{fig:S_height} depict however a rather complicated behavior. A slightly increasing trend of mean base shear-force and moment coefficients can be seen, but it is not perfectly monotonic. In particular, $C_F$ and $C_M$ for the baseline 115~m tower are slightly lower than those measured for the 100~m, 105~m and 110~m towers. A similar result is observed in the transcritical regime (towers with surface roughness) in smooth flow, while the trend becomes monotonic in the smooth-flow subcritical regime (towers without added roughness).
In order to understand if this pattern was influenced by the uncertainty in the coefficients highlighted in Section~\ref{S_tower} (although the measurements in Fig.~\ref{fig:S_height} were also repeated a few times), the same procedure that led to Fig.~\ref{fig:CF_CM_free_standing} was also applied to the 105~m and 125~m towers, dismounting and reassembling the test rig several times. The results are reported in Fig.~\ref{fig:CF_CM_height_scatter}, where it is clear that, despite the non-negligible dispersion of the data, the abovementioned behavior is clearly confirmed.

\begin{figure}
  \centering
	\includegraphics[angle=0, width=0.5\textwidth]{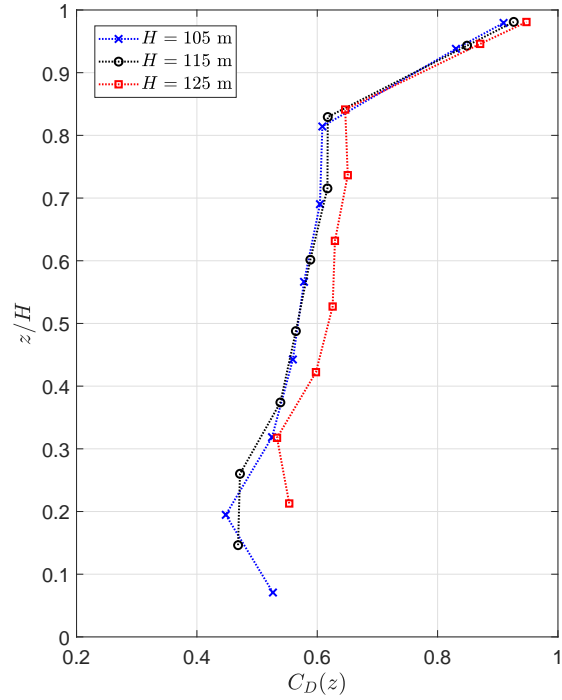}
	\caption{Distribution of the drag coefficient along the isolated tower (always normalized with the mean velocity pressure at a height corresponding to the tower top), obtained from pressure integration for three different tower heights.}
	\label{fig:Pressures_height}
\end{figure}

The complexity of these outcomes suggested a need for verification through pressure measurements. The ABS model equipped with pressure taps was then re-used, achieving equivalent full-scale towers of 105~m, 115~m and 125~m, by adjusting the position of the support below the wind tunnel floor. 
Furthermore, as this verification focused solely on determining the mean pressure coefficients, it was possible to maximize the number of measuring points in a simple but efficient way: since most taps in each pressure array were set 15~deg apart, with up to 24 taps in each array, the model was placed at four angular positions, that is 0, 90, 180 and 270~deg, and then rotated by 7.5~deg for each position (that is 7.5, 97.5, 187.5 and 277.5~deg), exploiting the polar symmetry of the tower and providing a larger number of measuring points.
Moreover, since multiple measurements were associated with each azimuthal position, the repeated data were averaged. Fig.~\ref{fig:Pressures_height} shows how integrated drag coefficient is distributed along the height of the towers, confirming the significant increase in nondimensional load for the tallest 125~m tower, suggesting the same drag coefficient distribution for the two smaller towers. Despite the expected sensitive of such a complicated behavior to small differences in the models for pressure and force measurements, these further experiments are deemed in reasonable agreement with the previous ones.

It can therefore be concluded that the definition of an accurate correction factor for height to be used for various groups of towers is a hard task, and possible engineering solutions should be handled with great attention.

\begin{figure}
  \centering
	\subfigure[\label{fig:G4_height}]%
  {\includegraphics[angle=0, width=0.45\textwidth]{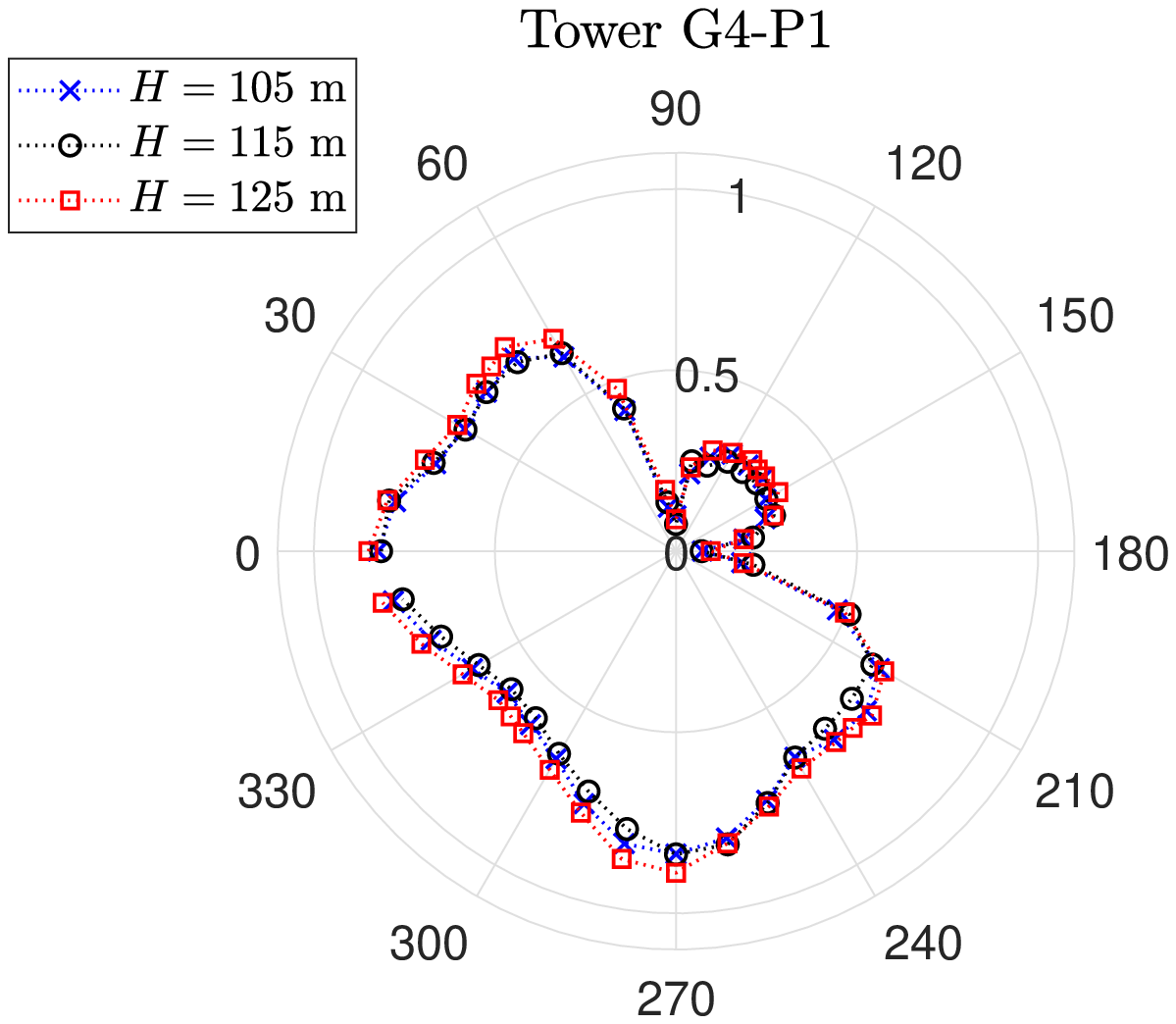}}\hspace{0.5cm}%
	\subfigure[\label{fig:R4-P2_height}]%
  {\includegraphics[angle=0, width=0.45\textwidth]{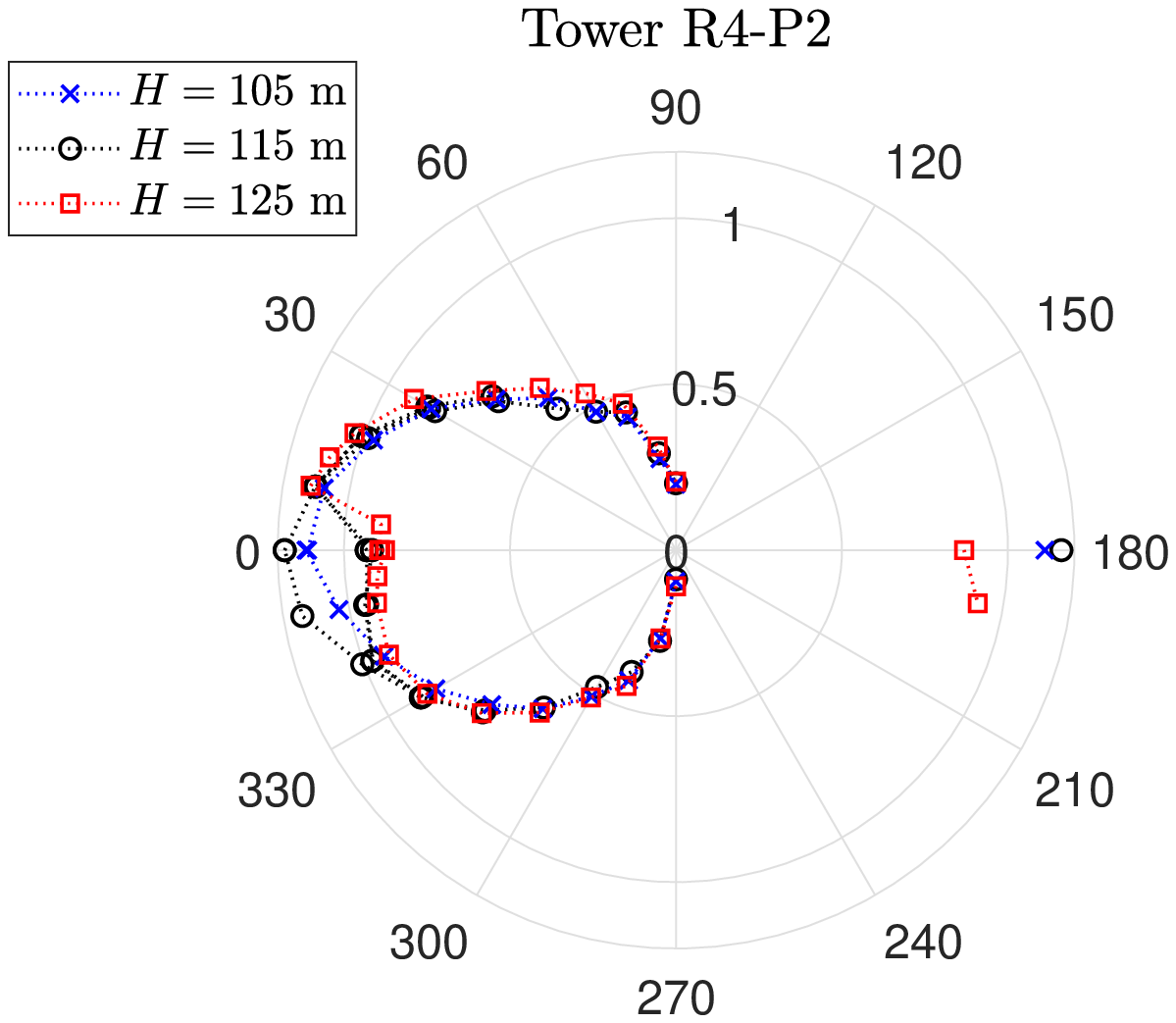}}%
	\caption{Mean resultant moment coefficient at the base of tower P1 in the double-row group G4 (a) and tower P2 in the single-row group R4 (b) for different angles of attack $\beta$ (in deg): comparison of results for three different tower heights.}
	\label{fig:G4_R4_height}
\end{figure}

However, the mean moment coefficients reported in Fig.~\ref{fig:G4_height} for group G4 confirm, to a certain extent, the previous outcomes. Indeed, the loads are non-negligibly higher for the 125~m-high tower compared to the baseline one; in contrast, $C_M$ changes only slightly when tower height is reduced from 115~m to 105~m.
The interpretation of the data is more challenging for the heavily loaded tower R4-P2 (Fig.~\ref{fig:R4-P2_height}), due to the biased flow results for the baseline and the 125~m-high tower but not for the shorter one.

Coming back to Fig.~\ref{fig:S_height}, a few additional comments are in order. The overall trend with the height of the mean base shear-force and moment coefficients complies with Eurocode~1 \citep{Eurocodice1} and ESDU \citep{ESDU1986, ESDU1987} recommendations, although the measured coefficients are significantly higher. After a closer inspection of the data provided by these documents, one can realize which is the origin of the discrepancy in the two cases. As Fig~\ref{fig:Cp_CD} shows, the reference drag coefficient provided by Eurocode~1 for the infinitely long circular cylinder at the target Reynolds number ($C_D = 0.729$ for ${\rm Re} = 2.1 \cdot 10^7$) is very close to that found in the experiments. In contrast, the end-effect factor takes smaller values than those measured in the wind tunnel (for the baseline 115~m-high tower, 0.72 vs. 0.86). This discrepancy is confirmed also by the measurements in smooth flow in both the transcritical and subcritical regimes.
The ESDU document \citep{ESDU1987} too provides coefficients to account for the finite height of the towers and the mean wind speed profile that are lower than those found in the experiments (0.76 for the baseline tower). However, as already noted in Section~\ref{High_Re_data}, the reference drag coefficient for an infinitely long circular cylinder ($C_D = 0.529$, according to ESDU \citep{ESDU1986}) is also significantly lower, thus explaining the apparent discrepancy compared to experiments.
Finally, the \cite{CICIND2002} document provides a single value for loads with no distinction based on a tower's slenderness ratio. However, this value is in very good agreement with the current experimental data.

\begin{figure}
  \centering
	\includegraphics[angle=0, width=0.75\textwidth]{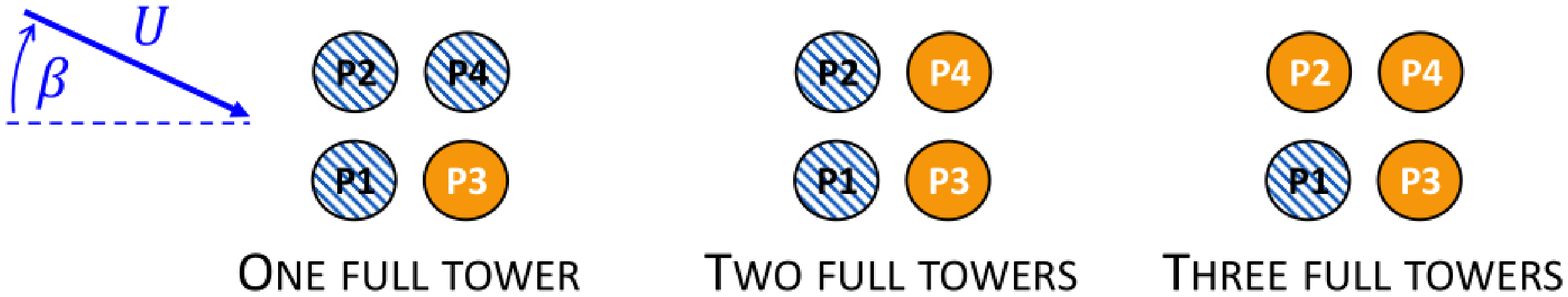}
	\caption{Schematic of possible partially assembled tower configurations for the double-row group G4. Hatched circles denote the partially assembled towers.}
	\label{fig:Half_towers_scheme}
\end{figure}

\begin{figure}
  \centering
	\subfigure
  {\includegraphics[angle=0, width=0.45\textwidth]{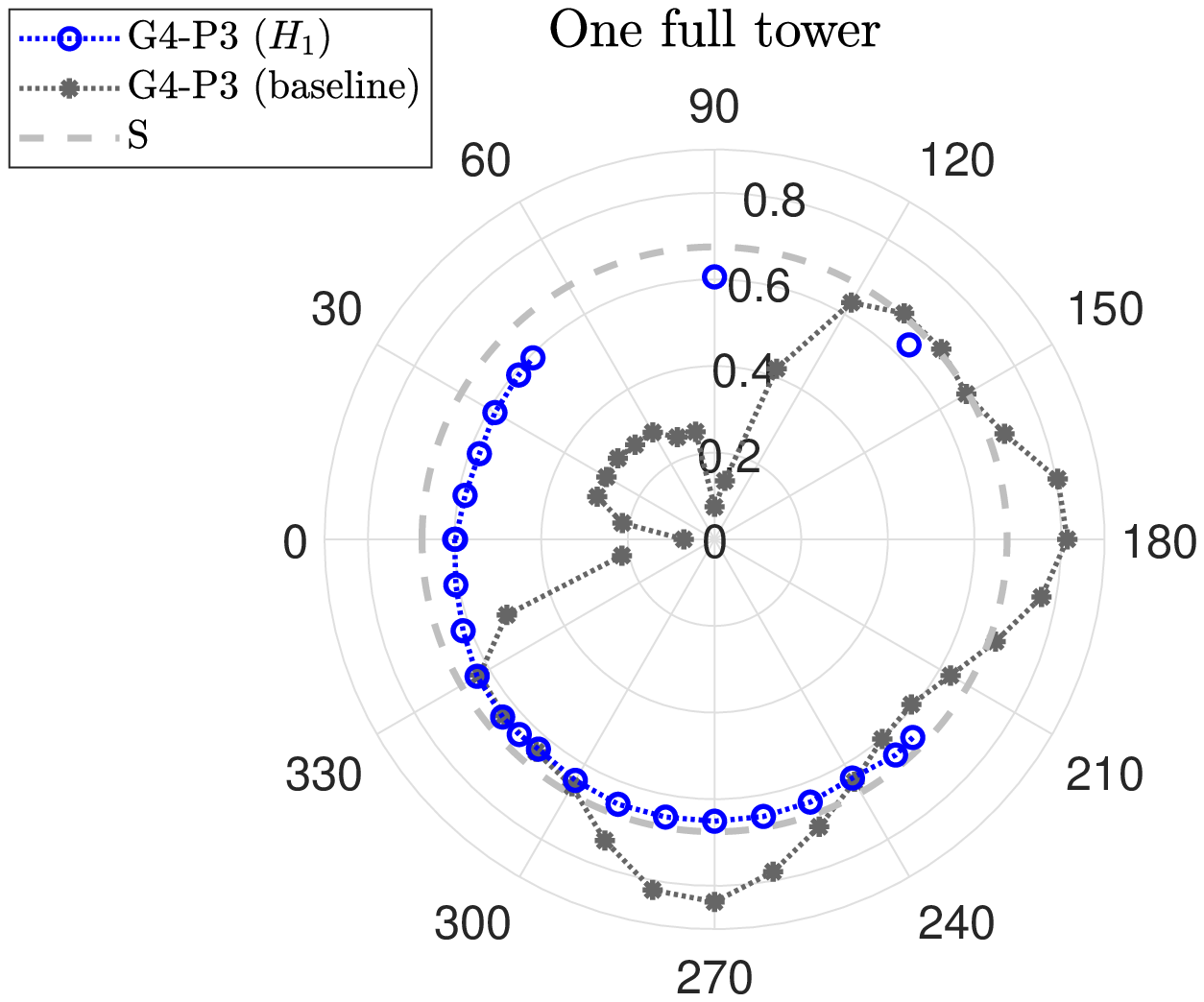}}\hspace{0.5cm}
	\subfigure
  {\includegraphics[angle=0, width=0.45\textwidth]{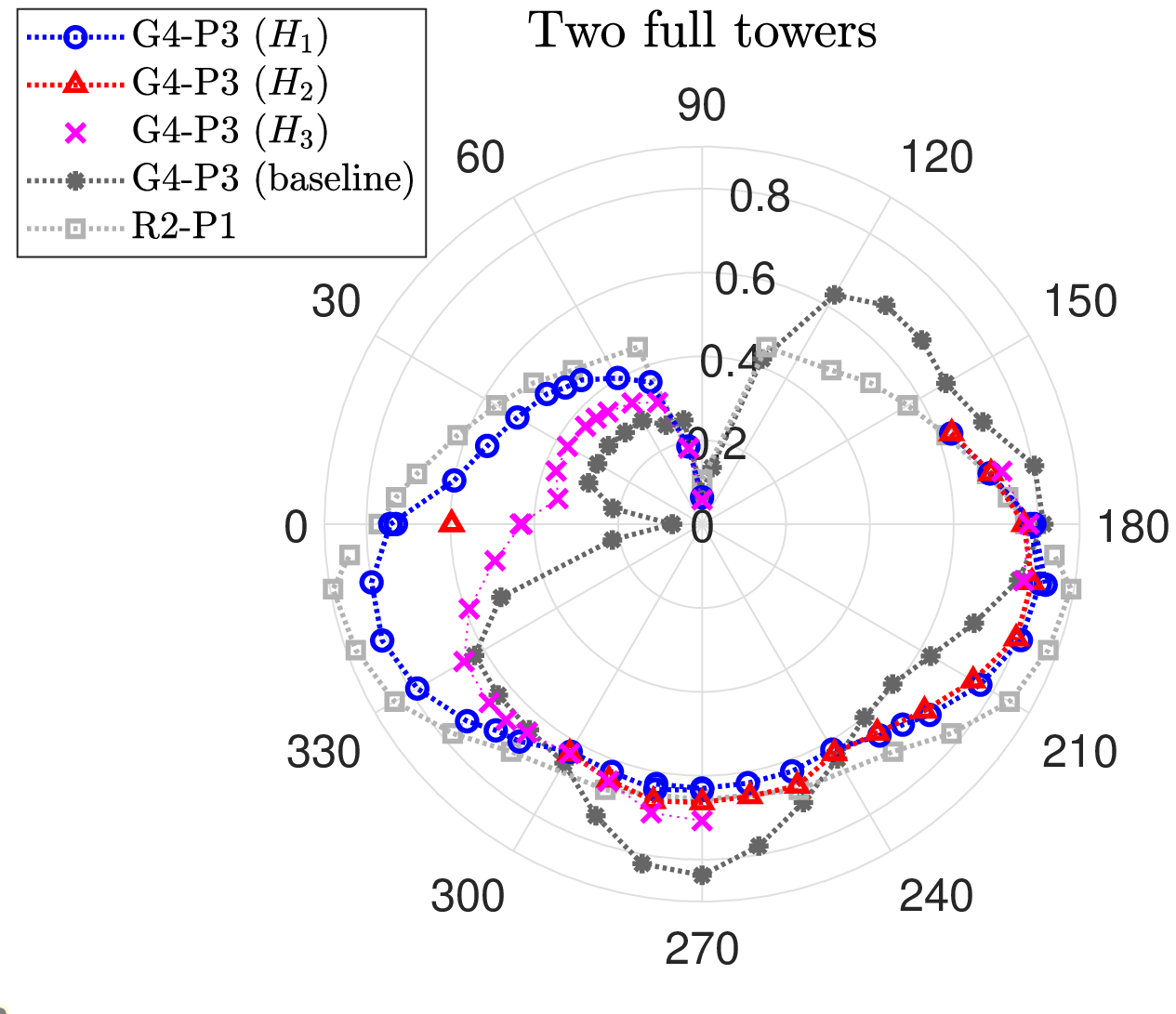}}\hspace{0.5cm}\\
	\subfigure
  {\includegraphics[angle=0, width=0.45\textwidth]{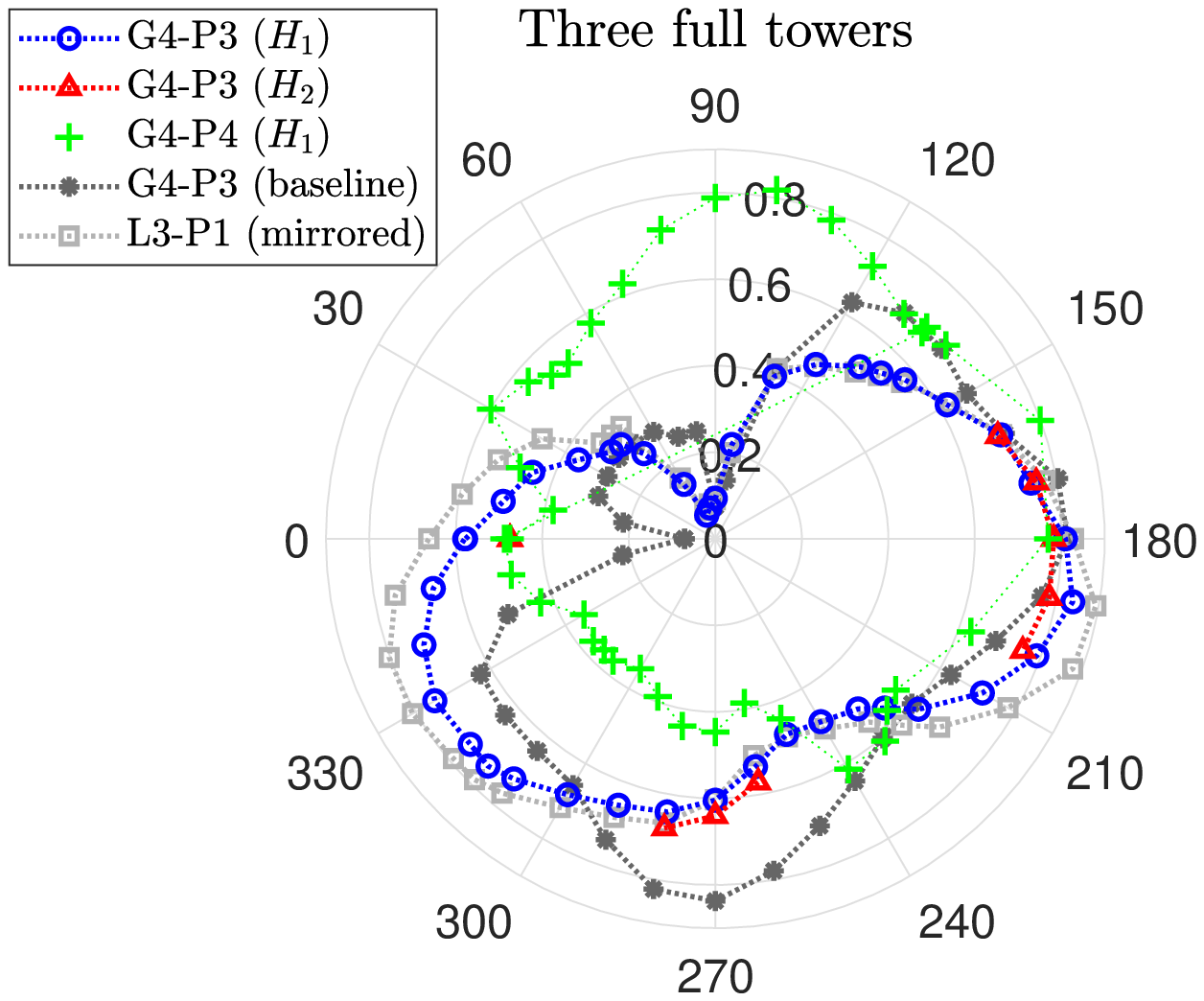}}\vspace{0.5cm}
	\caption{Mean resultant moment coefficient at the base of the full tower P3 (and P4 in one case) in the double-row group G4 in presence of partially assembled towers for different angles of attack $\beta$ (in deg). Diverse levels of completion of the latter are also considered.}
	\label{fig:Half_towers}
\end{figure}

\subsection{Partially assembled towers}

During pre-assembly of wind turbine towers on the quayside, temporary configurations are possible, where some towers are complete while others are only partially assembled.
It is therefore important that the loads in these cases are no higher than in the various group arrangements studied so far, which include full towers only.

The towers in this case study are composed of five segments, so partially assembled towers of three different heights were taken into account. There were $H_1=0.377 \cdot H$, $H_2=0.567 \cdot H$ and $H_3=0.756 \cdot H$, where again $H = 115$~m) were taken into account.
The common G4 group of four towers was considered for this case study, focusing on the loads at the base of tower P3 (also P4 in one configuration), where the group consisted of one, two or three full towers, as shown in Fig.~\ref{fig:Half_towers_scheme}.

The data, in terms of mean resultant moment coefficient $C_M$, are summarized in Fig.~\ref{fig:Half_towers}. The results obtained for some relevant previous configurations of full towers are also reported for the sake of comparison.
The most important consideration one can make is that the maximum mean load on the towers was never higher for partially assembled towers than for full towers only. Moreover, for the most critical wind directions for tower P3, namely for $\beta$ close to either 180~deg or 270~deg, the three heights considered for the partially assembled towers do not significantly influence the results.

Finally, although in the interest of brevity the results are not reported here, the overall loads acting on the group were measured in the case of two full towers (P3 and P4) and two partially assembled towers. The base resultant shear-force and moment coefficients increase with the number of segments of the incomplete towers and are always intermediate between those measured for the groups of just full towers R2 and G4.

\section{Conclusions}

Wind turbine towers pre-assembled on port quayside represent an interesting engineering problem, as they are so sensitive to wind load. For the first time, the current work reports on a broad wind tunnel campaign on various groups of finite-height towers exposed to a boundary layer flow, simulating the transcritical Reynolds number regime that characterizes the full-scale structures.
 
The main conclusions of the work can be highlighted as follows.
\begin{itemize}
	\item Simulation in the wind tunnel of the high Reynolds number regime at full scale is challenging but also crucial. The loads may be significantly overestimated if the subcritical flow regime naturally established in the wind tunnel is accepted for the tests. For the tall isolated towers considered in the current study, the load reduction due to the high Reynolds number (about 24\% for the baseline 115~m tower) is much higher than that due to the atmospheric boundary layer flow (11.4\% compared to the uniform and smooth flow case).
	However, based on the data available in the literature for the infinite circular cylinder, this study also underscores the uncertainty involved in defining the target behavior at very high Reynolds number.
	\item A regular behavior of the loads at the base of each tower is observed for groups composed of two lines of towers. Moreover, the base resultant shear force and overturning moment increase only negligibly for packs of more than eight towers.
	\item In contrast, multiple (either nearly-symmetric or biased) flow configurations were observed for towers arranged in a single row, for flow directions nearly perpendicular to the line of the centers. These high-load layouts therefore require careful great attention during wind tunnel tests. 
	\item Despite the many configurations investigated, a fairly general feature is that the overturning-moment coefficient is associated with a lever arm of the resultant aerodynamic force that is slightly larger than half of the height of the towers.
	\item Analysis of the dynamic loads on the towers revealed that the measured gust factors are very close to those estimated based on Eurocode~1.
	\item The mean resultant of the overall loads on a group of towers is nearly equal to the vector sum of the associated mean Cartesian components. This means that the mean forces and moments acting on the foundation slab can be accurately estimated from the loads measured on each tower in the group. Clearly, this cannot be done for the peak loads.
	\item The mean load coefficients at the base of the towers are not particularly sensitive to the assumed turbulent wind profile, at least for a moderate change in the terrain category. This conclusion complies with the limited impact of the atmospheric boundary layer flow on the load coefficients for the considered isolated towers.
	\item The wind loads on a realistic wind turbine tower with a fairly complex shape can reasonably be studied on a cylindrical tower having a calibrated equivalent diameter.
	\item The influence of tower height was studied in depth but the resultant measured behavior was not consistently monotonic, probably due to the complicated interplay of slenderness ratio, turbulent wind profile and local Reynolds number. Further investigation on the effects of tower height will be necessary in the future. However, the observed end-effect factors are generally higher than those provided by Eurocode~1 and ESDU's recommendations.
	\item The temporary presence of partially assembled towers in the groups does not seem to be crucial in terms of design wind loads.
\end{itemize}

\section*{Acknowledgments}

The authors gratefully acknowledge the help of Niccol\`o Barni and Bernardo Nicese in designing and assembling the ABS tower model equipped with pressure taps.
The authors would also like to thank Mikkel Traberg for promoting the wind tunnel campaign and the collaboration between CRIACIV and Siemens Gamesa Renewable Energy A/S.

\bibliographystyle{unsrtnat}
\bibliography{References_v7}  






\end{document}